\documentclass[reqno,11pt]{amsart}
\usepackage{hyperref}
\usepackage[mathscr]{eucal}
\usepackage{enumerate}
\numberwithin{equation}{section}

\newtheorem{Def}{Definition}[section]
\newtheorem{Thm}[Def]{Theorem}
\newtheorem{Prop}[Def]{Proposition}
\newtheorem{Lemma}[Def]{Lemma}

\newcommand{\beq}{\begin{equation}}
\newcommand{\eeq}{\end{equation}}
\newcommand{\Proof}{\begin{proof}}
\newcommand{\QED}{\end{proof} \noindent}

\newcommand{\msp}{\hspace{-.1cm}}

\newcommand{\mm}{\hspace{-.08cm}\cdot \hspace{-.08cm}}

\newcommand{\R}{\mathbb{R}}

\newcommand{\Gammati}{\tilde{\Gamma}}

\allowdisplaybreaks[1]

\title[RT-equations and Regularity Singularities]{Optimal metric regularity in General Relativity follows from the RT-equations by elliptic regularity theory in $L^p$-spaces} 

\author[M.\ Reintjes]{Moritz Reintjes}
\address{Fachbereich f\"ur Mathematik und Statistik \\ Universit\"at Konstanz \\ D-78467 \\ Germany}
\email{moritzreintjes@gmail.com}
\thanks{M. Reintjes is currently supported by the German Research Foundation, DFG grant FR822/10-1, and was supported by FCT/Portugal through (GPSEinstein) PTDC/MAT-ANA/1275/2014 and UID/MAT/04459/2013 from January 2017 until December 2018.}

\author[B.\ Temple]{Blake Temple \\ \\  October 31, 2020}
\address{Department of Mathematics\\ University of California\\ Davis, CA 95616\\ USA}
 \email{temple@math.ucdavis.edu}

\begin{document}

\begin{abstract}  
Shock wave solutions of the Einstein equations have been constructed in coordinate systems in which the gravitational metric is only Lipschitz continuous, but the connection $\Gamma$ and curvature $Riem(\Gamma)$ are both in $L^{\infty}$, the curvature being one derivative smoother than the curvature of a general Lipschitz metric.  At this low level of regularity, the physical meaning of such gravitational metrics remains problematic.  In fact, the Einstein equations naturally admit coordinates in which $\Gamma$ has the same regularity as $Riem(\Gamma)$ because the curvature transforms as a tensor, but the connection does not.  Here we address the mathematical problem as to whether the condition that  $Riem(\Gamma)$ has the same regularity as $\Gamma$, or equivalently the exterior derivatives $d\Gamma$ have the same regularity as $\Gamma$, is sufficient to allow for the existence of a coordinate transformation which perfectly cancels out the jumps in the leading order derivatives of $\delta\Gamma$, thereby raising the regularity of the connection and the metric by one order--a subtle problem.   We have now discovered, in a framework much more general than GR shock waves, that the regularization of non-optimal connections is determined by a nonlinear system of elliptic equations with matrix valued differential forms as unknowns, the \emph{Regularity Transformation equations}, or \emph{RT-equations}. In this paper we establish the first existence theory for the nonlinear RT-equations in the general case when $\Gamma, {\rm Riem}(\Gamma)\in W^{m,p}$, $m\geq1$,  $n<p< \infty$, where $\Gamma$ is any affine connection on an $n$-dimensional manifold. From this we conclude that for any such connection $\Gamma(x) \in W^{m,p}$ with ${\rm Riem}(\Gamma) \in W^{m,p}$, $m\geq1$, $n<p< \infty$, given in $x$-coordinates, there always exists a coordinate transformation $x\to y$ such that $\Gamma(y) \in W^{m+1,p}$.  This implies {\it all} discontinuities in $m'th$ derivatives of $\delta\Gamma$ {\it cancel out}, the transformation $x\to y$ raises the connection regularity by one order, and $\Gamma$ exhibits optimal regularity in $y$-coordinates.  The problem of optimal regularity for the {\it hyperbolic} Einstein equations is thus resolved by \emph{elliptic} regularity theory in $L^p$-spaces applied to the RT-equations.
\end{abstract}

\maketitle

\section{Introduction}   \label{Sec_intro}

Existence theorems for the Einstein equations are established in coordinate systems in which the equations take on a solvable form. In such coordinates the metric may not exhibit its optimal regularity, that is, two degrees smoother than its Riemann curvature tensor, or may lose its optimal regularity under time evolution \cite{GroahTemple}.  In this paper we give the first  proof of existence of solutions to the Regularity Transformation equations, (RT-equations), equations derived in \cite{ReintjesTemple_ell1} for the Jacobian and transformed connection of the coordinate transformations that map a gravitational metric in General Relativity (GR) to coordinates in which the metric displays its optimal regularity.\footnote{The results in this paper and \cite{ReintjesTemple_geo,ReintjesTemple_ell1,ReintjesTemple_ell4} are summarized in the RSPA paper \cite{ReintjesTemple_ell3}. The methods developed in this paper are the starting point for proving  optimal regularity and Uhlenbeck compactness for $L^\infty$ connections in \cite{ReintjesTemple_ell4}.}    This is a new approach to optimal metric regularity in GR because, rather than imposing an apriori coordinate ansatz, (like harmonic coordinates \cite{DeTurckKazdan,Anderson} or Gaussian normal coordinates \cite{Israel,SmollerTemple}), and trying to establish regularity of solutions of the Einstein equations in those coordinates, our premise here is that, in general, the coordinate systems of optimal regularity are too difficult to guess apriori, or the Einstein equations  too difficult to solve, and to find them, one has to discover and solve equations for the coordinates themselves.  In \cite{ReintjesTemple_ell1} the authors accomplished their goal of deriving such a system of equations, the {\it RT-equations}. The RT-equations are a system of elliptic PDE's derived from a geometric principle, the {\it Riemann-flat condition}, which the authors introduced in \cite{ReintjesTemple_geo}. Authors' motivation to study optimal metric regularity in GR began by asking whether shock wave solutions with Lipschitz continuous metric, proven to exist in Standard Schwarzschild coordinates, might actually be one order smoother in other coordinate systems in which the Einstein equations are too complicated to solve,  \cite{Reintjes,ReintjesTemple1,ReintjesTemple_wave,ReintjesTemple_geo,ReintjesTemple_ell1}.   This has led us to a much more general theory of optimal regularity for solutions of the Einstein equations based on the RT-equations. In Section \ref{Sec_appl} below we conjecture that without resolving the problem of optimal regularity, the existence theory for the initial value problem in GR is incomplete in each Sobolev Space, an issue at the foundation of the initial value problem in General Relativity.    

In this paper we apply elliptic regularity theory in $L^p$ spaces to give the first proof of existence of solutions to the RT-equations. From this we deduce the following theorem for geometry: 

\begin{Thm}  \label{Cor1}
Let $\Gamma$ be a connection and $Riem(\Gamma)$ its curvature tensor given by components $\Gamma^i_{jk}$ and $R^i_{jkl}$ in some coordinate system $x$ defined in an open set $\Omega \subset \R^n$. Assume all components satisfy $\Gamma^i_{jk}, R^i_{jkl} \in W^{m,p}(\Omega)$  for $m\geq 1$, $n<p< \infty$, $n\geq2$.  Then for each point $q \in \Omega$ there exists a neighborhood $\Omega_q \subset \Omega$ containing $q$, and a coordinate transformation $x \to y$ with $J^\mu_i \equiv\frac{\partial y^{\mu}}{\partial x^i}\in W^{m+1,p}(\Omega_q)$, such that, in $y$-coordinates, the components of $\Gamma$ are bounded in $W^{m+1,p}(\Omega_q)$.
\end{Thm}

Theorem \ref{Cor1} applies to general connections, including metric connections of arbitrary metric signature, and is applicable to solutions of the Einstein equations with arbitrary sources.  The result does not rely on special properties of the Einstein equations. It establishes that no {\it regularity singularities} exist when the curvature is in $W^{1,p}$ (c.f. \cite{ReintjesTemple1}).  Authors' current research program is to extend this existence theory for the RT-equations to Lipschitz continuous metrics, when $\Gamma,{\rm Riem}(\Gamma)\in L^\infty,$ and by this resolve the problem as to whether regularity singularities can be created by shock wave interaction in General Relativity.

Theorem \ref{Cor1} introduces a new point of view on solutions of the Einstein equations of General Relativity: It tells us that it is sufficient to solve the Einstein equations in coordinates in which the metric is only {\it one order} smoother than the curvature, allowing for equations which are only {\it first order} in metric components. Then by Theorem \ref{Cor1} we know local coordinate transformations always exist which smooth the metric by one order, to optimal regularity, two derivatives smoother than the curvature.  Since first order equations can be simpler than second order equations, Theorem 1.1 establishes that it is sufficient to solve the Einstein equations in coordinates in which the equations are simpler, and the solutions are weaker, and conclude in general that once the existence of weaker solutions is established, the stronger solutions with optimal regularity are guaranteed. Indeed, the Einstein equations naturally allow for solutions in which the metric regularity is only one level higher than that of its Riemann curvature, and hence not optimal. For example, given a solution of the Einstein equations of optimal regularity, say with metric in $W^{m+2,p}$ and its curvature in $W^{m,p}$, $m\geq0$, then applying a coordinate transformation with Jacobian in $W^{m+1,p}$, the resulting metric is no longer optimal, being in $W^{m+1,p}$, with connection dropping to $W^{m,p}$, and curvature remaining in $W^{m,p}$, \cite{ReintjesTemple_ell1}. Theorem \ref{Cor1} establishes that this can always be reversed in the case $m\geq 1$, $n<p< \infty$, which is essentially one derivative above the GR shock wave case ${\rm Riem}(\Gamma)\in L^{\infty}$. Theorem \ref{Cor1} guarantees that if in the time evolution of any such GR solution of optimal smoothness, the regularity breaks down by the metric losing one derivative relative to its curvature tensor, then this is only a breakdown in the coordinate system, not in the geometry.   This can be taken as a new regularity principle for the numerical simulation of solutions in GR. In particular, excluding non-optimal solutions from the initial value problem when they exist would lead to an incomplete picture of the solution space of the Einstein equations in every regularity class, and hence an incomplete picture of the underlying physics, (c.f. Section \ref{Sec_appl} below). 

As an application, Theorem \ref{Cor1} resolves the problem of optimal regularity for spherically symmetric solutions constructed in Standard Schwarzschild Coordinates (SSC) in the case $m\geq 1$, $n<p< \infty$ (c.f. Section \ref{Sec_appl}).  Non-optimal shock wave solutions constructed by the Glimm scheme in SSC, \cite{GroahTemple,Smoller}, have $\Gamma,Riem(\Gamma)\in L^{\infty}$, and solving the RT-equations in this case of lower regularity remains an open problem.   The example of SSC tells us that the spacetime metric can be expressed in a simpler, and more comprehensible form, within an atlas of coordinate systems in which the gravitational metric is one order less regular than optimal.   The RT-equations provide an explicit algorithm, amenable to numerics,  for constructing coordinate systems which display the optimal regularity of such metrics. In the case of shock waves, such transformations to coordinates of optimal regularity convert weak solutions of $G=\kappa T$, to strong solutions.

Theorem \ref{Cor1} resolves the problem of optimal metric regularity at the level of connections and curvatures in $W^{m,p}$, $m\geq1$, one order larger than the case $L^{\infty}$ (or $L^p$), applicable to shock wave theory in GR, \cite{GroahTemple,Reintjes,ReintjesTemple1,ReintjesTemple_wave,ReintjesTemple_geo,ReintjesTemple_ell1}. The case of $L^{\infty}$ (or $L^p$) curvature is the threshold between weak and strong solutions of the Einstein equations, c.f. \cite{ReintjesTemple1,ReintjesTemple_ell1}. However, even in the $L^{\infty}$ case, the equivalence between the existence of coordinate systems of optimal metric regularity and the existence of solutions of the RT-equations still applies.  There are two main obstacles to extending Theorem \ref{Cor1} to the case of $L^{\infty}$ curvature. First is the problem of Calderon-Zygmund singularities, the central issue in the $L^{\infty}$ case of elliptic regularity theory \cite{ReintjesTemple_ell1},  and second, the problem of handling nonlinear products in $L^p$. Obstacles to solving the RT-equations in the case of GR shock waves could lead to the discovery of new kinds of {\it regularity singularities} in GR \cite{ReintjesTemple1,ReintjesTemple_wave}. The $L^{\infty}$ case is the setting most intriguing to the authors, and the problem of extending solutions of the RT-equations to the lower regularity of $L^{\infty}$ (and $L^p$) will be addressed in forthcoming publications.\footnote{Since the writing of this paper, the $L^\infty$ case has been resolved in \cite{ReintjesTemple_ell4} by extending the existence theory for the RT-equations in this paper to the case of $L^\infty$ connections.} Theorem \ref{Cor1} demonstrates for the first time that determining optimal metric regularity by the RT-equations \emph{works} and the RT-equations bring elliptic regularity theory to bear on the problem of optimal regularity in General Relativity.

The point of departure for this paper is Theorem \ref{Thm_RT-eqn} below, proven in \cite{ReintjesTemple_ell1}, which establishes the equivalence of the Riemann-flat condition with the solvability of the RT-equations when $\Gamma$ and $d\Gamma\in W^{m,p}$, for $m\geq1$, $n<p< \infty$, (and hence ${\rm Riem}(\Gamma)\in W^{m,p}$ by Morrey's inequality \eqref{Morrey_textbook} applied to the identity $Riem(\Gamma)= d\Gamma + \Gamma \wedge \Gamma$, c.f. Section \ref{Sec_Prelim} below).  By this we mean the components of $\Gamma$ and $d\Gamma$ are  functions in $W^{m,p}$ in some given, but otherwise arbitrary, coordinate system $x$.  The Riemann-flat condition was derived in \cite{ReintjesTemple_geo} as a condition on a given connection $\Gamma$ equivalent to the existence of a local coordinate transformation which smooths the connection by one order.  The Riemann-flat condition states that there should exist a tensor $\Gammati$, one order smoother than $\Gamma$, such that $Riem(\Gamma-\Gammati)=0$.       It applies to connections down to the lowest regularity $\Gamma,d\Gamma\in L^{\infty}$, and in this case the theorem in \cite{ReintjesTemple_geo} states that there exists a coordinate transformation with Jacobian $J$ which smooths the components of $\Gamma$ to $C^{0,1}$ if and only if there exists a tensor $\Gammati\in C^{0,1}$ such that $Riem(\Gamma-\Gammati)=0$. It turns out that $\Gammati$ agrees with the smoothed connections in the new coordinates. 

The RT-equations, equations \eqref{PDE1} - \eqref{PDE4} below, were derived in \cite{ReintjesTemple_ell1}. The $J$ and $\Gammati$ components of the RT-equations come from two equivalent forms of the Riemann-flat condition, namely, $Riem(\Gamma-\Gammati)=0$ and $dJ=J(\Gamma-\Gammati)$. In the derivation, these two first order equations are converted into the RT-equations by use of the identity $\Delta \equiv d \delta + \delta d$ to re-express the first order equations as second order Poisson equations in the Laplacian $\Delta$ of the Euclidean coordinate metric. These two equations are augmented by a system of first order equations in auxiliary variables $A$ to arrange for the integrability condition $Curl(J)\equiv \partial_j J^\mu_i - \partial_i J^\mu_j =0$, c.f. \cite{ReintjesTemple_ell1}. The resulting unknowns in the RT-equations are then the matrix valued differential forms $(\Gammati,J,A)$ which have the following meaning: $J\equiv J^\mu_\nu$ is the Jacobian of the sought after coordinate transformation which smooths the connection, viewed as a matrix valued $0$-form;  $\Gammati$ is the unknown tensor one order smoother than $\Gamma$ such that $Riem(\Gamma-\Gammati)=0$, viewed as a matrix valued $1$-form, $\Gammati\equiv \Gammati^\mu_{\nu k}dx^k$; and $A\equiv A^\mu_\nu$ is an auxiliary matrix valued $0$-form introduced to impose $Curl( J)=0$. Also, $\vec{A} \equiv A^\mu_i dx^i$ and $\vec{J} \equiv J^\mu_i dx^i$ are vector valued $1$-forms, the  {\it vectorizations} of $A$ and $J$, introduced so that $Curl(J) = d\vec{J}$ and the integrability condition takes the form $d\vec{J}=0$, which allows us to augment the above two Riemann-flat conditions by an equation for $A$, resulting in the RT-equations, c.f. \cite{ReintjesTemple_ell1}. 
(Authors find the interplay between the interpretation of the Jacobian as a matrix valued $0$-form  $J$, to re-express the Riemann-flat condition, and its interpretation as a vector valued $1$-form $\vec{J}$, required to incorporate the integrability condition and close the RT-equations at the correct regularity, very interesting.) The point of departure for this paper is the following theorem, proven in \cite{ReintjesTemple_ell1}.

\begin{Thm} \label{Thm_RT-eqn}
Assume $\Gamma$ is defined in a fixed coordinate system $x$ on $\Omega$, where $\Omega\subset{\mathbb R}^n$ is a bounded open set with smooth boundary.   Assume that $\Gamma\in W^{m,p}(\Omega)$ and $d\Gamma\in W^{m,p}(\Omega)$ for $m\geq 1,$ $n<p< \infty$. Then the following equivalence holds: \vspace{.15cm} \newline 
If there exists a coordinate transformation $x \to y$ with Jacobian $J = \frac{\partial y}{\partial x} \in W^{m+1,p}(\Omega)$ such that the components of $\Gamma$ in $y$-coordinates are in $W^{m+1,p}(\Omega)$, then there exists $\Gammati \in W^{m+1,p}(\Omega)$ and $A\in W^{m,p}(\Omega)$ such that $(J,\Gammati,A)$ solve the elliptic system
\begin{eqnarray} 
\Delta \Gammati &=& \delta d \big( \Gamma - J^{-1} dJ \big) + d(J^{-1} A ), \label{PDE1} \\
\Delta J &=& \delta ( J \mm \Gamma ) - \langle d J ; \tilde{\Gamma}\rangle - A , \label{PDE2} \\
d \vec{A} &=& \overrightarrow{\text{div}} \big(dJ \wedge \Gamma\big) + \overrightarrow{\text{div}} \big( J\, d\Gamma\big) - d\big(\overrightarrow{\langle d J ; \tilde{\Gamma}\rangle }\big),   \label{PDE3}\\
\delta \vec{A} &=& v,  \label{PDE4}
\end{eqnarray}
with boundary data 
\begin{eqnarray}   
d\vec{J} =0 \ \ \text{on} \ \partial \Omega.  \label{BDD1}
\end{eqnarray}
Here $v\in W^{m-1,p}(\Omega)$ is a vector valued $0$-form free to be chosen. \vspace{.15cm} \newline 
Conversely, if there exists $J \in W^{m+1,p}(\Omega)$ invertible, $\Gammati \in W^{m+1,p}(\Omega)$  and $A\in W^{m,p}(\Omega)$ which solve \eqref{PDE1} - \eqref{BDD1} in $\Omega$, then for each $q\in\Omega$,  there exists a neighborhood $\Omega_q\subset\Omega$ of $q$ such that $J$ is the Jacobian of a coordinate transformation $x \to y$ on $\Omega_q$, and the components of $\Gamma$ in $y$-coordinates are in $W^{m+1,p}(\Omega_q)$.
\end{Thm}

We call system \eqref{PDE1} - \eqref{PDE4} the \emph{Regularity Transformation equations}, or \emph{RT-equations}. The principal parts are  the Laplacian $\Delta = \partial^2_{x^1} + ... + \partial^2_{x^n}$, the exterior derivative $d$, and the co-derivative $\delta$, all taken with respect to the \emph{Euclidean} metric in $x$-coordinates as an auxiliary Riemannian structure. By this, the RT-equations are elliptic. The operations $\vec{\cdot},$ $\overrightarrow{\rm div}$ and $\langle \cdot \; , \cdot \rangle$ are introduced in \cite{ReintjesTemple_ell1} as special operations on matrix valued differential forms, c.f. \eqref{def_matrixvalued_diff-form} - \eqref{colon} below. Note that the vector valued $0$-form $v$ (which is free to be chosen) has been introduced in \eqref{PDE4} so that (\ref{PDE3})-(\ref{PDE4}) takes the Cauchy-Riemann form $d\vec{A}=f$, $\delta\vec{A}=g$.   The consistency condition $df=0$ is met in (\ref{PDE3}) because the derivation of the RT-equations shows the right hand side is exact, (equation (\ref{PDE3}) is obtained by setting $d$ of the ``vectorized'' right hand side of (\ref{PDE2}) equal to zero, c.f. equation (3.40) in \cite{ReintjesTemple_ell1}), and $\delta g=0$ holds in (\ref{PDE4}) because $\delta v=0$ is an identity for vector valued $0$-forms $v$. 

The RT-equations apply to connections of arbitrary metric signature, and in particular to solutions of the Einstein equations with arbitrary sources. In this paper we establish the first existence theory for the RT-equations by proving existence of solutions when $\Gamma, d\Gamma\in W^{m,p}$, $m\geq1$, $n<p< \infty$.  The proof is based on a new iteration scheme which applies the linear theory of elliptic PDE's at each stage, to obtain solutions of the nonlinear RT-equations in the limit. The iteration scheme approximates the first two RT-equations \eqref{PDE1} and \eqref{PDE2} component-wise by Poisson equations and the third and fourth RT-equations \eqref{PDE3} - \eqref{PDE4} as Cauchy-Riemann type equations (of form $d\vec{A}=f$ and $\delta\vec{A}=g$) by replacing the unknowns on the right hand side of the RT-equations by the previous iterates; linear elliptic PDE theory applies to both equations and provided the estimates required to prove converegence, c.f. Section \ref{Sec_Prelim} below.  A key insight for the proof was to augment the RT-equations by ancillary elliptic equations in order to convert the non-standard boundary condition $Curl(J)=0$, which is of neither Neumann nor Dirichlet type, into Dirichlet data for $J$ at each stage of the iteration, c.f. Section \ref{Sec_extended_RT}. By this, each iterate can be constructed by applying standard existence theorems and elliptic regularity in $L^p$ spaces for the linear Poisson equation. Our main existence theorem is the following:

\begin{Thm}  \label{ThmMain}
Assume the components of $\Gamma, d\Gamma \in W^{m,p}(\Omega)$ for $m\geq 1$, $n<p< \infty$, $n\geq2$ in some coordinate system $x$. Then for each $q\in\Omega$ there exists a solution $(\Gammati,J,A)$ of the RT-equations \eqref{PDE1} - \eqref{BDD1} defined in a neighborhood $\Omega_q$ of $q$ such that 
$\Gammati\in W^{m+1,p}(\Omega_q),$  $J\in W^{m+1,p}(\Omega_q),$ $A\in W^{m,p}(\Omega_q)$.
\end{Thm}

Theorem \ref{Cor1} follows directly from Theorem \ref{Thm_RT-eqn} together with Theorem \ref{ThmMain}, and requires no further proof. (Observe that $d\Gamma\in W^{m,p}$ is equivalent to ${\rm Riem}(\Gamma)\in W^{m,p}$,  when $m\geq1$, $n<p< \infty$ by Morrey's inequality \eqref{Morrey_textbook} applied to $Riem(\Gamma)= d\Gamma + \Gamma \wedge \Gamma$, c.f. \cite{ReintjesTemple_ell1}.) The proof of Theorem \ref{ThmMain} is the subject of the remainder of this paper. This is a new application of the theory of elliptic regularity in $L^p$ spaces developed by Agmon, Nierenberg and others in the $'50$, at the time connecting the new theory of distributions to solutions of PDE's \cite{Agmon}. Interestingly, the analysis of the RT-equations {\it requires} $L^p$ spaces, and this cannot be replaced by the simpler $L^2$ theory because of non-linear products, nor by a Green's function approach which would require higher regularity. Most interesting to us is that one can address the problem of optimal regularity of solutions of the \emph{hyperbolic} Einstein equations by \emph{elliptic} regularity theory alone.

The structure of this paper is as follows: In Section \ref{Sec_Prelim} we give preliminaries and state the results we require from elliptic regularity theory in $L^p$ spaces. In Section \ref{Sec_extended_RT} we show how to augment the RT-equations by ancillary equations in order to reduce the boundary condition \eqref{BDD1} to standard Dirichlet data. In Section \ref{Sec_iteration} we set up the iteration scheme and we introduce a small parameter $\epsilon$ into the RT-equations to handle the non-linearities. In Section \ref{Sec_convergence} we outline the proof of convergence of our iteration scheme for the $\epsilon$ rescaled RT-equations. Section \ref{Sec_Proofs} contains the detailed proofs of the technical lemmas stated in Section \ref{Sec_convergence} from which the proof of convergence of the iteration scheme is deduced. In Section \ref{Sec_proof_ThmMain} we complete the proof of Theorem \ref{ThmMain}, by proving that the $\epsilon$ rescaled RT-equations can always be obtained by restricting to small neighborhoods. In Section \ref{Sec_appl} we discuss the initial value problem, and an application of Theorem \ref{Cor1} to spherically symmetric solutions of the Einstein equations in Standard Schwarzschild Coordinates.

\section{Preliminaries} \label{Sec_Prelim}

The point of departure for this paper is authors' prior paper \cite{ReintjesTemple_ell1}, and we refer the reader to this for more details on notation, motivation, and background. We now recall several definitions and identities from Section 2.1 in \cite{ReintjesTemple_ell1}. To begin, recall that we work in a fixed (but arbitrary) coordinate system $x$ defined on $n$-dimensional bounded open set $\Omega\subset \R^n$ with smooth boundary. The unknowns in the RT-equations are matrix valued differential forms.  By a matrix valued differential $k$-form $A$ we mean an $(n\times n)$-matrix whose components are $k$-forms, and we write
\beq \label{def_matrixvalued_diff-form}
A = A_{[i_1...i_k]} dx^{i_1} \wedge ... \wedge dx^{i_k} \equiv \sum_{i_1< ... < i_k} A_{i_1...i_k} dx^{i_1} \wedge ... \wedge dx^{i_k},
\eeq 
for $(n\times n)$-matrices  $A_{i_1...i_k}$  that are totally anti-symmetric in the indices $i_1,...,i_k \in \{1,...,n\}$. We define the wedge product of a matrix valued $k$-form $A$ with a matrix valued $l$-form $B = B_{j_1...j_l} dx^{j_1} \wedge ... \wedge dx^{j_l}$ as  
\begin{eqnarray} \label{def_wedge}
A \wedge B  
&\equiv & \frac{1}{l!k!} A_{i_1...i_k} \mm B_{j_1...j_l} \; dx^{i_1} \wedge ... \wedge dx^{i_k} \wedge dx^{j_1} \wedge ... \wedge dx^{j_l}, 
\end{eqnarray}
where the dot denotes standard matrix multiplication. Note the wedge product of a matrix valued $k$-form with itself is non-zero unless the component matrices commute, which is the main difference between matrix valued and scalar valued differential forms. The exterior derivative $d$ and its co-derivative $\delta$ are defined component-wise on matrix-components, with respect to the Euclidean metric in $x$-coordinates, so all properties of $d$ and $\delta$ on scalar forms carry over to matrix valued forms. In particular the Laplacian $\Delta \equiv d \delta + \delta d$ acts component-wise on matrix- and on $k$-form components and, in fact, $\Delta$ is identical to the Laplacian of the Euclidean metric in $x$-coordinates, $\Delta = \partial^2_{x^1} + ... + \partial^2_{x^n}$. The exterior derivative satisfies the product rule 
\beq \label{Leibniz_rule} 
d(A\wedge B) = dA \wedge B +(-1)^k  A \wedge dB  ,
\eeq
where $A\in W^{1,p}(\Omega)$ is a matrix valued $k$-form and $B \in W^{1,p}(\Omega)$ is a matrix valued $j$-form, (c.f. Lemma 3.3 of \cite{ReintjesTemple_ell1}), which implies for a matrix valued $0$-form $J$ that 
\beq \label{Leibnitz-rule}
d\big( J^{-1} \mm dJ \big) = d(J^{-1}) \wedge dJ = -  J^{-1} d J \wedge J^{-1} dJ.
\eeq
Regarding the co-derivative $\delta$, we require the following product rule
\beq \label{colon}
\delta (J\mm w ) = J \mm \delta w  + \langle d J ;  w \rangle
\eeq
where $J\in W^{2,p}(\Omega)$ is a matrix valued $0$-form, $w \in W^{2,p}(\Omega)$ a matrix valued $1$-form, and where $\langle \cdot\; ; \cdot \rangle $ is the matrix valued inner product defined on matrix valued $k$-forms $A$ and $B$ by,
\beq \label{def_inner-product}
\langle A\; ; B \rangle^\mu_\nu \equiv \sum_{i_1<...<i_k} A^\mu_{\sigma\: i_1...i_k} B^\sigma_{\nu\: i_1...i_k}.
\eeq
So $\langle A\; ; B \rangle$ converts two matrix valued $k$-forms into a matrix valued $0$-form. As shown in \cite{ReintjesTemple_ell1}, the Riemann curvature tensor is a matrix valued $2$-form and is given by $Riem(\Gamma)= d\Gamma + \Gamma \wedge \Gamma$ in $x$-coordinates.

The two operations which convert matrix valued differential forms to vector valued forms on the right hand side of (\ref{PDE4}) are $vec$ and $vec-divergence$.   First, $vec$ converts matrix valued $0$-forms into vector valued $1$-forms by the operation, (c.f. (2.20) of \cite{ReintjesTemple_ell1}), 
\beq \label{Def_vec-div}
\vec{A}^\mu =A^\mu_idx^i.
\eeq
The operation $vec-divergence$ converts matrix valued $k$-forms $A$ into vector valued $k$-forms $\overrightarrow{\text{div}}(A)$ by the operation  
\beq \nonumber
\overrightarrow{\text{div}}(A)^\alpha \equiv \sum_{l=1}^n \partial_l \big( (A^\alpha_l)_{i_1...i_k}\big) dx^{i_1}\wedge . . . \wedge dx^{i_k}.
\eeq
For a matrix valued $1$-form $w$ and a matrix valued $0$-form $J$, Lemma 2.4 of \cite{ReintjesTemple_ell1} gives the important identity
\beq \label{regularity-miracle}
d \big(\overrightarrow{\delta ( J \mm w )}\big) 
= \overrightarrow{\text{div}} \big(dJ \wedge w\big) + \overrightarrow{\text{div}} \big( J\mm dw\big) ,
\eeq
which is crucial for the regularity to close in the RT-equations, c.f. Section~1 in \cite{ReintjesTemple_ell1}.

We denote by $\|\cdot \|_{W^{m,p}(\Omega)}$ the standard $W^{m,p}$-norm, defined as the sum of the $L^p$-norms of derivatives up to order $m$ \cite{Evans}. (We often write $\partial^m$ for such derivatives in place of multi-index notation.)  When applied to matrix valued differential forms $\omega$,  $\|\omega\|_{W^{m,p}(\Omega)}$ denotes the sum of the $W^{m,p}$-norm applied to all matrix- and differential form-components of $\omega$. Theorems \ref{Thm_RT-eqn} and \ref{ThmMain} apply to connections in the space $W^{m,p}$ for $m\geq1$, $p>n$, because for these parameter values, Sobolev's theorem implies that $W^{1,p}(\Omega)$ is embedded in the space of H\"older continuous functions $C^{0,\alpha}(\overline{\Omega})$. Namely, for $p>n$ Morrey's inequality gives
\beq \label{Morrey_textbook}
\| f\|_{C^{0,\alpha}(\overline{\Omega})}  \leq C_M \|f\|_{W^{1,p}(\Omega)},
\eeq
where $\alpha \equiv 1 - \frac{n}{p}$ and $C_M>0$ is a constant only depending on $n$, $p$ and $\Omega$ \cite{Evans}.  Morrey's inequality (\ref{Morrey_textbook}) extends unchanged to components of matrix valued differential forms.

We finally summarize the estimate we use from elliptic theory. We assume throughout that $m\geq 1$, $1<p<\infty$, $n\geq 2$ and that $\Omega\subset \R^n$ is a bounded domain, simply connected and with smooth boundary. In fact, one could assume without loss of essential generality that $\Omega$ is a ball in $\R^n$. Our estimates are based on the following theorems, which extend to matrix valued and vector valued differential forms by component-wise application. 

\begin{Thm} \label{Thm_elliptic_reg} {\bf (Elliptic Regularity):} 
For $m\geq 1$, $1<p<\infty$, let $f\in W^{m-1,p}(\Omega)$ and $u_0 \in W^{m+1,p}(\Omega)$, which we both assume to be scalar functions. Assume $u \in W^{m+1,p}(\Omega)$ solves the Poisson equation $\Delta u = f$ with Dirichlet data $u_0$ in the sense that $u - u_0 \in W^{1,p}_0(\Omega)$.\footnote{The space $W^{1,p}_0(\Omega)$ denotes the closure of $C_0^\infty(\Omega)$, the space of smooth functions with compact support, with respect to the $W^{1,p}$-norm.}  Then there exists a constant $C>0$ depending only on $\Omega$, $m,n,p$ such that 
\beq \label{Poissonelliptic_estimate_Lp}
\| u \|_{W^{m+1,p}(\Omega)} \leq C \Big( \| f \|_{W^{m-1,p}(\Omega)} +  \| u_0 \|_{W^{m+1,p}(\Omega)} \Big).
\eeq 
\end{Thm}

Equation \eqref{Poissonelliptic_estimate_Lp} is the basic estimate of elliptic regularity theory in $L^p$ spaces, c.f. Lemma 9.17 in \cite{GilbargTrudinger} for a special case ($m=1$, $u_0=0$) which we extend in Section \ref{Sec_appendix_ell} to prove \eqref{Poissonelliptic_estimate_Lp}. Estimate \eqref{Poissonelliptic_estimate_Lp} is required to prove convergence of our iteration scheme introduced in Section \ref{Sec_iteration_scheme} below. Our analysis of the iteration scheme also requires an existence theory for Dirichlet problems for the Poisson-type equations \eqref{PDE1} - \eqref{PDE2}, which we apply component-wise. This is provided by the following existence theorem by component-wise application to the equations for $\Gammati$ and $J$, c.f. Theorems 9.15 and 9.19 in \cite{GilbargTrudinger}.

\begin{Thm}   \label{Thm_Poisson}
For $m\geq 1$, $1<p< \infty$, let $f\in W^{m-1,p}(\Omega)$ and $u_0 \in W^{m+1,p}(\Omega)$ both be scalar functions. Then there exists a unique solution $u \in W^{m+1,p}(\Omega)$ which solves the Poisson equation $$\Delta u = f \ \ \ \ \ \ \text{in} \ \ \ \Omega,$$ with Dirichlet data $u = u_0$ on $\partial\Omega$ in the sense that $u - u_0 \in W^{1,p}_0(\Omega)$.
\end{Thm}

The estimate corresponding to \eqref{Poissonelliptic_estimate_Lp} for first order equations is given by Gaffney's inequality, (c.f. Theorem 5.21 in \cite{Dac}), needed to address the first order equations \eqref{PDE3} - \eqref{PDE4} in the proof of convergence of our iteration scheme. 

\begin{Thm} {\bf (Gaffney Inequality):} 
Let $u \in W^{m+1,p}(\Omega)$ be a scalar valued $k$-form, $m\geq 0$, $1\leq k\leq n-1$, $n\geq2$, $1<p< \infty$. Then there exists a constant $C>0$ depending only on $\Omega$, $m,n,p$, such that\footnote{Note that the boundary term in \eqref{Gaffney} can be further estimated by the trace theorem as $\|u\|_{W^{m+1-\frac{1}{p},p}(\partial\Omega)} \leq C \|u\|_{W^{m+1,p}(\Omega)}$, providing an estimate closer in form to \eqref{Poissonelliptic_estimate_Lp}.}
\beq \label{Gaffney}
\|u\|_{W^{m+1,p}(\Omega)} \leq C \Big( \|du\|_{W^{m,p}(\Omega)} + \|\delta u\|_{W^{m,p}(\Omega)}+\| u\|_{W^{m+\frac{p-1}{p},p}(\partial\Omega)}\Big). 
\eeq  
\end{Thm}

Our analysis of an iteration scheme below requires an existence theory for the first order Cauchy-Riemann type equations \eqref{PDE3} and \eqref{PDE4} of the RT-equations \eqref{PDE1} - \eqref{PDE4}, the case when $\vec{A}$ is a $1$-form.   For this we are free to impose whatever boundary conditions are sufficient for a suitable existence theory.  The following special case of Theorem 7.4 in \cite{Dac} provides the existence theorem required to address the first order system, equations \eqref{PDE3} - \eqref{PDE4}, in our iteration scheme.

\begin{Thm} \label{Thm_CauchyRiemann}
$(i)$ Let $f\in W^{m,p}(\Omega)$ be a $2$-form with $df=0$, $m\geq 0$, $n\geq2$, $1<p< \infty$. For simplicity, assume further that $f=d v$ for some $1$-form $v \in W^{m,p}(\Omega)$. Then there exists a $1$-form $u=u_i\, dx^i \in W^{m+1,p}(\Omega)$ which solves
\beq \label{CauchyRiemann_eqn_Thm}
du=f  \hspace{.5cm}   \text{and} \hspace{.5cm}  \delta u =0 \hspace{1cm} \text{in} \ \ \ \Omega,
\eeq 
together with the boundary condition 
\beq \label{CauchyRiemann_bdd_Thm}
u \cdot N=0 \hspace{1cm} \text{on} \ \ \  \partial\Omega,
\eeq
where $N$ is the unit normal on $\partial\Omega$ and $u \cdot N \equiv u_i N^i$.  Moreover, there exists a constant $C>0$ depending only on $\Omega$, $m,n,p$, such that
\beq \label{Gaffney_2}
\| u \|_{W^{m+1,p}(\Omega)}  \leq C \| f \|_{W^{m,p}(\Omega)} .
\eeq
$(ii)$ Let $f\in W^{m,p}(\Omega)$ be a $1$-form with $df=0$. Then there exists a $0$-form $u \in W^{m+1,p}(\Omega)$ such that $du=f$ and $u$ satisfies estimate \eqref{Gaffney_2}.
\end{Thm}

\Proof
Part (i) is a special case of Theorem 7.4 in \cite{Dac} for $1$-forms with zero boundary conditions. Namely, our assumption $df=0$ together with zero boundary data, ($\omega_0 =0$, following notation in \cite{Dac}), directly gives condition (C1) of \cite[Thm 7.4]{Dac}. The first equation of condition (C2) of \cite[Thm 7.4]{Dac} follows trivially from our assumptions, ($g=0$ and $\omega_0=0$ in the notation of \cite{Dac}). The second equation in (C2), that $\int_\Omega \langle f; \Psi\rangle =0$ for any harmonic form $\Psi$ (i.e. $\delta \Psi=0$) with vanishing normal components (i.e. $N\cdot \Psi=0$) on the boundary ($\Psi \in \mathcal{H}_N$ in the notation of \cite{Dac}), follows by application of the integration by parts formula \cite[Thm 3.28]{Dac} for differential forms to $f=d v$,
$$
\int_{\Omega}\langle f;\Psi \rangle_{L^2} = - \int_{\Omega}\langle v;\delta \Psi \rangle_{L^2} + \int_{\partial\Omega}\langle v ;N\cdot \Psi \rangle =0.
$$ 
Theorem 7.4 in \cite{Dac} now yields the existence of a solution $u\in W^{m+1,p}(\Omega)$ to \eqref{CauchyRiemann_eqn_Thm} - \eqref{CauchyRiemann_bdd_Thm} satisfying estimate \eqref{Gaffney_2}.

Part (ii) of Theorem \ref{Thm_CauchyRiemann}, can be thought of as a version of Theorem \cite[Thm 7.4]{Dac}, in the special case of $0$-forms, which does not require condition (C2) by abandoning boundary data. That is, we seek a $0$-form $u$ solving the gradient equation $du=f$ such that estimate \eqref{Gaffney_2} holds. (No boundary data is required for our purposes). To begin the proof, observe that a solution $u \in W^{m+1,p}(\Omega)$ of $du=f$, in the case $m\geq 1$, is given by the path integral 
\beq \label{CauchyRiemann_int-formula}
u(x) = \int_{x_0}^x f \cdot d\vec{r} \; +\; u_0
\eeq
along any differentiable curve connecting $x_0$ and $x$, where $x_0 \in \Omega$ is some point we fix, and the constant $u_0$ is the value of $u$ at $x_0$, which is free to be chosen. Note, since $df=0$, the integral \eqref{CauchyRiemann_int-formula} is path independent, as can be shown by applying Stokes Theorem to integration of $df$ over the region enclosed by two curves connecting $x_0$ and $x$. To prove the sought after estimate we choose $u_0$ such that the average of $u$ is zero, then Poincar{\'e}'s inequality \cite[Eqn. (7.45)]{GilbargTrudinger} implies that $\| u\|_{L^p(\Omega)} \leq C\| f \|_{L^p(\Omega)}$ for a suitable constant $C>0$. Thus, since $\| du \|_{L^p(\Omega)} = \| f \|_{L^p(\Omega)}$ follows directly from $du=f$, we have 
\beq\label{CauchyRiemann_0estimate} 
\| u \|_{W^{1,p}(\Omega)}  \leq C \| f \|_{L^p(\Omega)}.
\eeq 
Estimate \eqref{Gaffney_2} follows by suitable differentiation of $du=f$ and application of estimate  \eqref{CauchyRiemann_0estimate}. Existence of a solution $u$ to $du=f$ in the case $m=0$ follows again from \eqref{CauchyRiemann_int-formula} by mollifying $f$, and using that this mollification is controlled by estimate \eqref{CauchyRiemann_0estimate}. This completes the proof of Theorem \ref{Thm_CauchyRiemann}. 
\QED

\section{Reduction to standard Dirichlet boundary data}   \label{Sec_extended_RT}

To prove Theorem \ref{ThmMain} we introduce an iteration scheme to construct approximate solutions of the RT-equations (\ref{PDE1})-(\ref{BDD1}), introduce a small parameter to handle the nonlinearities, and apply standard results on elliptic regularity in $L^p$ spaces to obtain convergence together with the sought after levels of smoothness.   One of the main technical issues is how to handle the non-standard boundary condition (\ref{BDD1}), which is neither standard Neumann nor Dirichlet data for the PDE (\ref{PDE2}) which determines $J$.   We now introduce a reformulation of the boundary condition (\ref{BDD1}) for the $J$ equation (\ref{PDE2}), (the only boundary condition specified by the RT-equations), as an equivalent implicit boundary condition, which has the advantage that it reduces to standard Dirichlet conditions for $J$ at each level of our iteration scheme introduced in Section \ref{Sec_iteration_scheme}.  

So assume $(\Gammati,J,A)$ is a solution of the RT-equations, and write (\ref{PDE1}) - (\ref{PDE4}) using the following compact notation:
\begin{eqnarray} 
\Delta \Gammati &=& \tilde{F}(\Gammati,J,A), \label{eqn11} \\
\Delta J &=& F(\Gammati,J) - A , \label{eqn22} \\
d \vec{A} &=& d\vec{F}(\Gammati,J)  \label{eqn33}\\
\delta \vec{A} &=& v,  \label{eqn44}
\end{eqnarray}
where $\vec{F}(\Gammati,J)$ is the vectorized version of $F(\Gammati,J)$, so that $d\vec{F}(\Gammati,J)$ is identical to the right hand side of \eqref{PDE3}, c.f. the derivation leading to equation (3.40) in \cite{ReintjesTemple_ell1}.  Now (\ref{eqn33}) implies the consistency condition 
$$
d\big(\vec{F}(\Gammati,J)-\vec{A}\big)=0,
$$ 
so that we can solve 
\beq \label{definepsi}
\begin{cases}
d\Psi =\vec{F}(\Gammati,J)-\vec{A}, \\
\delta\Psi = 0,
\end{cases}
\eeq
for a vector valued function $\Psi$, (c.f. Theorem 7.4 in \cite{Dac}). Let $y$ then be any solution of  
\begin{eqnarray}\label{definey}
\Delta y=\Psi.
\end{eqnarray}
Now we claim that in place of the Poisson equation (\ref{PDE2}) for $J$ with the boundary condition (\ref{BDD1}), it suffices to solve the boundary value problem\footnote{Assigning $\vec{J}$ on $\partial\Omega$ is the same as assigning $J$ on $\partial\Omega$ because both contain the same component functions, and as in theorem \ref{Thm_Poisson} boundary data is assigned in the sense that $\vec{J}-dy \in W^{1,p}_0(\Omega)$.}
\begin{eqnarray}
&\Delta J=F(\Gammati,J)-A\ \ {\rm in}\ \ \Omega,& \label{thepoint1}\\
&\vec{J}=dy\ \ {\rm on}\ \ \partial\Omega.& \label{thepoint2}
\end{eqnarray} 
To see this, observe that 
\begin{eqnarray}\label{thepoint2b}
\Delta dy=d\Delta y=d\Psi=\vec{F}-\vec{A}=\Delta\vec{J},
\end{eqnarray} 
which uses that, after taking $vec$ on both sides of the $J$-equation (\ref{thepoint1}), the operation $vec$ commutes with $\Delta$ on the left hand side $(\ref{thepoint1})$ because the Laplacian acts component-wise. 
Thus,
\begin{eqnarray} \nonumber
&\Delta(\vec{J}-dy)=0\ \ {\rm in}\ \ \Omega,&\\
&\vec{J}-dy=0\ \  {\rm on}\ \ \partial\Omega,&  
\end{eqnarray} 
which implies by uniqueness of solutions of the Laplace equation that $\vec{J}=dy$ in $\Omega$. Since second derivatives commute, we conclude that 
\beq   \label{thepoint3}
d\vec{J}=Curl(\vec{J})=0 \ \ \ \ \text{in} \ \ \Omega,
\eeq 
on solutions of (\ref{thepoint1}) - (\ref{thepoint2}), as claimed.   The point of using (\ref{thepoint2}) in place of (\ref{BDD1}) is that $dy$ can be determined at the $k$-th step of an iteration scheme in which the $(k+1)$-st iterate is determined by (\ref{thepoint1}) - (\ref{thepoint2}), c.f. Section \ref{Sec_iteration}.   In this setting, (\ref{thepoint2}) is standard Dirchlet data for $J$.   The equivalence between the boundary conditions \eqref{BDD1} and \eqref{thepoint2} is recorded in the following theorem.      
 
\begin{Prop} \label{Thm_RT-eqnAgain}
Assume $J \in W^{m+1,p}(\Omega)$ is invertible, and assume $J$, $\Gammati \in W^{m+1,p}(\Omega)$  and $A\in W^{m,p}(\Omega)$ solve \eqref{PDE1} - \eqref{PDE4}, where $m\geq 1$, $p>n$.  Then the boundary condition  \eqref{BDD1} holds if and only if 
\begin{eqnarray} \label{BDD2}
\vec{J}=dy\ \  {\rm on}\ \ \partial\Omega,  
\end{eqnarray}     
for some $y$ satisfying \eqref{definey}.
\end{Prop}

\Proof
The argument between equations \eqref{eqn11} and \eqref{thepoint3} proves that the boundary data \eqref{BDD2} implies that  $d\vec{J}=0$ holds everywhere in $\Omega$. By Sobolev imbedding (for $p>n$), $d\vec{J}$ is H\"older continuous on the closure of $\Omega$, (c.f. \eqref{Morrey_textbook} below), so that we can restrict $d\vec{J}$ to the boundary $\partial \Omega$ which gives the sought after boundary condition \eqref{BDD1}.  

To prove the inverse implication, assume that $(J,\Gammati,A)$ solves the RT-equations \eqref{PDE1} - \eqref{PDE4} with boundary data \eqref{BDD1}. Lemma 3.7 in \cite{ReintjesTemple_ell1} then implies that   $d\vec{J} = 0$ in $\Omega$ so that one can integrate $J$ to some coordinate function $y$, i.e. $dy=\vec{J}$. Defining $\Psi \equiv \Delta y$, it follows from $J$ solving \eqref{thepoint1} that
\beq \nonumber
d\Psi= d\Delta y = \Delta dy = \Delta \vec{J} = \overrightarrow{\Delta J} \overset{\eqref{thepoint1}}{=} \vec{F}-\vec{A}.
\eeq 
Thus $\Psi$ satisfies \eqref{definepsi}, while \eqref{definey} holds by the above definition of $\Psi$. So restriction of $dy=\vec{J}$ to $\partial\Omega$ gives the sought after boundary data \eqref{BDD2}. This completes the proof of Proposition \ref{Thm_RT-eqnAgain}.
\QED

\section{The iteration scheme} \label{Sec_iteration}

In this section we introduce our iteration scheme for approximating solutions of the RT-equations. We begin by setting up our iteration scheme in terms of the extended RT-equations \eqref{PDE1} - \eqref{PDE4} and \eqref{definepsi} - \eqref{definey} with standard Dirichlet data \eqref{thepoint2} in a non-technical way. In Section \ref{Sec_rescaling}, we introduce a small parameter $\epsilon>0$ into the RT-equations, (by smallness of the coordinate neighborhood), which allows us to estimate the non-linearities on the right hand side of the RT-equations and prove convergence of the iterates for sufficiently small $\epsilon>0$ in Section \ref{Sec_convergence}. In Section \ref{Sec_iteration_scheme}, we introduce our iteration scheme in terms of the $\epsilon$-rescaled RT-equations and prove its well-posedness. Throughout the remainder of this paper we take
$$
v \equiv 0
$$ 
in \eqref{PDE4}, fixing the freedom to choose $v\in W^{m-1,p}(\Omega)$. We assume a given connection $\Gamma$ of suitable regularity, defined in a given coordinate system $x$ in an open and bounded set $\Omega \subset \R^n$ with smooth boundary. To define the iteration by induction, it suffices to start with given $(\Gammati_0,J_0)$, show how to construct $A_1$, $\Gammati_1$ and $J_1$ from $(\Gammati_0,J_0)$. This then tells us how to construct $(A_{k+1},\Gammati_{k+1},J_{k+1})$ from $(\Gammati_{k},J_{k})$ for each $k\geq1$ by recursion. 

So assume $\Gammati_k$ and $J_k$ are given for some $k\geq 0$.  Define $A_{k+1}$ as the solution of 
\beq \label{iteration_eqnA}
\begin{cases}
d\vec{A}_{k+1} = d\vec{F}(\Gammati_k,J_k), \cr
\delta \vec{A}_{k+1}=0,
\end{cases}
\eeq    
for $A_{k+1} \mm N=0$ on $\partial \Omega$, where $N$ is the unit normal vector of $\partial \Omega$ which is multiplied to the matrix $A_{k+1}$.  Our iteration does not require to assume $A_k$, since $A_{k+1}$ is defined in terms of $\Gammati_k$ and $J_k$ alone. Note, our choice of boundary data in \eqref{iteration_eqnA} and $v=0$ was made so that  Theorem \ref{Thm_CauchyRiemann} applies to give existence.

To introduce the Dirichlet data for $J_{k+1}$, we first define the auxiliary variables $\psi_{k+1}$ and $y_{k+1}$, for which we again do not require the previous iterates $\psi_{k}$ and $y_{k}$. So use the identity $d(\vec{F}(\Gammati_k,J_k)-\overrightarrow{A_{k+1}})=0$ of \eqref{iteration_eqnA} to solve 
\beq \label{iteration_eqnPsi}
\begin{cases}
d\Psi_{k+1}=\vec{F}(\Gammati_k,J_k)-\overrightarrow{A_{k+1}}, \cr 
\delta\Psi_{k+1}=0,
\end{cases}
\eeq
and then solve 
\beq \label{iteration_eqn_y}
\Delta y_{k+1}=\Psi_{k+1},
\eeq
where for \eqref{iteration_eqnPsi} and \eqref{iteration_eqn_y} any convenient boundary condition can be implemented; we only require the boundary data for the elliptic estimates \eqref{Poissonelliptic_estimate_Lp} and \eqref{Gaffney}.   
 
Now, define $J_{k+1}$ to be the solution of the following {\it standard Dirichlet} boundary problem,   
\begin{eqnarray} \label{iteration_eqn_J}
\Delta J_{k+1} &=& F(\Gammati_{k},J_k)-\overrightarrow{A_{k+1}},\\
\overrightarrow{J_{k+1}} &=& dy_{k+1} \ \ {\rm on}\ \ \partial\Omega, \label{iteration_eqn_y_bdd}
\end{eqnarray} 
and, to obtain $\Gammati_{k+1}$, solve 
\beq \label{iteration_eqn_Gammati}
\Delta \Gammati_{k+1} = \tilde{F}(\Gammati_k,J_k,A_{k+1}),
\eeq
where the boundary data for $\Gammati_{k+1}$ is free to be chosen. 

The implicit boundary condition (\ref{BDD2}) reduced to \eqref{iteration_eqn_y_bdd}, which is standard Dirichlet data at each step of the iteration. As in Proposition \ref{Thm_RT-eqnAgain}, the iterates so defined imply $Curl(J_{k+1})=~0$ for each $k\geq 0$, as proven below in Lemma \ref{Lemma_curl} for the iterates of the rescaled RT-equations. We defined here an iteration scheme for the RT-equations in terms of solutions of the Dirichlet problem for the linear Poisson equation and Cauchy-Riemann equations.

\subsection{The rescaled equations}  \label{Sec_rescaling}

We now introduce a small parameter $\epsilon>0$ and derive an $\epsilon$-rescaled version of the RT-equations, which allows us to handle the non-linearities. To introduce a small parameter $\epsilon$, assume the components of $\Gamma$ are given in $x$-coordinates in an open and bounded set $\Omega \subset \R^n$ with smooth boundary, and assume $\Gamma$ and $d\Gamma$ are both bounded in $W^{m,p}(\Omega)$.   Let $\Gamma^*$ be a connection in $x$-coordinates satisfying 
\begin{eqnarray}
\|\Gamma^*\|_{W^{m,p}(\Omega)} \: + \: \|d\Gamma^*\|_{W^{m,p}(\Omega)}<C_0,\label{Gamma-bound}
\end{eqnarray} 
for $m\geq 1$ and $C_0$ a fixed constant. Now, we assume without loss of generality that $\Gamma$ scales with $\epsilon>0$ according to the definition
\begin{eqnarray}
\Gamma=\epsilon\Gamma^*. \label{small_Gamma}
\end{eqnarray} 
The assumptions \eqref{Gamma-bound} and \eqref{small_Gamma} can be made without loss of generality regarding the \emph{local} problem of optimal metric regularity. Namely, given any connection $\Gamma' \in W^{m,p}(\Omega)$ with $d\Gamma'$ bounded in $W^{m,p}(\Omega)$, we define $\Gamma^*$ as the restriction of $\Gamma'$ to the ball of radius $\epsilon$ with its components transformed as scalars to the ball or radius $1$ (which we take to be $\Omega$), while $\Gamma$ is taken to be the connection resulting from transforming $\Gamma'$ \emph{as a connection} under the same coordinate transformation, c.f. the proof of Theorem \ref{ThmMain} in Section \ref{Sec_proof_ThmMain}.  

We further assume the scaling ansatz 
\beq \label{ansatz_scaling}
J=I+\epsilon \, J^* , \hspace{1cm} 
\tilde{\Gamma}=\epsilon\: \tilde{\Gamma}^*, \hspace{1cm} 
A = \epsilon A^*.
\eeq 
Since we only need to prove {\it existence} of a solution for our purposes, assumption \ref{ansatz_scaling} is again made without loss of generality for the problem of optimal metric regularity. Now, to derive the RT-equations tuned to the $\epsilon$-scaling, substitute \eqref{small_Gamma} and \eqref{ansatz_scaling} into the RT-equations \eqref{PDE1} - \eqref{PDE4} for $v \equiv 0$ and divide by $\epsilon$, we then obtain an equivalent set of equations as recorded in the following lemma.

\begin{Lemma}   \label{Lemma_rescaled_RT-eqn}
 Let $u \equiv \left(\msp\begin{array}{c}  \tilde{\Gamma}^* \cr J^* \end{array}\msp\right)$ and $a \equiv A^*$, 
and define
\begin{align}  
& F_u(u,a) \equiv  \left(\msp \begin{array}{c} \delta d \Gamma^*  -  \delta d\big( J^{-1} \mm dJ^* \big)  +  d (J^{-1} a)  \cr \delta \Gamma^*   + \epsilon\: \delta ( J^* \mm \Gamma^* ) - \epsilon\: \langle d J^* ; \tilde{\Gamma}^*\rangle - a \end{array} \msp\right) ,     \label{Def_Fu}  \\  
& F_a(u)  \equiv   \overrightarrow{\text{div}} \big(d\Gamma^*\big) + \epsilon\: \overrightarrow{\text{div}} \big( J^* \mm d\Gamma^*\big) + \epsilon\: \overrightarrow{\text{div}} \big(dJ^* \wedge \Gamma^*\big)  -   \epsilon\: d\big(\overrightarrow{\langle d J^* ; \tilde{\Gamma}^*\rangle }\big).  \label{Def_Fa}
\end{align}
Then, substituting \eqref{small_Gamma} and \eqref{ansatz_scaling} into the RT-equations \eqref{PDE1} - \eqref{PDE4} for $v \equiv 0$ and dividing by $\epsilon$, we obtain the equivalent set of equations
\begin{eqnarray}
\Delta u = F_u(u,a), \label{pde_u}  \\
\begin{cases} \label{pde_a}
d\vec{a} = F_a(u)  \\
\delta\vec{a} = 0  .
\end{cases}
\end{eqnarray}
\end{Lemma}

\Proof
Substituting \eqref{small_Gamma} and \eqref{ansatz_scaling} into the RT-equations \eqref{PDE1} - \eqref{PDE4} with $v \equiv 0$, and dividing by $\epsilon$, we obtain
\begin{eqnarray} 
\Delta \tilde{\Gamma}^*  &=& \delta d \Gamma^* - \delta d\big( J^{-1} \mm dJ^* \big) +  d (J^{-1} A^*),  \label{pde1} \\ 
\Delta J^* &=&  \delta ( J \mm \Gamma^* ) -  \epsilon\: \langle d J^* ; \tilde{\Gamma}^*\rangle - A^*  \label{pde2} \\
 d \vec{A}^* &=& \epsilon\: \overrightarrow{\text{div}} \big(dJ^* \wedge \Gamma^*\big) +  \overrightarrow{\text{div}} \big( J\mm d\Gamma^*\big) - \epsilon\: d\big(\overrightarrow{\langle d J^* ; \tilde{\Gamma}^*\rangle }\big) . \label{pde3} \\
 \delta \vec{A}^* &=& 0. \label{pde4}
\end{eqnarray}
Now, equations \eqref{pde1} - \eqref{pde4} together with the definitions of $u,a$ and \eqref{Def_Fu} - \eqref{Def_Fa} imply the sought after equations \eqref{pde_u} - \eqref{pde_a}.
\QED

We often refer to \eqref{pde_u} - \eqref{pde_a} as the ``rescaled RT-equations''. We further introduce the following useful notation,
\begin{eqnarray}   \label{F_notation}
F_{\Gammati}(u,a) &\equiv & \delta d \Gamma^* + da -  \delta d\big( J^{-1} \mm dJ^* \big)  +  d (J^{-1} a) \cr
F_J(u) & \equiv & \delta \Gamma^*   + \epsilon\: \delta ( J^* \mm \Gamma^* ) - \epsilon\: \langle d J^* ; \tilde{\Gamma}^*\rangle ,
\end{eqnarray}
so that $F_u(u,a) = (F_{\Gammati}(u,a),F_J(u)-a)$ and by equation (3.40) in \cite{ReintjesTemple_ell1} we have $F_a(u) = d \overrightarrow{F_J}$. The rescaled RT-equations \eqref{pde_u} - \eqref{pde_a} can then be written equivalently as
\begin{eqnarray}
\Delta u = \left(\hspace{-.2cm} \begin{array}{c} F_{\Gammati}(u,a) \cr  F_J(u)-a  \end{array} \hspace{-.2cm}\right), \label{pde_u_alt}  \\
\begin{cases} \label{pde_a_alt}
d\vec{a} = d \overrightarrow{F_J(u)} \\
\delta\vec{a} = 0 .
\end{cases}
\end{eqnarray}
We use the alternative form \eqref{pde_u_alt} - \eqref{pde_a_alt} to set up the iteration scheme below. 
As proven in Section \ref{Sec_proof_ThmMain}, Theorem \ref{ThmMain} now follows from the following theorem, the proof of which is the topic of Sections \ref{Sec_iteration_scheme} - \ref{Sec_Proofs}.

\begin{Thm}  \label{Thm2}
Let $\Gamma^*, d\Gamma^* \in W^{m,p}(\Omega)$ satisfy \eqref{Gamma-bound} and let $m\geq 1$, $p>n\geq2$. Then there exists $\epsilon_*$ such that, if $\epsilon<\epsilon_*$, then there exists $u \in W^{m+1,p}(\Omega)$ and $a \in W^{m,p}(\Omega)$ which solve the RT-equation \eqref{pde_u} - \eqref{pde_a} with boundary data \eqref{BDD1}. 
\end{Thm}

In Section \ref{Sec_convergence} we summarize the proof of Theorem \ref{Thm2} which is based on the iteration scheme in Section \ref{Sec_iteration_scheme}. The details of the proof are postponed to Section  \ref{Sec_Proofs}.

\subsection{The Iteration scheme for the rescaled equations}    \label{Sec_iteration_scheme}

In this section we define the iteration scheme $(u_{k},a_{k})$, $k\geq0$, for approximating solutions of (\ref{pde_u})-(\ref{pde_a}), and set up the framework for proving convergence of the scheme in the appropriate Sobolev spaces for $\epsilon$ sufficiently small.  Define $(u_{k+1},a_{k+1})$ by induction as follows. Start the induction by assuming 
$$u_0=a_0=0.$$   
Then, given $u_k \in W^{m+1,p}(\Omega)$ and $a_k \in W^{m,p}(\Omega)$ for $k\geq0$, we define $a_{k+1} \in W^{m,p}(\Omega)$ by solving
\beq \label{iterate_a}
\begin{cases} 
d (\overrightarrow{a_{k+1}}) = F_a(u_k), \cr 
\delta (\overrightarrow{a_{k+1}})=0,
\end{cases}
\eeq
with Dirichlet boundary data
\beq \label{iterate_a_bdd}  
a_{k+1} \mm N=0,
\eeq 
where $N$ is the unit normal on the boundary $\partial \Omega$, and $a_{k+1}$ is a matrix valued $0$-form. (Our boundary data \eqref{iterate_a_bdd} and the equation $\delta (\overrightarrow{a_{k+1}})=0$ are chosen so that the existence theory in \cite{Dac} applies.) Next, in order to arrange for the non-standard boundary condition \eqref{BDD1}, we introduce the vector valued $0$-form $\psi_{k+1} \in W^{m,p}(\Omega)$ as a solution of
\beq \label{iterate_psi}                 
d \psi_{k+1} = \overrightarrow{F_J(u_k)} - \overrightarrow{a_{k+1}}, 
\eeq
which satisfies the estimate $\|\psi_{k+1}\|_{W^{m,p}} \leq C \|\overrightarrow{F_J(u_k)} - \overrightarrow{a_{k+1}}\|_{W^{m-1,p}}$ for some constant $C>0$ independent of $k$, c.f. estimate \eqref{Gaffney_2} of Theorem \ref{Thm_CauchyRiemann}; no boundary data is required.\footnote{Recall that $d\overrightarrow{F_J(u_k)} = F_a(u_k)$, as explained below \eqref{F_notation}, which is required for solvability as explained in the proof of Lemma \ref{Lemma_existence_iterates} below.}   In terms of $\psi_{k+1}$ we next define the vector valued function $y_{k+1} \in W^{m+2,p}(\Omega)$ as the solution of
\beq \label{iterate_y}
\begin{cases} 
\Delta y_{k+1} = \psi_{k+1}, \cr 
y_{k+1}\big|_{\partial\Omega} = 0.
\end{cases}
\eeq
The vector valued functions $\psi_{k+1}$ and $y_{k+1}$ are auxiliary variables which we introduce to assign standard Dirichlet data for the Poisson equation which defines $u_{k+1} = (J^*_{k+1},\Gammati_{k+1})$. Namely, we define $u_{k+1} \in W^{m+1,p}(\Omega)$ as the solution of 
\beq \label{iterate_u}
\Delta u_{k+1}=F_u(u_k,a_{k+1}),
\eeq
with Dirichlet boundary data    
\begin{eqnarray} 
\Gammati^*_{k+1} &=& 0 \hspace{1cm} \text{on} \ \ \partial\Omega, \label{bdd1} \\
J^*_{k+1} &=& dy_{k+1} \hspace{.5cm} \text{on} \ \ \partial\Omega.  \label{bdd2}
\end{eqnarray}
Equations \eqref{iterate_a} - \eqref{bdd2} define the iteration scheme underlying our existence theory for the RT-equations. The next lemma shows that the standard Dirichlet data \eqref{bdd1} - \eqref{bdd2} suffices to impose the non-standard boundary condition \eqref{BDD1} and obtain integrability of the Jacobian at each step of the iteration.

\begin{Lemma} \label{Lemma_curl}
Any solution $u_{k+1} = (J^*_{k+1},\Gammati^*_{k+1}) \in W^{m+1,p}(\Omega)$ of \eqref{iterate_u} with boundary data \eqref{bdd1} - \eqref{bdd2} satisfies
\beq \label{curl_J_iterates}
d\overrightarrow{J^*_{k+1}} \equiv Curl(J^*_{k+1})=0
\eeq
in $\Omega$, which automatically implies the boundary condition \eqref{BDD1}. Then $J_{k+1} \equiv I + \epsilon J^*_{k+1}$ is integrable and defines the Jacobian of the coordinate transformation $x \to x +\epsilon y_{k+1}(x)$, where $y_{k+1}$ is defined in \eqref{iterate_y}.
\end{Lemma}

\Proof
We compute that
\begin{eqnarray} \nonumber
\Delta (dy_{k+1}) 
= d (\Delta y_{k+1}) 
\overset{\eqref{iterate_y}}{=} d \psi_{k+1} 
\overset{\eqref{iterate_psi}}{=}   F_J(u_k) - a_{k+1}
\overset{\eqref{iterate_u}}{=} \Delta J^*_{k+1}, 
\end{eqnarray}
which implies that
\beq \label{techeqn1_Lemma_curl}
\Delta \big(J^*_{k+1} - dy_{k+1} \big) =0.
\eeq
Now, since $J^*_{k+1} - dy_{k+1}$ vanishes on $\partial\Omega$ by \eqref{bdd2}, we conclude that $J^*_{k+1} = dy_{k+1}$ in $\Omega$. This implies \eqref{curl_J_iterates}, and since 
\beq \nonumber
d(x + \epsilon y_{k+1}) = I + \epsilon J^*_{k+1} = J_{k+1}, 
\eeq 
we conclude that $J_{k+1}$ is the Jacobian of the coordinate transformation $x \to x +\epsilon y_{k+1}(x)$. This completes the proof.
\QED

Our strategy for completing the proof of Theorem \ref{Thm2} is to first state the main technical lemmas  in Lemmas \ref{Lemma_existence_iterates} - \ref{Lemma_induction_consistency} together with Proposition \ref{Lemma_decay} to follow, use them to prove Theorem \ref{Thm2}, and postpone the proofs of these lemmas to Sections \ref{Sec_source-estimates} - \ref{Sec_decay}. We end this section by stating the first technical lemma which addresses the well-posedness of the iteration scheme \eqref{iterate_a} - \eqref{bdd2}.   

\begin{Lemma} \label{Lemma_existence_iterates}
Assume $u_k\in W^{m+1,p}(\Omega)$ is given, for $m\geq 1$, $p>n\geq 2$, and that $\epsilon>0$ satisfies
\beq \label{epsilon_bound_0}
\epsilon \leq  \epsilon(k) \equiv \frac{1}{4 C_M \|u_{k}\|_{W^{m+1,p}}},
\eeq
where $C_M>0$ is the constant from Morrey's inequality \eqref{Morrey_textbook}. Then there exists $a_{k+1}\in W^{m,p}(\Omega)$ which solves \eqref{iterate_a} - \eqref{iterate_a_bdd}, there exists the auxilliary iterates $\psi_{k+1} \in W^{m,p}(\Omega)$ and $y_{k+1} \in W^{m+2,p}(\Omega)$ which solve \eqref{iterate_psi} - \eqref{iterate_y}, and there exists $u_{k+1}\in W^{m+1,p}(\Omega)$ which solves \eqref{iterate_u} with boundary data \eqref{bdd1} - \eqref{bdd2}. Moreover, these iterates satisfy the elliptic estimates  
\begin{eqnarray}
\|a_{k+1}\|_{W^{m,p}(\Omega)}  &\leq & C_e \; \|F_a(u_k)\|_{W^{m-1,p}(\Omega)}, \label{existence_est1} \\
\|u_{k+1}\|_{W^{m+1,p}(\Omega)} &\leq & C_e \; \|F_u(u_k,a_{k+1})\|_{W^{m-1,p}(\Omega)}, \label{existence_est2}
\end{eqnarray}
and the auxiliary iterates satisfy       
\begin{eqnarray}
\|\psi_{k+1}\|_{W^{m,p}(\Omega)}  &\leq & C_e \; \|F_u(u_k,a_{k+1})\|_{W^{m-1,p}(\Omega)}, \label{existence_est3} \\
\|y_{k+1}\|_{W^{m+2,p}(\Omega)} &\leq & C_e \; \|F_u(u_k,a_{k+1})\|_{W^{m-1,p}(\Omega)}, \label{existence_est4}
\end{eqnarray}
for some constant $C_e >0$ depending only on $m,n,p$ and $\Omega$. 
\end{Lemma}

The proof of Lemma \ref{Lemma_existence_iterates}, given in Section \ref{Sec_existence}, is based  solely on the $L^p$ elliptic estimate \eqref{Poissonelliptic_estimate_Lp} and Gaffney's inequality \eqref{Gaffney}.

\section{Convergence of Iterates and Proof of Theorem \ref{Thm2}}  \label{Sec_convergence}

In this section we state the main lemmas and propositions required for the proof of Theorem \ref{Thm2}, and assuming these, give the proof of Theorem \ref{Thm2}. Proofs of the supporting lemmas and propositions are postponed until Section \ref{Sec_Proofs} below. The proof of Theorem \ref{Thm2} follows directly from the existence result of Lemma \ref{Lemma_existence_iterates} together with Proposition \ref{Lemma_decay} alone, the latter providing estimates for the differences between subsequent iterates. The main steps in the proof of Proposition \ref{Lemma_decay} are contained in Lemmas \ref{Lemma_sources_difference} and \ref{Lemma_induction_consistency}.  To outline the proof here we state these lemmas in this section, and their proofs are given in  Sections \ref{Sec_differences} - \ref{Sec_Cons_Ind_Assump}.

To begin, observe that Lemma \ref{Lemma_existence_iterates} yields a sequence of iterates $(u_k,a_k)_{k\in \mathbb{N}}$. In order to establish convergence of this sequence in $W^{m+1,p}(\Omega)\times W^{m,p}(\Omega)$, we require estimates on the differences 
\beq
\begin{aligned} \label{diff1}
\overline{a_{k}} &\equiv  a_k- a_{k-1},\cr 
\overline{u_{k}} &\equiv  u_{k}-u_{k-1} ,
\end{aligned}
\eeq 
in terms of the corresponding differences of source terms, 
\beq
\begin{aligned}
\overline{F_a(u_{k})} &\equiv  F_a(u_{k}) - F_a(u_{k-1}), \cr 
\overline{F_{u}(u_{k},a_{k+1})} &\equiv  F_{u}(u_{k},a_{k+1}) - F_{u}(u_{k-1},a_{k}).\label{diff2}
\end{aligned}
\eeq 
The next technical lemma provides estimates of \eqref{diff1} in terms of \eqref{diff2}. The proof of Lemma \ref{Lemma_sources_difference} is given in Section \ref{Sec_differences}. 

\begin{Lemma} \label{Lemma_sources_difference}
Assume $0<  \epsilon\leq  \min \big(\epsilon(k),\epsilon(k-1)\big)$, that is, $\epsilon$ satisfies \eqref{epsilon_bound_0} in terms of $u_k$ and $u_{k-1}$. Then 
\begin{eqnarray} 
\|\overline{F_u(u_k,a_{k+1})}\|_{W^{m-1,p}} & \leq &   C_u(k) \Big( \epsilon\: \|\overline{u_{k}}\|_{W^{m+1,p}} + \|\overline{a_{k+1}}\|_{W^{m,p}} \Big), \label{bound_diff_Fu_intro} \label{bound_diff_Fu}  \\
\|\overline{F_a(u_{k})}\|_{W^{m-1,p}} & \leq & \epsilon\: C_a(k) \,  \|\overline{u_{k}}\|_{W^{m+1,p}} , \label{bound_diff_Fa}
\end{eqnarray} 
where
\begin{eqnarray} 
C_u(k) &\equiv & C_s \big( 1 +  \| u_k \|_{W^{m+1,p}} + \| {u_{k-1}}\|_{W^{m+1,p}} + \| a_{k+1} \|_{W^{m,p}}  \big), \label{diff_Cu}\\
C_a(k) & \equiv & C_s \big( 1 +  \| u_k \|_{W^{m+1,p}} + \|  {u_{k-1}}\|_{W^{m+1,p}} \big), \label{diff_Ca}
\end{eqnarray}
where $C_s$ is a constant that only depends on $m,n,p,\Omega$ and the constant $C_0$ of \eqref{Gamma-bound}.
\end{Lemma}

The next lemma establishes the induction step for our proof that the iteration scheme converges in the appropriate spaces, by bounding $C_u$ and $C_a$ independent of $k$ for $\epsilon >0$ sufficiently small. Recall, $C_0$ is the constant bounding $\Gamma^*$ and $d\Gamma^*$ in \eqref{Gamma-bound} and $C_e$ is the constant introduced in Lemma \ref{Lemma_existence_iterates}. We assume from now on and without loss of generality that $C_e>1$, which allows us to simplify the $\epsilon$-bound \eqref{epsilon_bound} below. The proof of Lemma \ref{Lemma_induction_consistency} is given in Section \ref{Sec_Cons_Ind_Assump}. 

\begin{Lemma} \label{Lemma_induction_consistency}              
Assume the induction hypothesis
\begin{eqnarray}   \label{hypothesis_induction}
\|u_k\|_{W^{m+1,p}(\Omega)} \leq 4\, C_0 C_e^2,
\end{eqnarray}
for some $k\in \mathbb{N}$ and let $C_e>1$. If        
\beq \label{epsilon_bound}
\epsilon \leq  \epsilon_1 \equiv  \min\Big( \tfrac{1}{4C^2_e C_s(1+ 2C_e C_0 + 4C_e^2 C_0)},\tfrac{1}{16 C_M C_0 C_e^2}  \Big),
\eeq 
then $0 <  \epsilon_1 \leq  \epsilon(k+l)$ for all $l \in \mathbb{N}$, (c.f. \eqref{epsilon_bound_0}), and the subsequent iterates satisfy the bounds 
\begin{eqnarray}
\|a_{k+l}\|_{W^{m,p}} &\leq & 2 C_0C_e,  \ \ \  \ \forall \: l \in \mathbb{N}, \label{uniformbound_a} \\
\|u_{k+l} \|_{W^{m+1,p}} &\leq & 4 C_0 C_e^2, \ \ \  \ \forall \: l \in \mathbb{N}. \label{uniformbound_u}
\end{eqnarray}
\end{Lemma}

In Section \ref{Sec_decay}, we prove the following proposition, which is based on combining Lemmas \ref{Lemma_sources_difference} and \ref{Lemma_induction_consistency} together with the elliptic estimates \eqref{existence_est1} - \eqref{existence_est2}. This is the main step needed to prove convergence of the iteration scheme.

\begin{Prop} \label{Lemma_decay}                         
Assume the induction hypothesis \eqref{hypothesis_induction} and $C_e>1$. If $0<  \epsilon \leq  \epsilon_1$, so $\epsilon$ satisfies \eqref{epsilon_bound}, then there exists a constant $C_d>0$ such that 
\begin{eqnarray}
\|\overline{a_{k+1}}\|_{W^{m,p}} 
&\leq & \epsilon\: C_d \:\|\overline{u_{k}}\|_{W^{m+1,p}} , \label{decay_a} \\
\|\overline{u_{k+1}}\|_{W^{m+1,p}}  
&\leq & \epsilon\: C_d \:\|\overline{u_{k}}\|_{W^{m+1,p}}, \label{decay_u}
\end{eqnarray} 
and $C_d>0$ depends only on $m$, $n$, $p$, $\Omega$ and $C_0$.
\end{Prop}

At this stage of the argument it is important to note that the auxiliary iterates $\psi_k$ and $y_k$ are not coupled to the equations for $a_k$ and $u_k$, except through the boundary data \eqref{bdd2}. The only purpose of the auxiliary variables $\psi_k$ and $y_k$ is to impose that the Jacobian $J_k = I + \epsilon J^*_k$ be curl free in each step of the iteration, and this only requires that $dy_k$ as the boundary data to $u_k$ satisfies the estimated
\begin{eqnarray} \nonumber 
\| y_{k+1} - y_k \|_{W^{m+2,p}(\Omega)} 
& \leq & C \| \overline{F_u(u_k,a_{k+1})} \|_{W^{m-1,p}(\Omega)},
\end{eqnarray}
where $C>0$ depends only on $m$, $n$, $p$, $\Omega$; which is established in the proof of Proposition \ref{Lemma_decay} below.\footnote{Convergence of $dy_k$ would follow directly from the convergence of $a_{k}$ and $u_k$ and could be proven easily by our methods, but this is not needed for the proof of Theorem \ref{Thm2} and is therefore omitted.} 

Assuming Lemmas \ref{Lemma_existence_iterates} - \ref{Lemma_induction_consistency} and Proposition \ref{Lemma_decay}, we now prove the following theorem which gives convergence of the iteration scheme. This directly implies, and hence completes, the proof of Theorem \ref{Thm2}.

\begin{Thm}   \label{Thm3}
Let $\Gamma^*, d\Gamma^* \in W^{m,p}(\Omega)$ satisfy \eqref{Gamma-bound} for $m\geq 1$, $p>n\geq 2$. Assume $\epsilon >0$ satisfies
\beq \label{epsilon_bound_2}
\epsilon < \epsilon_2 \equiv \min\Big( \epsilon_1, \frac{1}{C_d}  \Big),
\eeq 
where $\epsilon_1$ is defined in \eqref{epsilon_bound} and $C_d>0$ is the constant in \eqref{decay_a} - \eqref{decay_u}. Then the sequence of iterates $(u_k,a_k)_{k\in \mathbb{N}}$ defined by \eqref{iterate_a} - \eqref{bdd2} converges in $W^{m+1,p}(\Omega) \times W^{m,p}(\Omega)$, and the corresponding limits 
\begin{eqnarray} \nonumber
& u \equiv \lim\limits_{k\rightarrow\infty} u_k \ \in W^{m+1,p}(\Omega), & \cr
&a \equiv \lim\limits_{k\rightarrow\infty} a_k \ \in W^{m,p}(\Omega), &
\end{eqnarray}
solve the RT-equations \eqref{pde_u} - \eqref{pde_a} with boundary data \eqref{BDD1}.
\end{Thm}

\Proof
Assume Lemma \ref{Lemma_existence_iterates} and Proposition \ref{Lemma_decay} hold. Then, given two iterates $u_k,u_l \in W^{m+1,p}(\Omega)$, ($k\geq l$), estimate \eqref{decay_u} implies 
\begin{eqnarray}
\|u_{k}-u_{l}\|_{W^{m+1,p}} 
& \leq & \sum_{j=l+1}^{k}\|\overline{u_j}\|_{W^{m+1,p}} \leq \sum_{j=l+1}^{k}(\epsilon C)^j .
\end{eqnarray} 
By \eqref{epsilon_bound_2}, the above geometric series converges as $k \rightarrow \infty$. This implies that $(u_k)_{k\in \mathbb{N}}$ is a Cauchy sequence in the Banach space $W^{m+1,p}(\Omega)$. Therefore, $(u_k)_{k\in \mathbb{N}}$ converges to some $u$ in $W^{m+1,p}(\Omega)$. Similarly, \eqref{decay_a} implies
\begin{eqnarray}
\|a_{k}-a_{l}\|_{W^{m,p}} 
& \leq & \sum_{j=l+1}^{k}\|\overline{a_j}\|_{W^{m,p}} \leq \sum_{j=l+1}^{k}(\epsilon C)^j ,
\end{eqnarray} 
which in light of \eqref{epsilon_bound_2} is also a convergent geometric series. This implies convergence of $(a_k)_{k \in \mathbb{N}}$ to some $a$ in $W^{m,p}(\Omega)$. 

Now the limit $(u,a)$ solves \eqref{pde_u} and \eqref{pde_a} because each term in the equations \eqref{iterate_a} and \eqref{iterate_u} converge to the corresponding terms in \eqref{pde_u} and \eqref{pde_a} in the $L^p$-norm on $\Omega$. By Lemma \ref{Lemma_curl}, $(u,a)$ satisfies the boundary condition \eqref{BDD1}, since $Curl(J^*_k)=0$ in $\Omega$ for all $k\in \mathbb{N}$, and this property is maintain under the limit. Thus the limit satisfies $Curl(J^*)=0$ in $\Omega$ which implies the sought after boundary condition \eqref{BDD1} by restriction to the boundary.
\QED

Theorem \ref{Thm3} is a refined restatement of Theorem \ref{Thm2}, so this completes the proof of Theorem \ref{Thm2}.  It remains to give the proofs of Lemmas \ref{Lemma_existence_iterates} - \ref{Lemma_induction_consistency} and Proposition \ref{Lemma_decay}, which is accomplished in Sections \ref{Sec_source-estimates} - \ref{Sec_differences}.

\section{Proofs of technical Lemmas and Propositions} \label{Sec_Proofs}

\subsection{Estimates on the non-linear sources}  \label{Sec_source-estimates}  

In this section we prove the basic estimates for the non-linear sources on the right hand side of equations \eqref{pde_a} - \eqref{pde_u}, which are required for the proofs of Lemmas \ref{Lemma_existence_iterates} and \ref{Lemma_induction_consistency}. Our main tool is Morrey's inequality \eqref{Morrey_textbook}, which allows us to bound the supremum norm of (scalar) functions $f \in W^{1,p}(\Omega,\R)$ by 
\beq \label{Morrey}
\| f\|_{L^\infty(\Omega)}  \leq C_M \|f\|_{W^{1,p}(\Omega)},
\eeq
when $p>n$.  Below, we use \eqref{Morrey} together with the boundedness of $\Omega$ to estimate $L^p$-norms of products of functions $f \in W^{1,p}(\Omega,\R)$ and $g \in L^p(\Omega,\R)$ by
\beq \nonumber
\|fg\|_{L^p(\Omega)} \leq \|f\|_{L^\infty(\Omega)} \|g\|_{L^p(\Omega)} \leq C_M \|f\|_{W^{1,p}(\Omega)} \|g\|_{L^p(\Omega)}  .
\eeq 
In fact, $W^{1,p}(\Omega)$ is closed under multiplication for $p>n$. That is, 
\begin{eqnarray} \nonumber
\|fg\|_{W^{1,p}(\Omega)}   
&\leq & \|fg\|_{L^p} + \|g Df\|_{L^p} + \|f Dg\|_{L^p}   \cr
&\leq & \|fg\|_{L^p} + \|Df\|_{L^p} \|g\|_{L^\infty} + \|Dg\|_{L^p} \|g\|_{L^\infty}  \cr
&\leq & 3 C_M \|f\|_{W^{1,p}(\Omega)} \|g\|_{W^{1,p}(\Omega)} ,
\end{eqnarray} 
where $f,g\in W^{1,p}(\Omega,\R)$. Before we derive the basic source estimates in Lemma \ref{Lemma_source_estimate_Fa_Fu}, we establish bounds on the inverse Jacobian for $\epsilon >0$ sufficiently small.
 
\begin{Lemma} \label{Lemma_J_inverse}  \label{Lemma2_J_inverse} 
Let $J= I + \epsilon J^*$ for some $J^*\in W^{m+1,p}(\Omega)$, where $m\geq 0$ and $p>n$. 
Assume $\epsilon>0$ satisfies the bound
\beq \label{Lemma2_J_epsilon}
\epsilon \leq  \frac{1}{2 C_M \|J^*\|_{W^{m+1,p}}} ,
\eeq                                 
where $C_M>0$ is the constant from Morrey's inequality \eqref{Morrey}. Then $J$ is invertible and there exists a matrix valued $0$-form $J^{-*}\in W^{m+1,p}(\Omega)$ such that 
\beq \label{J_inverse_eqn}
J^{-1}= I + \epsilon  J^{-*}
\eeq
and such that
\beq \label{J_inverse_bound}
\| J^{-*}\|_{W^{m+1,p}} \leq \: C_- \: \|J^*\|_{W^{m+1,p}},
\eeq
where $C_->0$ is a constant depending only on $m,n,p,$ and $\Omega$.
\end{Lemma} 

Note that in our iteration scheme $\epsilon \leq  \epsilon(k)$ always guarantees for \eqref{Lemma2_J_epsilon}, because $\|J^*_k\|_{W^{m+1,p}(\Omega)} \leq \|u_k\|_{W^{m+1,p}(\Omega)}$, c.f. \eqref{epsilon_bound_0}.     

\Proof
The $\epsilon$-bound \eqref{Lemma2_J_epsilon} implies that $J=I + \epsilon J^*$ is invertible, since the supremum-norm of the Hilbert Schmidt norm of $J$ (taken point-wise) is bounded below by
\beq \nonumber
\| J \|_{L^\infty} = \| I + \epsilon J^* \|_{L^\infty}  
\geq  \| I \|_{L^\infty} -  \epsilon \| J^* \|_{L^\infty} 
\overset{\eqref{Morrey}}{\geq} 1 -  \epsilon \: C_M \| J^* \|_{W^{1,p}} 
\overset{\eqref{Lemma2_J_epsilon}}{>}    \frac12,
\eeq
keeping in mind that $\| I \|_{L^\infty} = 1$. Now, since $J\in W^{m+1,p}(\Omega)$ for $p>n$, Morrey's inequality implies that $J$ is H\"older continuous, so $J^{-1}$ is H\"older continuous as well. Substituting ansatz \eqref{J_inverse_eqn} into $J J^{-1} = I$ and solving for $J^{-*}$, we obtain that
\beq \label{techeqn1_J_inverse}
J^{-*} = - J^{-1} J^*,
\eeq
which implies existence and continuity of $J^{-*}$.

To prove estimate \eqref{J_inverse_bound}, that $J^{-*} \in W^{m+1,p}(\Omega )$, we proceed by induction in $m\geq 0$. To derive \eqref{J_inverse_bound} in the case $m=0$, we first use \eqref{J_inverse_eqn} to write \eqref{techeqn1_J_inverse} equivalently as 
$$
J^{-*} = - J^*- \epsilon J^{-*}  J^*.
$$ 
We now apply Morrey's inequality \eqref{Morrey} and the $\epsilon$-bound \eqref{Lemma2_J_epsilon} to estimate  
\begin{eqnarray} \nonumber
\|J^{-*}\|_{L^\infty} 
&\leq & \|J^*\|_{L^\infty} + \epsilon\: \|J^*\|_{L^\infty} \| J^{-*}\|_{L^\infty} \cr
&\overset{\eqref{Morrey}}{\leq}& \|J^*\|_{L^\infty} + \epsilon\:C_M \|J^*\|_{W^{1,p}} \| J^{-*}\|_{L^\infty} \cr
&\overset{\eqref{Lemma2_J_epsilon}}{\leq}& \|J^*\|_{L^\infty} + \frac 12 \| J^{-*}\|_{L^\infty}.
\end{eqnarray}
Subtraction of the last term gives 
\beq \label{techeqn3_J_inverse}
\|J^{-*}\|_{L^\infty} \ \leq \ 2 \|J^*\|_{L^\infty} \ \leq \ 2\: C_M \|J^*\|_{W^{1,p}},
\eeq
where we used Morrey's inequality \eqref{Morrey} in the last step. Since $\Omega$ is bounded, we conclude with the estimate
\beq \label{techeqn3b_J_inverse}
\|J^{-*}\|_{L^p}  \leq  \|J^{-*}\|_{L^\infty}   {\rm vol}(\Omega)  \overset{\eqref{techeqn3_J_inverse}}{\leq} 2  C_M{\rm vol}(\Omega)\: \|J^*\|_{W^{1,p}}  ,
\eeq
which proves \eqref{J_inverse_bound} for the case $m=0$ for $C_- = 2\: {\rm vol}(\Omega) C_M$.

We now show that $J^{-*} \in W^{1,p}(\Omega)$ and derive estimate \eqref{J_inverse_bound} for $m=1$. To begin, let $D_h$ denote the difference quotient in $x^j$-direction, (so that $D_h(f)$ converges to $\partial_j f$ as $h\rightarrow 0$ for $f \in W^{1,p}(\Omega)$). Now, since
\begin{eqnarray}  \label{techeqn2a_J_inverse}
0 \ = \ D_h(J^{-1} J)|_x \ = \ D_h(J^{-1})|_x \cdot J(x+h) + J^{-1}(x) \cdot D_h(J)|_x,
\end{eqnarray}
we have
\beq \label{techeqn2_J_inverse}
D_h(J^{-1})|_x = - J^{-1}(x) \cdot D_h(J)|_x \cdot  J^{-1}(x+h).
\eeq
The right hand side of \eqref{techeqn2_J_inverse} converges to $\partial_j f$ in $L^p(\Omega)$ as $h\rightarrow 0$, since
\begin{eqnarray}  \nonumber
\big\| J^{-1}  \big(D_h(J) - \partial_j J \big) J(\cdot +h) \big\|_{L^p} 
&\leq &   \| J^{-1} \|^2_{L^\infty} \big\| D_h(J) - \partial_j J \big\|_{L^p} 
\end{eqnarray}
converges to zero as $h \rightarrow 0$ by $L^p$-convergence of $D_h(J)$ to $\partial_j J$ for $J \in W^{1,p}(\Omega)$ and by boundedness of $\| J^{-1} \|_{L^\infty}$ independent of $h$ in light of \eqref{techeqn3_J_inverse}. Thus the left hand side of \eqref{techeqn2_J_inverse} converge in $L^p(\Omega)$ and the limit function is indeed the weak derivative of $J^{-1}$, which is given explicitly by 
\beq \label{techeqn2b_J_inverse}
\partial_j J^{-1} = - J^{-1} \cdot \partial_j(J) \cdot  J^{-1}.
\eeq 
This implies that $J^{-1} \in W^{1,p}(\Omega)$ and, in light of \eqref{J_inverse_eqn}, $J^{-*} \in W^{1,p}(\Omega)$. 

To derive estimate \eqref{J_inverse_bound} for $m=1$, substitute $J^{-1}= I + \epsilon J^{-*}$ on the left hand side of  \eqref{techeqn2b_J_inverse} and $J= I + \epsilon J^{*}$ on the right hand side, which gives
\begin{eqnarray} \nonumber
\epsilon \partial_j J^{-*} 
&=&\epsilon  J^{-1}\partial_j (J^{*}) \: J^{-1} ,
\end{eqnarray}
so dividing by $\epsilon$ and substituting $J^{-1}= I + \epsilon J^{-*}$ yields
\beq \label{techeqn2c_J_inverse}
\partial_j J^{-*} 
= \partial_j J^{*}  +\epsilon \: J^{-*}\partial_j (J^{*})  +\epsilon\: \partial_j (J^{*}) \: J^{-*} + \epsilon^2 J^{-*}\partial_j (J^{*}) \: J^{-*}.
\eeq
This  leads to the estimate  
\begin{eqnarray} \nonumber
\|\partial_j J^{-*}\|_{L^p} 
&\leq &  \big( 1 + \epsilon \: \| J^{-*} \|_{L^\infty} \big)^2  \|\partial_j J^{*}\|_{L^p}  \cr
&\overset{\eqref{techeqn3_J_inverse}}{\leq} & \big( 1 + 2\epsilon \: \| J^{*} \|_{L^\infty} \big)^2  \|\partial_j J^{*}\|_{L^p} \cr
&\overset{\eqref{Morrey}}{\leq} & \big( 1 + 2 C_M \epsilon \: \| J^{*} \|_{W^{1,p}} \big)^2  \|\partial_j J^{*}\|_{L^p} \cr
&\overset{\eqref{Lemma2_J_epsilon}}{\leq} & 4 \:  \|\partial_j J^{*}\|_{L^p} ,
\end{eqnarray}
which in combination with \eqref{techeqn3b_J_inverse} implies the sought after estimate  
\beq \nonumber
\| J^{-*}\|_{W^{1,p}} \leq  C_-  \|J^{*}\|_{W^{1,p}} ,
\eeq
where $C_- \equiv 4 + {\rm vol}(\Omega) C_M$. This proves \eqref{J_inverse_bound} in the case $m=1$.

To prove the general case, let $k \geq 1$, and assume that $J\in W^{k+1,p}(\Omega)$, $J^{-*}\in W^{k,p}(\Omega)$ and that \eqref{J_inverse_bound} holds for $m=k-1$, i.e.,
\beq \label{J_inverse_bound_IndAss}
\| J^{-*}\|_{W^{k,p}} \leq \: C'_- \: \|J^*\|_{W^{k,p}}.
\eeq
For the induction step, we need to show that \eqref{J_inverse_bound} holds for $k+1$. For this, we take $k$-th order derivatives of \eqref{techeqn2c_J_inverse} and find that
\begin{align} 
\partial^{k+1}(J^{-*})  
\ = \ & \partial^{k+1} ( J^*) + \epsilon\: \partial^k \Big( J^{-*} \partial( J^*) + \partial( J^*) \: J^{-*}  \Big)  \cr &+ \epsilon^2 \partial^k \big( J^{-*} \partial( J^*) \: J^{-*} \big), \label{Jinv_techeqn_3}
\end{align}
where $\partial^k$ denotes $k-th$ order partial derivatives, not necessarily all in the same direction. (Note that the right hand side of \eqref{techeqn2c_J_inverse} contains no derivatives of $J^{-*}$ so that we do not need to use difference quotients in \eqref{Jinv_techeqn_3}.) From \eqref{Jinv_techeqn_3}, it follows that $\|\partial^{k+1}(J^{-*})\|_{L^p}$ is bounded by $\|\partial^{k+1}(J^{*})\|_{L^p}$, by terms linear in $\epsilon$ which are of the form
\begin{eqnarray} \nonumber
\epsilon \| \partial^k ( J^{-*})  \partial( J^*)    \|_{L^p} 
&\leq &   \epsilon \| \partial^k ( J^{-*}) \|_{L^p} \| \partial( J^*)\|_{L^\infty} \cr
&\overset{\eqref{J_inverse_bound_IndAss}}{\leq} & C'_- \epsilon  \| ( J^{*}) \|_{W^{k,p}} \| \partial( J^*)\|_{L^\infty}  \cr
&\overset{\eqref{Lemma2_J_epsilon}}{\leq} &  C'_- \frac{1}{2C_M} \| \partial( J^*)\|_{L^\infty}  \cr
&\overset{\eqref{Morrey}}{\leq} & \frac12 C'_- \| J^{*} \|_{W^{k+1,p}},
\end{eqnarray}
or of the form
\begin{eqnarray} \nonumber
\epsilon \|  J^{-*}  \partial^{k+1}( J^*) \|_{L^p} 
&\leq &  \epsilon \|  J^{-*}\|_{L^\infty} \| \partial^{k+1}( J^*)\|_{L^p} \cr
&\overset{\eqref{techeqn3_J_inverse}}{\leq}&  \epsilon \|  J^{*}\|_{L^\infty} \| \partial^{k+1}( J^*)\|_{L^p} \cr
&\overset{\eqref{Morrey}}{\leq}&  \epsilon \:  C_M \| ( J^{*}) \|_{W^{1,p}} \| \partial^{k+1}( J^*)\|_{L^p}  \cr
&\overset{\eqref{Lemma2_J_epsilon}}{\leq}& \frac12 \: \|  J^{*} \|_{W^{k+1,p}},
\end{eqnarray}
or the more regular terms containing mixed derivatives, and by $\epsilon^2$-term in \eqref{Jinv_techeqn_3} which can be bounded in a similar fashion. Namely, denoting with $\mathcal{L}$ the $L^p$-norm of terms not containing the critical derivative $\partial^k J^{-*}$,  the $\epsilon^2$-term in \eqref{Jinv_techeqn_3} can be estimated by 
\begin{align}
\epsilon^2 \|\partial^k \big( J^{-*} &\partial( J^*) \: J^{-*} \big)\|_{L^p(\Omega)}  \cr
& \leq  \epsilon^2 \|\partial^k J^{-*} \|_{L^p(\Omega)} \|\partial( J^*)\|_{L^\infty(\Omega)} \| J^{-*}\|_{L^\infty(\Omega)} + \mathcal{L} \cr
&\overset{\eqref{Morrey}}{\leq}   \epsilon^2 C_M^2 \|\partial^k J^{-*} \|_{L^p(\Omega)} \|\partial( J^*)\|_{W^{1,p}(\Omega)} \| J^{-*}\|_{W^{1,p}(\Omega)} + \mathcal{L} \cr
&\overset{\eqref{J_inverse_bound_IndAss}}{\leq}   \epsilon^2 C_M^2 \| J^{*} \|^3_{W^{k,p}(\Omega)} + \mathcal{L} \cr
&\overset{\eqref{Lemma2_J_epsilon}}{\leq}   \| J^{*} \|_{W^{k,p}(\Omega)} + \mathcal{L}, \nonumber
\end{align}
while the term $\mathcal{L}$ can be estimated similarly by using  \eqref{Morrey}, \eqref{J_inverse_bound_IndAss} and \eqref{Lemma2_J_epsilon}. In summary, we showed that 
\beq \nonumber
\|\partial^{k+1}(J^{-*})\|_{L^p} \leq C_-  \| J^{*} \|_{W^{k+1,p}(\Omega)} ,
\eeq  
from which we conclude that \eqref{J_inverse_bound} holds for $k+1$, taking $C_-$ as the largest constant that appears in the above estimates (for $m\geq 1$ fixed). Recursion of the above argument proves \eqref{J_inverse_bound} in the general case $m\geq 1$. This completes the proof of Lemma \eqref{Lemma2_J_inverse}.   
\QED 
 
We now prove the basic estimates for the non-linear source terms on the right hand side of equations \eqref{pde_u} - \eqref{pde_a}, which are required for the proofs of Lemmas \ref{Lemma_existence_iterates} and \ref{Lemma_induction_consistency}.

\begin{Lemma}\label{Lemma_source_estimate_Fa_Fu}
Let $\Gamma^*,d\Gamma^* \in W^{m,p}(\Omega)$ for $m\geq 1$ and $p>n$, bounded by $C_0$ as in \eqref{Gamma-bound}, and assume $u \in W^{m+1,p}(\Omega)$ and $a\in W^{m,p}(\Omega)$. Then, if $\epsilon >0$ satisfies the bound \eqref{Lemma2_J_epsilon}, then there exists a constant $C_s>0$ depending only on $C_0$, $m$, $p$ and $\Omega$, such that          
\begin{eqnarray}                      
\|F_u(u,a)\|_{W^{m-1,p}} &\leq &  \epsilon\:  C_s \big( 1  + \|a\|_{W^{m,p}}  + \|u\|_{W^{m+1,p}} \big) \|u\|_{W^{m+1,p}} \cr  &&  +\: C_0 +  \|a\|_{W^{m,p}} \label{bound_Fu}\\
\|F_a(u)\|_{W^{m-1,p}} &\leq & C_0 + \epsilon \: C_s \big( 1 + \|u\|_{W^{m+1,p}} \big) \|u\|_{W^{m+1,p}} .   \label{bound_Fa}
\end{eqnarray}
\end{Lemma} 

\Proof  
We focus on proving the lemma in the case $m=1$, since higher derivative estimates for $m>1$ then follow by an analogous argument. Note that, because $\epsilon >0$ is assumed to satisfy \eqref{Lemma2_J_epsilon}, Lemma \ref{Lemma_J_inverse} applies and yields the existence of the inverse $J^{-1} = I + \epsilon J^{-*}$ together with the estimate \eqref{J_inverse_bound} on $ J^{-*}$.

We first derive \eqref{bound_Fu} in the case $m=1$. From \eqref{Def_Fu} we find that 
\begin{align}  \label{techeqn0_estimates_Fu}
\| F_u(u,a) \|_{L^p} &\leq   \| \delta d \Gamma^* \|_{L^p} + \|\delta \Gamma^* \|_{L^p}  + \|a\|_{L^p} + \| d a \|_{L^p} + \epsilon \|\delta ( J^* \mm \Gamma^* )\|_{L^p}  \hspace{.7cm} \cr 
&\ \ \ + \epsilon \:\| d (J^{-*} a) \|_{L^p}  + \epsilon \|\langle d J^* ; \tilde{\Gamma}^*\rangle \|_{L^p}  + \|\delta d\big( J^{-1} \mm dJ^* \big)\|_{L^p}  ,
\end{align}
where we used that $d (J^{-1} a) = \epsilon d(J^{-*} a) + da$ by \eqref{J_inverse_eqn}. We now estimate the right hand side term by term. By our incoming assumption \eqref{Gamma-bound} on the spacetime connection we have
\beq \label{techeqn1_estimates_Fu}
\| \delta d \Gamma^* \|_{L^p} + \|\delta \Gamma^* \|_{L^p} \leq \|d \Gamma^* \|_{W^{1,p}} + \| \Gamma^* \|_{W^{1,p}} \leq C_0,
\eeq
and clearly we have
\beq \label{techeqn3a_estimates_Fu}
 \|a\|_{L^p} + \| d a \|_{L^p} \leq   \|a\|_{W^{1,p}} .
\eeq
Applying the Leibniz-rule \eqref{Leibniz_rule}, we find that      
\beq \nonumber
\delta ( J^* \mm \Gamma^* ) = \langle d J^* ; \Gamma^* \rangle + J \mm \delta \Gamma^*,
\eeq
which leads to the bound
\begin{eqnarray} \label{techeqn2_estimates_Fu}
\|\delta ( J^* \mm \Gamma^* )\|_{L^p}  
& \leq & \| d J^*\|_{L^\infty}  \| \Gamma^* \|_{L^p} + \| J \|_{L^\infty} \| \delta \Gamma^* \|_{L^p} \cr
& \overset{\eqref{Morrey}}{\leq} & C_M \Big(\| d J^*\|_{W^{1,p}} \| \Gamma^* \|_{L^p} + \| J \|_{W^{1,p}} \| \delta \Gamma^* \|_{L^p} \Big) \cr
& \leq & C_M  \| \Gamma^* \|_{W^{1,p}}  \| J^*\|_{W^{2,p}} ,
\end{eqnarray}
where the H\"older continuity of $J^* \in W^{2,p}(\Omega)$ allowed us to estimate the $L^p$-norm of products in terms of the $L^\infty$-norm on $dJ^*$ and $J^*$, which we further estimated using Morrey's inequality \eqref{Morrey}.    
Similarly, the H\"older continuity of $a\in W^{1,p}(\Omega)$ and of $J^{-*}$ together with the bound \eqref{J_inverse_bound} on $J^{-*}$ lead to 
\begin{eqnarray} \label{techeqn3_estimates_Fu}
 \| d (J^{-*} a) \|_{L^p} 
&\leq &   \| d( J^{-*})\|_{L^\infty} \| a \|_{L^p} +  \|  J^{-*} \|_{L^\infty} \| d a\|_{L^p}  \cr
& \overset{\eqref{Morrey}}{\leq} & C_M\: \|a\|_{W^{1,p}} \Big( \| d( J^{-*})\|_{W^{1,p}} + \| J^{-*}\|_{W^{1,p}} \Big) \cr 
& \overset{\eqref{J_inverse_bound}}{\leq} &  C_-\: C_M \: \|a\|_{W^{1,p}} \: \| J^{*}\|_{W^{2,p}}.
\end{eqnarray}
In a similar fashion, we obtain
\beq \label{techeqn4_estimates_Fu}
\|\langle d J^* ; \tilde{\Gamma}^*\rangle \|_{L^p}  
\leq \| d J^* \|_{L^p} \| \tilde{\Gamma}^* \|_{L^\infty}
\leq C_M \: \| J^* \|_{W^{1,p}} \| \tilde{\Gamma}^* \|_{W^{1,p}} ,
\eeq 
where we applied again Morrey's inequality \eqref{Morrey}. For the last term in \eqref{techeqn0_estimates_Fu}, use the Leibniz rule \eqref{Leibnitz-rule} together with $d^2=0$ and formula \eqref{J_inverse_eqn} for $J^{-1}$, to compute 
\beq \label{techeqn5a_estimates_Fu}
d\big( J^{-1} \mm dJ^* \big) =  \epsilon \: d J^{-*} \wedge dJ^*  ,
\eeq
which leads to the estimate
\begin{eqnarray} \label{techeqn5_estimates_Fu}
     \|\delta d\big( J^{-1} \mm dJ^* \big)\|_{L^p} 
& \leq & \epsilon\:  \|\delta ( d J^{-*} \wedge dJ^* )\|_{L^p} \cr
& \leq & \epsilon\: \| J^{-*} \|_{W^{2,p}} \| dJ^* \|_{L^\infty} + \epsilon\: \| dJ^{-*} \|_{L^\infty} \| J^{*} \|_{W^{2,p}} \cr
& \leq & \epsilon\: 2C_- C_M \: \| J^{*} \|_{W^{2,p}} \| J^{*} \|_{W^{2,p}} ,
\end{eqnarray}
where we applied Morrey's inequality \eqref{Morrey} together with the bound \eqref{J_inverse_bound} on $J^{-*}$ in the last step. Combing now the estimates \eqref{techeqn1_estimates_Fu} - \eqref{techeqn5_estimates_Fu} to bound the right hand side in \eqref{techeqn0_estimates_Fu}, we obtain the sought after estimate \eqref{bound_Fu}.

Estimate \eqref{bound_Fu} for the general case $m\geq 1$ follows by a straightforward adaptation of the argument \eqref{techeqn0_estimates_Fu} - \eqref{techeqn5_estimates_Fu} to the $W^{m-1,p}$-norm, using H\"older continuity of $m-1$-derivatives of $u$, $a$, $\Gamma^*$ or $d\Gamma^*$ to estimate products in terms of products of the $L^p$-norm and the $L^\infty$-norm of such derivatives. For instance, estimate \eqref{techeqn5_estimates_Fu} extends as follows:
\begin{align} \label{techeqn5b_estimates_Fu}
\|\delta d&\big( J^{-1} \mm dJ^* \big)\|_{W^{m-1,p}} 
 \overset{\eqref{techeqn5a_estimates_Fu}}{\leq}  \epsilon\:  \|\delta ( d J^{-*} \wedge dJ^* )\|_{W^{m-1,p}} \cr
& \overset{(*)}{\leq}  \epsilon\: \|d J^{-*} \|_{W^{m,p}} \: C_M\| dJ^* \|_{W^{m,p}} + \epsilon\: C_M \| dJ^{-*} \|_{W^{m,p}} \: \| dJ^{*} \|_{W^{m,p}} \cr
& \overset{\eqref{J_inverse_bound}}{\leq}  \epsilon\: 2C_- C_M \: \| J^{*} \|_{W^{m+1,p}} \| J^{*} \|_{W^{m+1,p}} ,
\end{align}
where in the first term in $(*)$ results from applying Morrey's inequality \eqref{Morrey} to estimate derivatives of order less than $m-1$ of $dJ^*$ (which are H\"older continuous), while the second term in $(*)$ results from applying \eqref{Morrey} to derivatives of order less than $m-1$ of $dJ^{-*}$. Extending \eqref{techeqn0_estimates_Fu} - \eqref{techeqn4_estimates_Fu} analogously to \eqref{techeqn5b_estimates_Fu} proves the sought after estimate \eqref{bound_Fu} for the general case $m\geq 1$.

We now prove \eqref{bound_Fa} in the case $m=1$. From our definition of $F_a$ in \eqref{Def_Fa} we find that 
\begin{eqnarray}  \label{techeqn0_estimates_Fa}
\|F_a(u)\|_{L^p}  &\leq & \|\overrightarrow{\text{div}} \big(d\Gamma^*\big) \|_{L^p} + \epsilon\: \| \overrightarrow{\text{div}} \big( J^* \mm d\Gamma^*\big) \|_{L^p} \cr 
&+& \epsilon\: \| \overrightarrow{\text{div}} \big(dJ^* \wedge \Gamma^*\big) \|_{L^p} + \epsilon\: \| d\big(\overrightarrow{\langle d J^* ; \tilde{\Gamma}^*\rangle }\big)\|_{L^p} .   
\end{eqnarray}
We now estimate each term on the right hand side of \eqref{techeqn0_estimates_Fa} separately. Our incoming assumption \eqref{Gamma-bound} immediately gives
\begin{eqnarray} \label{techeqn1_estimates_Fa}
\|\overrightarrow{\text{div}} \big(d\Gamma^*\big) \|_{L^p} \leq \| d\Gamma^* \|_{W^{1,p}}   \leq C_0.
\end{eqnarray}
Applying Morrey's inequality \eqref{Morrey} to bound the supremum-norm of $J^*$ and $d\Gamma^*$ leads to 
\begin{eqnarray} \label{techeqn3_estimates_Fa}
\| \overrightarrow{\text{div}} \big( J^* \mm d\Gamma^*\big) \|_{L^p} 
&\leq & \| J^* \|_{W^{1,p}}   \| d\Gamma^* \|_{L^\infty} + \| J^*\|_{L^\infty} \|d\Gamma^*\|_{W^{1,p}} \cr
&\overset{\eqref{Gamma-bound}}{\leq} & 2C_M \: C_0\: \|J^*\|_{W^{1,p}} \cr
&\leq & 2 C_M \: C_0\: \|J^*\|_{W^{2,p}}.
\end{eqnarray} 
Likewise, using \eqref{Morrey} to bound the supremum-norm of $dJ^*$ and $\Gamma^*$, we obtain
\begin{eqnarray} \label{techeqn2_estimates_Fa}
\|\overrightarrow{\text{div}} \big(dJ^* \wedge \Gamma^*\big) \|_{L^p}   
& \leq & \| dJ^* \|_{W^{1,p}}   \| \Gamma^* \|_{L^\infty} + \| dJ^*\|_{L^\infty} \|\Gamma^*\|_{W^{1,p}} \cr
& \overset{\eqref{Morrey}}{\leq} & C_M \big( \| dJ^* \|_{W^{1,p}}   \| \Gamma^* \|_{W^{1,p}} +  \| dJ^* \|_{W^{1,p}}   \| \Gamma^* \|_{W^{1,p}} \big) \cr
& \overset{\eqref{Gamma-bound}}{\leq} &  2\: C_M \: C_0\: \|J^*\|_{W^{2,p}} .
\end{eqnarray}
Finally, we estimate the non-linear term by
\begin{eqnarray} \label{techeqn4_estimates_Fa}
\| d\big(\overrightarrow{\langle d J^* ; \tilde{\Gamma}^*\rangle }\big)\|_{L^p}
& \leq & \| dJ^* \|_{W^{1,p}} \| \Gammati^* \|_{L^\infty} + \| dJ^* \|_{L^\infty} \| \Gammati^* \|_{W^{1,p}} \cr
& \overset{\eqref{Morrey}}{\leq} & C_M \big(  \| dJ^* \|_{W^{1,p}} \| \Gammati^* \|_{W^{1,p}} + \| dJ^* \|_{W^{1,p}} \| \Gammati^* \|_{W^{1,p}} \big) \cr
& \leq & 2\: C_M \|u\|^2_{W^{2,p}}  ,
\end{eqnarray}
recalling that $u\equiv (J^*,\Gammati^*)$. Combing \eqref{techeqn1_estimates_Fa} - \eqref{techeqn4_estimates_Fa} to bound the right hand side in \eqref{techeqn0_estimates_Fa} we obtain the sought after estimate \eqref{bound_Fa} in the case $m=1$.

Estimate \eqref{bound_Fa} for the general case $m\geq 1$ follows by extending \eqref{techeqn0_estimates_Fa} - \eqref{techeqn4_estimates_Fa} to the $W^{m-1,p}$-norm in a fashion similar to \eqref{techeqn5b_estimates_Fu}. Taking $C_s>0$ as the maximum over all constants in \eqref{techeqn1_estimates_Fu} - \eqref{techeqn4_estimates_Fa} and the constants arising from higher derivatives estimates completes the proof.
\QED

\subsection{Well-posedness of iteration scheme - Proof of Lemma \ref{Lemma_existence_iterates}}    \label{Sec_existence}

We now prove Lemma \ref{Lemma_existence_iterates}, which gives well-posedness of the iteration scheme and the basic elliptic estimates \eqref{existence_est1} - \eqref{existence_est4}. For this, assume $u_k\in W^{m+1,p}(\Omega)$ is given, for $m\geq 1$, $p>n\geq 2$, and assume $\epsilon$ satisfies \eqref{epsilon_bound_0}, that is $0<  \epsilon \leq  \epsilon(k)$.  Lemma \ref{Lemma_existence_iterates} states that there exists $a_{k+1}\in W^{m,p}(\Omega)$ which solves \eqref{iterate_a} - \eqref{iterate_a_bdd}, there exists $\psi_{k+1} \in W^{m,p}(\Omega)$ and $y_{k+1} \in W^{m+2,p}(\Omega)$ which solve \eqref{iterate_psi} - \eqref{iterate_y}, and there exists $u_{k+1}\in W^{m+1,p}(\Omega)$ which solves \eqref{iterate_u} with boundary data \eqref{bdd1} - \eqref{bdd2}, and these solutions satisfy the elliptic estimates \eqref{existence_est1} - \eqref{existence_est4}. \\

\noindent {\it Proof of Lemma \ref{Lemma_existence_iterates}.} 
First note that assumption \eqref{epsilon_bound_0}, that $0< \epsilon \leq  \epsilon(k)$ implies that $\epsilon$ satisfies the bound \eqref{Lemma2_J_epsilon}, so that the source estimates of Lemma \ref{Lemma_source_estimate_Fa_Fu} apply and yield $F_a(u_k) \in W^{m-1,p}(\Omega)$ and $F_u(u_k,a_{k+1}) \in W^{m-1,p}(\Omega)$.

We begin the proof by proving existence of a solution $a_{k+1}$ to the first order system \eqref{iterate_a} - \eqref{iterate_a_bdd} by applying Theorem \ref{Thm_CauchyRiemann} (i). For this, first note that the conditions of Theorem \ref{Thm_CauchyRiemann} (i) are met. In particular, the condition $df=0$ of Theorem \ref{Thm_CauchyRiemann} (i)  is satisfied by \eqref{iterate_a}, since $F_a(u_k)$ is the exterior derivative $d$ of a vector valued differential form, namely,
\beq \label{existence_iterates_techeqn1}
F_a(u) = d \big(\overrightarrow{\delta ( J \mm \Gamma )}\big) - d\big(\overrightarrow{\langle d J ; \tilde{\Gamma}\rangle }\big),
\eeq 
c.f. equation (3.40) in \cite{ReintjesTemple_ell1}. This shows that \eqref{iterate_a} - \eqref{iterate_a_bdd} satisfies the assumption of Theorem \ref{Thm_CauchyRiemann} (i). Regarding regularity, our incoming assumption $u_k\in W^{m+1,p}(\Omega)$ together with the source estimates of Lemma \ref{Lemma_source_estimate_Fa_Fu} show that $F_a(u_k) \in W^{m-1,p}(\Omega)$. We conclude that Theorem \ref{Thm_CauchyRiemann} applies to \eqref{iterate_a} - \eqref{iterate_a_bdd} and yields the existence of a solution $a_{k+1} \in W^{m,p}(\Omega)$. Moreover, by Gaffney's inequality \eqref{Gaffney_2}, this solution satisfies
\begin{align} \nonumber
\| \vec{a}_{k+1} \|_{W^{m,p}(\Omega)} 
\leq  C\: \| F_a(u_k)\|_{W^{m-1,p}(\Omega)} ,
\end{align}
for some constant $C>0$ depending only on $\Omega$, $m,n,p$, which is the sought after estimate \eqref{existence_est1}; (note that the range of $m$ in Theorem \ref{Thm_CauchyRiemann} starts at $m=0$, while here we assume $m\geq 1$). 

To prove the existence of a $\psi_{k+1}$ solving $d \psi_{k+1} = \overrightarrow{F_J(u_k)} - \overrightarrow{a_{k+1}}$, i.e., equation \eqref{iterate_psi}, we first show the consistency condition that the exterior derivative on the right hand side of \eqref{iterate_psi}, interpreted as a vector valued $1$-form, vanishes. For this recall from \eqref{F_notation} that $F_a(u) = d \overrightarrow{F_J}$, so that equation \eqref{iterate_a} for $a_{k+1}$ implies the sought after consistency condition
\beq \label{existence_iterates_techeqn2}
d \big(\overrightarrow{F_J(u_k)} - \overrightarrow{a_{k+1}} \big) =  F_a(u) - d\overrightarrow{a_{k+1}} \overset{\eqref{iterate_a}}{=} 0.
\eeq
Moreover, the source estimate \eqref{bound_Fu} in combination with $a_{k+1} \in W^{m,p}(\Omega)$ imply that $\overrightarrow{F_u(u_k,a_{k+1})} = \overrightarrow{F_J(u_k)} - \overrightarrow{a_{k+1}} \in W^{m-1,p}(\Omega)$.  Thus, Theorem \ref{Thm_CauchyRiemann} (ii) yields existence of a vector valued $0$-form $\psi_{k+1} \in W^{m,p}(\Omega)$ solving \eqref{iterate_psi} such that estimate \eqref{existence_est3} holds,
$$
\|\psi_{k+1}\|_{W^{m,p}(\Omega)}  \leq  C \; \|F_u(u_k,a_{k+1})\|_{W^{m-1,p}(\Omega)},
$$
keeping in mind that $F_u(u_k,a_{k+1}) = F_J(u_k) - a_{k+1}$. 

The existence of a solution $y_{k+1} \in W^{m+2,p}(\Omega)$ to \eqref{iterate_y} follows from the existence theorem for the Dirichlet problem of the Poisson equation with $L^p$ sources, Theorem \ref{Thm_Poisson}, keeping in mind that $F_u(u_k,a_{k+1}) \in W^{m-1,p}(\Omega)$ by Lemma \ref{Lemma_source_estimate_Fa_Fu}. We now prove estimate \eqref{existence_est4}. Applying the elliptic estimate \eqref{Poissonelliptic_estimate_Lp} component-wise, $\Delta y_{k+1} = \psi_{k+1}$ and $y_{k+1}=0$ on $\partial\Omega$, c.f. \eqref{iterate_y}, we obtain
\begin{eqnarray} \nonumber
\|y_{k+1}\|_{W^{m+2,p}(\Omega)} 
&\leq &    C \:  \| \psi_{k+1} \|_{W^{m,p}(\Omega)} \\
& \overset{\eqref{existence_est3}}{\leq}&   C \: \|F_u(u_k,a_{k+1})\|_{W^{m-1,p}(\Omega)} ,\label{existence_iterates_techeqn3}
\end{eqnarray}
where we absorbed the constant from the estimate on $\|\psi_{k+1}\|_{W^{m,p}}$ into the universal constant $C>0$. This is the sought after estimate \eqref{existence_est4}.

Finally, we prove existence of a solution $u_{k+1} \in W^{m+1,p}(\Omega)$ of \eqref{iterate_u} with boundary data \eqref{bdd1} - \eqref{bdd2}. Since $F_u(u_k,a_{k+1}) \in W^{m-1,p}(\Omega)$, existence of a solution $u_{k+1} \in W^{m+1,p}(\Omega)$ of the Poisson equation \eqref{iterate_u} with the Dirichlet boundary data \eqref{bdd1} - \eqref{bdd2} follows from Theorem \ref{Thm_Poisson}. To prove estimate \eqref{existence_est2}, we apply the basic elliptic estimate \eqref{Poissonelliptic_estimate_Lp} to $\Delta u_{k+1} = F_u(u_k,a_{k+1})$ with Dirichlet data $(\Gammati^*_{k+1},J^*_{k+1} - dy_{k+1}) \in W^{1,p}_0(\Omega)$, and thereby obtain 
\begin{align} \label{elliptic_estimate_techeqn1}
\|u_{k+1}&\|_{W^{m+1,p}(\Omega)} 
\leq  C  \Big( \|F_u(u_k,a_{k+1})\|_{W^{m-1,p}(\Omega)} + \|dy_{k+1}\|_{W^{m+1,p}(\Omega)} \Big).
\end{align}
Using estimate \eqref{existence_iterates_techeqn3} to bound $\|dy_{k+1}\|_{W^{m+1,p}(\Omega)}$ in \eqref{elliptic_estimate_techeqn1}, we finally obtain
\beq \nonumber
\|u_{k+1}\|_{W^{m+1,p}(\Omega)} \leq  C_e \; \|F_u(u_k,a_{k+1})\|_{W^{m-1,p}(\Omega)},
\eeq
which is the sought after estimate \eqref{existence_est2}, where we take $C_e$ as the maximum over all constants in the above estimates. This completes the proof.
\hfill $\Box$

\subsection{Estimates on Differences of Iterates - Proof of Lemma \ref{Lemma_sources_difference}}   \label{Sec_differences}

We introduce the notation 
\begin{eqnarray} \nonumber
\overline{\tilde{\Gamma}^*_k} &\equiv & \tilde{\Gamma}^*_k - \tilde{\Gamma}^*_{k-1}, \cr
\overline{J^*_k} &\equiv & J^*_k - J^*_{k-1},
\end{eqnarray}
so $\overline{u_k}  = (\overline{J^*_k},\overline{\tilde{\Gamma}^*_k})$. Let $J^{-1}_k$ be the inverse of $J_k\equiv I + \epsilon J^*_k$ and let $J^{-1}_{k-1}$ be the inverse of $J_{k-1}\equiv I + \epsilon J^*_{k-1}$, and denote with $J^{-*}_k$ the matrix valued $0$-form that satisfies $J^{-1}_k = I +\epsilon J^{-*}_k$ and likewise $J^{-1}_{k-1} = I +\epsilon J^{-*}_{k-1}$. We begin by deriving a bound on $\overline{J^{-*}_k} \equiv J^{-*}_k - J^{-*}_{k-1}$.

\begin{Lemma} \label{Lemma_diff_J_inverse} 
Assume $u_k, u_{k-1} \in W^{m+1,p}(\Omega)$ for $m\geq 0$, $p>n$, and assume $0<  \epsilon\leq  \min \big(\epsilon(k),\epsilon(k-1)\big)$, so $\epsilon$ satisfies \eqref{epsilon_bound_0} in terms of $u_k$ and $u_{k-1}$. Then $J_k$ and $J_{k-1}$ are invertible with $J^{-1}_k = I +\epsilon J^{-*}_k \in W^{m+1,p}(\Omega)$ and $J^{-1}_{k-1} = I +\epsilon J^{-*}_{k-1} \in W^{m+1,p}(\Omega)$, and there exists a constant $C_-'>0$ depending only on $m$, $n$, $p$, $\Omega$, such that
\beq \label{J_inverse_diff_bound}
\|\overline{ J^{-*}_k} \|_{W^{m+1,p}} \leq C_-' \|\overline{J^*_k}\|_{W^{m+1,p}}.
\eeq                   
\end{Lemma}                      
                           
\Proof
To begin, note that the $\epsilon$-bound \eqref{epsilon_bound_0}, $0< \epsilon \leq  \epsilon(k)$, implies that $\epsilon$ satisfies \eqref{Lemma2_J_epsilon} for $J^*=J^*_k$, so that Lemma \ref{Lemma2_J_inverse} implies that $J_k$ is invertible with $J_k^{-1} =  I +\epsilon J^{-*}_k$ and $J^{-*}_k \in W^{m+1,p}(\Omega)$. Likewise, the $\epsilon$-bound \eqref{epsilon_bound_0} for $u_{k-1}$ implies that $J_{k-1}$ is invertible with $J^{-1}_{k-1} = I +\epsilon J^{-*}_{k-1} \in W^{m+1,p}(\Omega)$.

Now, substituting $J_k=I + \epsilon J^*_k$ and $J^{-1}_k = I + \epsilon J^{-*}_k$ into the identity
\beq \nonumber
0 = J_k J^{-1}_k - J_{k-1} J^{-1}_{k-1},
\eeq 
and solving for $\overline{J^{-*}_k}  \equiv J^{-*}_k - J^{-*}_{k-1}$, we find after  dividing by $\epsilon$ that
\begin{eqnarray} 
\overline{J^{-*}_k}  
&= & -\overline{J^{*}_k} - \epsilon\: \big( J^{*}_k J^{-*}_k - J^{*}_{k-1}J^{-*}_{k-1}  \big)     \cr
&=&  - \overline{J^{*}_k} - \epsilon \: \big( \overline{J^{*}_k}\cdot J^{-*}_k + J^{*}_{k-1}\cdot \overline{J^{-*}_k}     \big) .  \label{J_inverse_diff_id} 
\end{eqnarray}
Thus, taking the $L^p$ norm of \eqref{J_inverse_diff_id} and applying Morrey's inequality \eqref{Morrey}, gives
\begin{eqnarray}\nonumber
\big\| \overline{J^{-*}_k} \big\|_{L^p}  
&\leq & \big\| \overline{J^{*}_k} \big\|_{L^p}  + \epsilon\: \big\|\overline{J^{*}_k} J^{-*}_k\big\|_{L^p} + \epsilon\: \big\|J^{*}_{k-1} \overline{J^{-*}_k}\big\|_{L^p} \cr
&\leq & \big\| \overline{J^{*}_k} \big\|_{L^p}  + \epsilon\: \big\|\overline{J^{*}_k}\big\|_{L^p} \| J^{-*}_k\|_{L^\infty} + \epsilon\: \|J^{*}_{k-1} \|_{L^\infty} \big\| \overline{J^{-*}_k}\big\|_{L^p} \cr
& \overset{\eqref{Morrey}}{\leq}& \big\| \overline{J^{*}_k} \big\|_{L^p}  + \epsilon\: C_M \big\|\overline{J^{*}_k}\big\|_{L^p} \| J^{-*}_k\|_{W^{1,p}} + \epsilon\: C_M \|J^{*}_{k-1} \|_{W^{1,p}} \big\| \overline{J^{-*}_k}\big\|_{L^p}.
\end{eqnarray}
Using for the last term that the $\epsilon$-bound \eqref{epsilon_bound_0} for $u_{k-1}$ implies that
\beq \nonumber
\epsilon\: C_M \|J^{*}_{k-1} \|_{W^{1,p}} \leq  \frac12,
\eeq
we find after subtraction of $\frac12 \big\| \overline{J^{-*}_k} \big\|_{L^p}$ that
\begin{eqnarray} \label{techeqn0_J_inv_diff}
\frac12 \big\| \overline{J^{-*}_k} \big\|_{L^p} 
&\leq & \big( 1  + \epsilon\: C_M \| J^{-*}_k\|_{W^{1,p}} \big) \big\| \overline{J^{*}_k} \big\|_{L^p} \cr
&\overset{\eqref{J_inverse_bound}}{\leq} & \big( 1  + \epsilon\: C_- C_M \| J^{*}_k\|_{W^{1,p}} \big) \big\|\overline{J^{*}_k}\big\|_{L^p} .
\end{eqnarray}
Now, using the $\epsilon$-bound \eqref{epsilon_bound_0} for $u_{k}$, we find that
\beq \nonumber
\epsilon\: C_M \| J^{*}_k\|_{W^{1,p}} \leq  \frac{1}{2} ,
\eeq
which in light of \eqref{techeqn0_J_inv_diff} gives 
\beq \label{techeqn1_J_inv_diff}
\big\| \overline{J^{-*}_k} \big\|_{L^p} \leq  (2C_M +C_-) \big\| \overline{J^{*}_k} \big\|_{L^p}.
\eeq 

To prove \eqref{J_inverse_diff_bound} for $m=1$, we first differentiate \eqref{J_inverse_diff_id} and find
\beq \label{techeqn1b_J_inv_diff}
\partial_j  \overline{J^{-*}_k}
= - \partial_j \overline{J^{*}_k}  - \epsilon \Big( \partial_j \overline{J^{*}_k} \mm J^{-*}_k + \partial_j J^{*}_{k-1} \mm \overline{J^{-*}_k}  + \overline{J^{*}_k} \; \partial_j  J^{-*}_k +  J^{*}_{k-1}  \partial_j \overline{ J^{-*}_k} \Big), 
\eeq
which implies for the gradient $d\overline{J^{-*}_k}$ the estimate
\begin{eqnarray} \nonumber
\| d\overline{J^{-*}_k}  \|_{L^{p}}
&\leq& \| d\overline{J^{*}_k} \|_{L^p}   + \epsilon\: \| J^{-*}_k \|_{L^\infty} \|\overline{J^{*}_k}\|_{W^{1,p}}
+ \epsilon\: \| J^{*}_{k-1} \|_{W^{1,p}} \| \overline{J^{-*}_k} \|_{L^\infty} 
\cr && + \epsilon\: \| J^{-*}_k \|_{W^{1,p}} \| \overline{J^{*}_k} \|_{L^\infty} + \epsilon\: \| J^{*}_{k-1} \|_{L^\infty}  \| \overline{J^{-*}_k} \|_{W^{1,p}} ,
\end{eqnarray}
where we bounded undifferentiated terms by their $L^\infty$-norm and differentiated terms by their $W^{1,p}$-norm. Applying Morrey's inequality \eqref{Morrey} we obtain the further estimate
\beq \nonumber
\| d\overline{J^{-*}_k}  \|_{L^{p}}
\leq  \| d\overline{J^{*}_k} \|_{L^p}   + \epsilon\:2 C_M \Big( \| J^{-*}_k \|_{W^{1,p}} \|\overline{J^{*}_k}\|_{W^{1,p}} +  \| J^{*}_{k-1} \|_{W^{1,p}} \| \overline{J^{-*}_k} \|_{W^{1,p}} \Big)   ,
\eeq
and applying the bound \eqref{J_inverse_bound} on $J^{-*}_k$ and $J^{-*}_{k-1}$ yields 
\begin{eqnarray} \nonumber
\| d\overline{J^{-*}_k}  \|_{L^{p}}  
&\leq & \| d\overline{J^{*}_k} \|_{L^p}   + 2\epsilon\: C_M \: C_- \| J^{*}_k \|_{W^{1,p}} \|\overline{J^{*}_k}\|_{W^{1,p}} \cr
& & + 2\epsilon\: C_M \| J^{*}_{k-1} \|_{W^{1,p}} \| \overline{J^{-*}_k} \|_{W^{1,p}} 
\end{eqnarray}
so that the $\epsilon$-bound \eqref{epsilon_bound_0} for $u_k$ and $u_{k-1}$ implies
\begin{eqnarray} \label{techeqn2_J_inv_diff}
\| d\overline{J^{-*}_k}  \|_{L^{p}}  
&\leq & \| d\overline{J^{*}_k} \|_{L^p}   + \frac12 C_-  \|\overline{J^{*}_k}\|_{W^{1,p}} + \frac12 \| \overline{J^{-*}_k} \|_{W^{1,p}}  .
\end{eqnarray}
Adding $\|\overline{J^{-*}_k}  \|_{L^{p}}$ to both sides of \eqref{techeqn2_J_inv_diff} and using estimate \eqref{techeqn1_J_inv_diff} to bound $\|\overline{J^{-*}_k}  \|_{L^{p}}$ on the right hand side,
we find
\begin{eqnarray} \nonumber
\|\overline{J^{-*}_k}  \|_{W^{1,p}}  
& \leq &  3( C_M +C_- +1) \|\overline{J^{*}_k} \|_{W^{1,p}} + \frac12 \| \overline{J^{-*}_k} \|_{W^{1,p}} .
\end{eqnarray} 
So subtraction of the second term on the right hand side finally yields
\begin{eqnarray} \nonumber
\|\overline{J^{-*}_k}  \|_{W^{1,p}}  
& \leq &  6( C_M +C_- +1) \|\overline{J^{*}_k} \|_{W^{1,p}} ,
\end{eqnarray} 
which is the sought after bound \eqref{J_inverse_diff_bound} for $m=1$ and $C_-' = 6( C_M +C_- +1)$. 

To derive \eqref{J_inverse_diff_bound} for $m\geq 2$, we proceed by induction. For this, assume  \eqref{J_inverse_diff_bound} holds for some $1 \leq l \leq m$, i.e.
\beq \label{J_inverse_diff_bound_IndAss}
\|\overline{ J^{-*}_k} \|_{W^{l,p}} \leq C_-' \|\overline{J^*_k}\|_{W^{l,p}},
\eeq 
and assume $J^{-1}_k = I +\epsilon J^{-*}_k \in W^{m+1,p}(\Omega)$ and $J^{-1}_{k-1} = I +\epsilon J^{-*}_{k-1} \in W^{m+1,p}(\Omega)$, (c.f. Lemma \ref{Lemma_J_inverse}). We need to show that \eqref{J_inverse_diff_bound_IndAss} holds for $l+1$. For this, denote with $\partial^{l+1}$ a combination of partial derivatives of $l+1$-st order (not necessarily in the same direction), i.e. $\partial^{l+1}$ denotes partial differentiation corresponding to a specific multi-index. Now, taking $\partial^{l+1}$ of \eqref{J_inverse_diff_id}, we obtain
\beq \nonumber
\partial^{l+1} \overline{J^{-*}_k}  
= - \partial^{l+1} \overline{J^{*}_k} - \epsilon \: \partial^{l+1} \big( \overline{J^{*}_k}\cdot J^{-*}_k \big) - \epsilon \: \partial^{l+1} \big( J^{*}_{k-1}\cdot \overline{J^{-*}_k} \big)   , 
\eeq
which gives the estimate
\beq \label{techeqn3_J_inv_diff}
\|\partial^{l+1} \overline{J^{-*}_k} \|_{L^p}
 \leq  \| \partial^{l+1} \overline{J^{*}_k} \|_{L^p} + \epsilon \| \partial^{l+1} \big( \overline{J^{*}_k}\mm J^{-*}_k \big) \|_{L^p}  + \epsilon \| \partial^{l+1} \big( J^{*}_{k-1}\mm \overline{J^{-*}_k} \big) \|_{L^p}.
\eeq
The first term on the right hand side is bounded by the $W^{l+1,p}$-norm of $\overline{J^{*}_k} .$ Using Morrey's inequality \eqref{Morrey} to estimate product terms and using \eqref{J_inverse_bound} to bound the $W^{l+1,p}$-norm of $J^{-*}_k$, we estimate the second term on the right hand side of \eqref{techeqn3_J_inv_diff} by
\begin{align} \label{techeqn4_J_inv_diff}
\epsilon\:  \| \partial^{l+1} \big( \overline{J^{*}_k}\cdot J^{-*}_k \big) \|_{L^p}
& \overset{\eqref{Morrey}}{\leq} \epsilon  C_M \: (l+1)!\: \| \overline{J^{*}_k} \|_{W^{l+1,p}} \| J^{-*}_k \|_{W^{l+1,p}}   \cr
& \overset{\eqref{J_inverse_bound}}{\leq}  \epsilon  C_M C_-\: (l+1)!\: \| \overline{J^{*}_k} \|_{W^{l+1,p}} \| J^{*}_k \|_{W^{l+1,p}}   \cr
& \overset{\eqref{epsilon_bound_0}}{\leq}  C_-\: (l+1)!\: \| \overline{J^{*}_k} \|_{W^{l+1,p}} , 
\end{align}
where the factor $(l+1)!$ takes account for repeated lower derivative terms resulting form the product rule on the left hand side and is non-optimal. Similarly, using in addition the induction assumption \eqref{J_inverse_diff_bound_IndAss}, we obtain 
\begin{align} \label{techeqn5_J_inv_diff}
\epsilon & \| \partial^{l+1} \big( J^{*}_{k-1}\cdot \overline{J^{-*}_k} \big) \|_{L^p}   \cr
& \overset{(*)}{\leq}    \epsilon  C_M \: (l+1)!\: \|J^{*}_{k-1}\|_{W^{l+1,p}} \| \overline{J^{-*}_k} \|_{W^{l,p}} +  \epsilon  \| J^{*}_{k-1} \|_{L^\infty} \|\partial^{l+1} \overline{J^{-*}_k} \|_{L^p}     \cr
& \overset{\eqref{Morrey}}{\leq}   \epsilon  C_M \: (l+1)!\: \|J^{*}_{k-1}\|_{W^{l+1,p}} \| \overline{J^{-*}_k} \|_{W^{l,p}}  +  \epsilon C_M  \| J^{*}_{k-1} \|_{W^{1,p}} \|\partial^{l+1} \overline{J^{-*}_k} \|_{L^p}     \cr
& \overset{\eqref{epsilon_bound_0}}{\leq}   (l+1)!\: \| \overline{J^{-*}_k} \|_{W^{l,p}} +  \frac12 \|\partial^{l+1} \overline{J^{-*}_k} \|_{L^p}  \cr
& \overset{\eqref{J_inverse_diff_bound_IndAss}}{\leq}   (l+1)! C_-' \: \| \overline{J^{*}_k} \|_{W^{l,p}} +  \frac12 \|\partial^{l+1} \overline{J^{-*}_k} \|_{L^p}  
\end{align}
where the second term in $(*)$ results form the contribution of $(l+1)$-st order derivatives on $\overline{J^{-*}_k}$. Now, estimating the right hand side in \eqref{techeqn3_J_inv_diff} by \eqref{techeqn4_J_inv_diff} and \eqref{techeqn5_J_inv_diff}, we find 
\beq \nonumber
\|\partial^{l+1} \overline{J^{-*}_k} \|_{L^p}
 \leq   \|  \overline{J^{*}_k} \|_{W^{l+1,p}} + 2C_-\: (l+1)!\: \| \overline{J^{*}_k} \|_{W^{l+1,p}} +  \frac12 \|\partial^{l+1} \overline{J^{-*}_k} \|_{L^p} 
\eeq
so that subtraction of the last term yields 
\beq \label{techeqn6_J_inv_diff}
\|\partial^{l+1} \overline{J^{-*}_k} \|_{L^p}
 \leq  2 \|  \overline{J^{*}_k} \|_{W^{l+1,p}} + 4C_-\: (l+1)!\: \| \overline{J^{*}_k} \|_{W^{l+1,p}} .
\eeq
Repeating the argument \eqref{techeqn3_J_inv_diff} - \eqref{techeqn6_J_inv_diff} for each multi-index $\partial^{l+1}$ gives a suitable estimate on the $L^p$-norm of all combinations of $(l+1)$-st order derivatives. Adding then the $W^{l,p}$-norm of $\overline{J^{-*}_k}$ to both sides of that estimate, and applying the induction assumption \eqref{J_inverse_diff_bound_IndAss} to bound the $W^{l,p}$-norm of $\overline{J^{-*}_k}$ on the resulting right hand side, the sought after estimate \eqref{J_inverse_diff_bound} for $l+1$ follows. This completes the induction and the proof of Lemma \ref{Lemma_diff_J_inverse}.                   
\QED

\noindent {\it \textbf{Proof of Lemma \ref{Lemma_sources_difference}.}}
We now estimate the difference of the source functions and thereby prove Lemma \ref{Lemma_sources_difference}, which states that, if $$0<  \epsilon\leq  \min \big(\epsilon(k),\epsilon(k-1)\big),$$ (that is, \eqref{epsilon_bound_0} holds), then there exists a constant $C_s>0$ depending only on $m,n,p,\Omega$ and $C_0>0$, such that \eqref{bound_diff_Fu} - \eqref{bound_diff_Fa} hold, i.e.
\begin{eqnarray} \nonumber
\|\overline{F_u(u_k,a_{k+1})}\|_{W^{m-1,p}} & \leq &   C_u(k) \Big( \epsilon\: \|\overline{u_{k}}\|_{W^{m+1,p}} + \|\overline{a_{k+1}}\|_{W^{m,p}} \Big), \cr
\|\overline{F_a(u_{k})}\|_{W^{m-1,p}} & \leq & \epsilon\: C_a(k) \,  \|\overline{u_{k}}\|_{W^{m+1,p}} ,
\end{eqnarray} 
where
\begin{eqnarray} \nonumber
C_u(k) &\equiv & C_s \big( 1 +  \| u_k \|_{W^{m+1,p}} + \| {u_{k-1}}\|_{W^{m+1,p}} + \| a_{k+1} \|_{W^{m,p}}  \big), \cr
C_a(k) & \equiv & C_s \big( 1 +  \| u_k \|_{W^{m+1,p}} + \|  {u_{k-1}}\|_{W^{m+1,p}} \big). 
\end{eqnarray}

We only prove Lemma \ref{Lemma_sources_difference} for the critical case $m=1$, since the cases $m\geq 2$ follow by an analogous reasoning, (see also \eqref{techeqn5b_estimates_Fu} for an example of extending source estimate to higher derivatives). Note that, because $\epsilon >0$ is assumed to satisfy \eqref{epsilon_bound_0}, Lemma \ref{Lemma_diff_J_inverse} applies and gives estimate \eqref{J_inverse_bound} on $\overline{J^{-*}_k}$.

We begin by proving \eqref{bound_diff_Fa}. From the definition of $F_a$ in \eqref{Def_Fa}, using that the source term $d\Gamma^*$ cancels out in $\overline{F_a(u_k)}$, we obtain 
\beq  \label{techeqn0_diff_Fa}
\big\|\overline{F_a(u_k)}\big\|_{L^p}  
\leq   \epsilon\: \big\| \overrightarrow{\text{div}} \big(d\overline{J^*_k} \wedge \Gamma^*\big) \big\|_{L^p} + \epsilon\: \big\| \overrightarrow{\text{div}} \big(\overline{J^*_k} \cdot d\Gamma^*\big) \big\|_{L^p}  + \epsilon\: \big\| d\big(\overrightarrow{\overline{\langle d J^*_k ; \tilde{\Gamma}^*_k\rangle} }\big)\big\|_{L^p} .   
\eeq
We estimate the linear terms using Morrey's inequality \eqref{Morrey} and resulting H\"older continuity, and obtain 
\begin{eqnarray} \label{techeqn1_diff_Fa}
\big\| \overrightarrow{\text{div}} \big(d\overline{J^*_k} \wedge \Gamma^*\big) \big\|_{L^p}
&\leq & \| d\overline{J^*_k}\|_{W^{1,p}} \|\Gamma^* \|_{L^\infty} + \| d\overline{J^*_k}\|_{L^\infty} \|\Gamma^* \|_{W^{1,p}}  \cr
& \overset{\eqref{Morrey}}{\leq} & C_M\: \| \overline{J^*_k}\|_{W^{2,p}} \|\Gamma^* \|_{W^{1,p}}  \cr
&\overset{\eqref{Gamma-bound}}{\leq} & C_M\: C_0\: \| \overline{u_k}\|_{W^{2,p}}
\end{eqnarray}
and
\begin{eqnarray} \label{techeqn2_diff_Fa}
\big\| \overrightarrow{\text{div}} \big(\overline{J^*_k} \mm d\Gamma^*\big) \big\|_{L^p}  
&\leq & \| \overline{J^*_k}  \|_{W^{1,p}} \| d\Gamma^* \|_{L^\infty} + \|  \overline{J^*_k} \|_{L^\infty} \|  d\Gamma^*\|_{W^{1,p}}  \cr
& \overset{\eqref{Morrey}}{\leq} & C_M \:\| \overline{J^*_k}\|_{W^{1,p}} \|d\Gamma^* \|_{W^{1,p}}  \cr
&\overset{\eqref{Gamma-bound}}{\leq}& C_M\: C_0\: \| \overline{u_k}\|_{W^{2,p}}.
\end{eqnarray}
For the non-linear term we first compute
\begin{eqnarray} \nonumber
d\big(\overrightarrow{\overline{\langle d J^*_k ; \tilde{\Gamma}^*_k\rangle} }\big) 
&=& d\big(\overrightarrow{\langle d (J^*_k- J^*_{k-1}) ; \tilde{\Gamma}^*_k\rangle}\big) + d\big(\overrightarrow{\langle d J^*_{k-1} ; (\tilde{\Gamma}^*_k - \tilde{\Gamma}^*_{k-1})\rangle}\big)  \cr
&=& d\big(\overrightarrow{\langle d \overline{J^*_k} ; \tilde{\Gamma}^*_k\rangle}\big) + d\big(\overrightarrow{\langle d J^*_{k-1} ; \overline{\tilde{\Gamma}^*_k}\rangle}\big)
\end{eqnarray}
and then estimate 
\begin{eqnarray} \nonumber
\big\| d\big(\overrightarrow{\langle d \overline{J^*_k} ; \tilde{\Gamma}^*_k\rangle}\big) \big\|_{L^p} 
&\leq &  \| d \overline{J^*_k} \|_{W^{1,p}} \| \tilde{\Gamma}^*_k \|_{L^\infty} + \|  d \overline{J^*_k} \|_{L^\infty} \|  \tilde{\Gamma}^*_k  \|_{W^{1,p}}    \cr
 &\overset{\eqref{Morrey}}{\leq} & C_M \: \|\tilde{\Gamma}^*_k \|_{W^{1,p}}   \| \overline{J^*_k} \|_{W^{2,p}}, 
\cr
\big\|   d\big(\overrightarrow{\langle d J^*_{k-1} ; \overline{\tilde{\Gamma}^*_k}\rangle}\big) \big\|_{L^p}
 & \leq & \| d J^*_{k-1} \|_{W^{1,p}} \| \overline{\tilde{\Gamma}^*_k} \|_{L^\infty} + \|  d J^*_{k-1} \|_{L^\infty} \|  \overline{\tilde{\Gamma}^*_k}  \|_{W^{1,p}}  \cr
 &\overset{\eqref{Morrey}}{\leq} & C_M \:\|J^*_{k-1} \|_{W^{2,p}}  \| \overline{\tilde{\Gamma}^*_k} \|_{W^{1,p}} ,
\end{eqnarray}
which combined yields
\begin{align} \label{techeqn3_diff_Fa}
\big\| d\big(\overrightarrow{\overline{\langle d J^*_k ; \tilde{\Gamma}^*_k\rangle} }\big)\big\|_{L^p} 
&\leq &  C_M \: \big(2C_0 + \|u_{k-1} \|_{W^{2,p}} + \|u_{k} \|_{W^{2,p}} \big) \| \overline{u_k} \|_{W^{2,p}} .
\end{align}
Combining \eqref{techeqn1_diff_Fa} - \eqref{techeqn3_diff_Fa} with \eqref{techeqn0_diff_Fa} yields the sought after bound \eqref{bound_diff_Fa}.

We now prove \eqref{bound_diff_Fu}. From definition \eqref{Def_Fu}, using that the source terms $\delta d \Gamma^*$ and $\delta\Gamma^*$ cancel and substituting $d(J^{-1}_k a_k)= da_k + \epsilon d(J^{-*}_k a_k)$,  we find that 
\begin{align}  \label{techeqn0_diff_Fu}
\| \overline{F_u(u_k,a_{k+1})} \|_{L^p}  \
\leq \  \|\overline{a_{k+1}}\|_{W^{1,p}} + \epsilon\: \|\delta ( \overline{J^*_k} \mm \Gamma^* )\|_{L^p} 
+ \epsilon\: \| d (\overline{J^{-*}_k a_{k+1}}) \|_{L^p}  \cr + \epsilon\: \|\overline{\langle d J^*_k ; \tilde{\Gamma}^*_k\rangle} \|_{L^p}   + \epsilon\: \|\delta d\big( \overline{J^{-*}_k \mm dJ^*_k} \big)\|_{L^p}  ,
\end{align}
where we used for the last term that $d^2=0$ gives
\beq \nonumber
d\big( \overline{J^{-1}_k \mm dJ^*_k} \big) = \epsilon\: d\big( \overline{J^{-*}_k \mm dJ^*_k} \big) .
\eeq
Now, for the linear term in \eqref{techeqn0_diff_Fu} we obtain
\begin{eqnarray} \label{techeqn1_diff_Fu}
\|\delta ( \overline{J^*_k} \mm \Gamma^* )\|_{L^p}  
&\leq &  \| \overline{J^*_k} \|_{W^{1,p}} \|\Gamma^*\|_{L^\infty} + \| \overline{J^*_k} \|_{L^\infty} \|\Gamma^*\|_{W^{1,p}}  \cr
& \overset{\eqref{Morrey}}{\leq} &  C_M \: \| \overline{J^*_k} \|_{W^{1,p}} \|\Gamma^*\|_{W^{1,p}}  \cr
& \overset{\eqref{Gamma-bound}}{\leq} & C_M\: C_0\:  \| \overline{u_k} \|_{W^{1,p}}. 
\end{eqnarray}
For the first non-linear term we compute 
\beq \nonumber
\overline{J^{-*}_k \cdot a_{k+1}}  = \overline{J^{-*}_k} \cdot a_{k+1} + J^{-*}_{k-1} \cdot \overline{a_{k+1}},
\eeq
so that
\beq \nonumber
d(\overline{J^{-*}_k \cdot a_{k+1}})  = d(\overline{J^{-*}_k}) \cdot a_{k+1} + \overline{J^{-*}_k} \cdot da_{k+1} + d(J^{-*}_{k-1}) \cdot \overline{a_{k+1}} + J^{-*}_{k-1} \cdot d(\overline{a_{k+1}}),
\eeq
from which we obtain the estimate
\begin{align} \label{techeqn3_diff_Fu}
\| d (\overline{J^{-*}_k a_{k+1}})  \|_{L^p}   
 \leq &  \|d(\overline{J^{-*}_k}) \cdot a_{k+1}\|_{L^p} + \|\overline{J^{-*}_k} \cdot d(a_{k+1})\|_{L^p}  \cr 
 &+ \|d(J^{-*}_{k-1}) \cdot \overline{a_{k+1}}\|_{L^p} +\| J^{-*}_{k-1} \cdot d(\overline{a_{k+1}})\|_{L^p} \cr
\leq & \|d(\overline{J^{-*}_k}) \|_{L^p} \|  a_{k+1}\|_{L^\infty} + \|\overline{J^{-*}_k} \|_{L^\infty} \|d(a_{k+1})\|_{L^p}  \cr 
& + \|d(J^{-*}_{k-1}) \|_{L^p} \|\overline{a_{k+1}}\|_{L^\infty} +\| J^{-*}_{k-1} \|_{L^\infty} \|d(\overline{a_{k+1}})\|_{L^p} , \ \ \ \ \ \ \   
\end{align}
so that Morrey's inequality \eqref{Morrey} and \eqref{J_inverse_diff_bound}, (the bound on $\overline{J^{-1}_k}$), yield
\begin{align}
\| d (\overline{J^{-*}_k a_{k+1}}) & \|_{L^p}   
\overset{\eqref{Morrey}}{\leq}   C_M\big( \|\overline{J^{-*}_k} \|_{W^{1,p}} \| a_{k+1}\|_{W^{1,p}} + \| J^{-*}_{k-1} \|_{W^{1,p}} \| \overline{a_{k+1}} \|_{W^{1,p}} \big) \cr
&\overset{\eqref{J_inverse_diff_bound}}{\leq}  C_M \big( \|\overline{J^{*}_k} \|_{W^{1,p}} \| a_{k+1}\|_{W^{1,p}} +C\: \| J^{*}_{k-1} \|_{W^{1,p}} \| \overline{a_{k+1}} \|_{W^{1,p}} \big)  \cr
& \ \leq  C_M\: \big(   \| a_{k+1}\|_{W^{1,p}} + C \| u_{k-1} \|_{W^{1,p}}  \big) \big( \|\overline{u_k} \|_{W^{1,p}} +  \| \overline{a_{k+1}} \|_{W^{1,p}} \big)    .
\end{align}
For the second non-linear term, similar to the argument leading to \eqref{techeqn3_diff_Fa}, we first compute
\beq \nonumber
\overline{\langle d J^*_k ; \tilde{\Gamma}^*_k\rangle} = \langle \overline{d J^*_k} ; \tilde{\Gamma}^*_k\rangle + \langle d J^*_{k-1} ; \overline{\tilde{\Gamma}^*_k}\rangle
\eeq 
and then estimate
\begin{eqnarray} \label{techeqn4_diff_Fu}
\|\overline{\langle d J^*_k ; \tilde{\Gamma}^*_k\rangle} \|_{L^p}  
& \overset{\eqref{Morrey}}{\leq} & C_M \big( \|\overline{J^*_k} \|_{W^{1,p}} \| \tilde{\Gamma}^*_k\|_{W^{1,p}} + \| J^*_{k-1}\|_{W^{1,p}} \|\overline{\tilde{\Gamma}^*_k}\|_{W^{1,p}}\big) \cr
& \leq &  C_M\: \big(  \|u_k\|_{W^{1,p}} + \| u_{k-1}\|_{W^{1,p}} \big) \|\overline{u_k} \|_{W^{1,p}}   . 
\end{eqnarray}
For the last non-linear term we first compute 
\beq \nonumber
\overline{J^{-*}_k \mm dJ^*_k} = \overline{J^{-*}_k} \cdot dJ^*_k + J^{-*}_{k-1} \cdot \overline{dJ^*_k},
\eeq
so that the Leibniz rule \eqref{Leibnitz-rule} and $d^2=0$ yield 
\begin{eqnarray}   \nonumber
d\big( \overline{J^{-*}_k \mm dJ^*_k} \big) &=& \overline{d J^{-*}_k} \wedge dJ^*_k + dJ^{-*}_{k-1} \wedge \overline{dJ^*_k}.
\end{eqnarray}
From this, we obtain the (higher derivative) estimate
\begin{eqnarray} \label{techeqn5_diff_Fu}
\|\delta d\big( \overline{J^{-*}_k \mm dJ^*_k} \big)\|_{L^p}   
& \leq & \|\overline{d J^{-*}_k} \|_{W^{1,p}}  \| dJ^*_k \|_{L^\infty} + \|\overline{d J^{-*}_k} \|_{L^\infty}  \| dJ^*_k \|_{W^{1,p}}    \cr
& & + \| dJ^{-*}_{k-1} \|_{L^\infty} \| \overline{dJ^*_k} \|_{W^{1,p}}  + \| dJ^{-*}_{k-1} \|_{W^{1,p}} \| \overline{dJ^*_k} \|_{L^\infty}    \cr
& \overset{\eqref{Morrey}}{\leq} & 2 C_M \big( \|\overline{d J^{-*}_k} \|_{W^{1,p}}  \| dJ^*_k \|_{W^{1,p}} + \| dJ^{-*}_{k-1} \|_{W^{1,p}} \| \overline{dJ^*_k} \|_{W^{1,p}} \big)  \cr
&\leq & 2 C_M \big( \|\overline{J^{-*}_k} \|_{W^{2,p}}  \| J^*_k \|_{W^{2,p}} + C\: \| J^{-*}_{k-1} \|_{W^{2,p}} \| \overline{J^*_k} \|_{W^{2,p}}  \big) \cr
&\overset{\eqref{J_inverse_diff_bound}}{\leq} & 2 C_M C  \big( \| J^*_k \|_{W^{2,p}} +  \| J^{*}_{k-1} \|_{W^{2,p}}    \big) \| \overline{ J^{*}_k}\|_{W^{2,p}}   \cr
& \leq & 2C_M C \big( \| u_k \|_{W^{2,p}} +  \| u_{k-1} \|_{W^{2,p}}    \big) \| \overline{u_k}\|_{W^{2,p}} .
\end{eqnarray}
Combining \eqref{techeqn1_diff_Fu} - \eqref{techeqn5_diff_Fu} with \eqref{techeqn0_diff_Fu} yields the sought after estimate \eqref{bound_diff_Fu}. Taking $C_s>0$ as the maximum over all constants \eqref{techeqn0_diff_Fa} - \eqref{techeqn5_diff_Fu} and the constant in \eqref{bound_Fa} and \eqref{bound_Fu}, completes the proof of Lemma \ref{Lemma_sources_difference}.
\hfill $\Box$

\subsection{Consistency of Induction Assumption - Proof of Lemma \ref{Lemma_induction_consistency}} \label{Sec_Cons_Ind_Assump}

We now prove Lemma \ref{Lemma_induction_consistency}, which shows that the induction assumption \eqref{hypothesis_induction} is maintained in each step of the iteration. Lemma \ref{Lemma_induction_consistency} states that, if $$0< \epsilon\leq  \epsilon_1,$$ i.e. \eqref{epsilon_bound} holds, and if the induction assumption \eqref{hypothesis_induction} holds, namely
\beq \nonumber  
\|u_k\|_{W^{m+1,p}(\Omega)} \leq 4\, C_0 C_e^2,
\eeq 
then \eqref{uniformbound_a} - \eqref{uniformbound_u} holds, that is,
\begin{eqnarray}
\|a_{k+l}\|_{W^{m,p}} &\leq & 2 C_0C_e, \label{uniformbound_a_again} \\
\|u_{k+l} \|_{W^{m+1,p}} &\leq & 4 C_0 C_e^2,  \label{uniformbound_u_again}
\end{eqnarray}
for all $l \in \mathbb{N}$, and the induction assumption \eqref{hypothesis_induction} holds for each subsequent iterate.

\Proof
To begin observe that the $\epsilon$-bound \eqref{epsilon_bound}, i.e. 
\beq \nonumber
0<  \epsilon \leq  \epsilon_1 \equiv  \min\Big( \tfrac{1}{4C^2_e C_s(1+ 2C_e C_0 + 4C_e^2 C_0)},\tfrac{1}{16 C_M C_0 C_e^2}  \Big),
\eeq 
together with the induction assumption \eqref{hypothesis_induction} imply  
\beq \label{ind_consistency_techeqn0}
\epsilon \leq  \frac{1}{4 C_M \cdot 4C_0 C_e^2} \overset{\eqref{hypothesis_induction}}{\leq } \frac{1}{4 C_M \|u_k\|_{W^{m+1,p}(\Omega)}} = \epsilon(k) ,
\eeq     
which is the $\epsilon$ bound \eqref{epsilon_bound_0} of Lemma \ref{Lemma_existence_iterates}, so that existence of iterates and the elliptic estimates \eqref{existence_est1} - \eqref{existence_est2} hold.  Moreover, since $\|J^*_k\|_{W^{m+1,p}(\Omega)} \leq \|u_k\|_{W^{m+1,p}(\Omega)}$, \eqref{ind_consistency_techeqn0} implies that 
\beq \label{ind_consistency_techeqn01}
\epsilon \leq  \epsilon(k)  \leq  \frac{1}{2 C_M \|J^*_k\|_{W^{m+1,p}(\Omega)}},
\eeq
which is the $\epsilon$-bound \eqref{Lemma2_J_epsilon}  of Lemma \ref{Lemma_source_estimate_Fa_Fu} in terms of $J^*=J^*_k$. Thus the source estimates  \eqref{bound_Fu} - \eqref{bound_Fa} of Lemma \ref{Lemma_source_estimate_Fa_Fu} apply and yield that the right hand sides of the elliptic estimates \eqref{existence_est1} - \eqref{existence_est2} are indeed finite. 

We now derive the uniform bound \eqref{uniformbound_a_again}. From the elliptic estimate \eqref{existence_est1} together with the source estimate \eqref{bound_Fa}, we find that
\begin{eqnarray} \nonumber
\| a_{k+1}\|_{W^{m,p}} 
& \overset{\eqref{existence_est1}}{\leq} & C_e \| F_a(u_k) \|_{W^{m-1,p}} \cr
 & \overset{\eqref{bound_Fa}}{\leq} & C_e \big( C_0 + \epsilon\: C_s ( 1 + \| u_k \|_{W^{m+1,p}} ) \| u_k \|_{W^{m+1,p}} \big) ,
\end{eqnarray}
so application of the induction assumption \eqref{hypothesis_induction} gives
\begin{eqnarray} \label{ind_consistency_techeqn1}
\| a_{k+1}\|_{W^{m,p}} 
 \leq C_e C_0 + \epsilon\: 4 C_e^2 C_s  \big( 1 + 4 C_e^2 C_0 \big) C_e C_0.
\end{eqnarray}
Now, by the $\epsilon$-bound \eqref{epsilon_bound}, we have
\beq \nonumber
\epsilon \leq  \frac{1}{4C^2_e C_s(1+ 2C_e C_0 + 4C_e^2 C_0)} \leq  \frac{1}{4C^2_e C_s(1+ 4C_e^2 C_0)},
\eeq
so that substituting the above $\epsilon$-bound into \eqref{ind_consistency_techeqn1} yields
\beq \nonumber
\| a_{k+1}\|_{W^{m,p}} \leq 2 C_0C_e ,
\eeq 
which is the sought after bound \eqref{uniformbound_a_again} for $l=1$.

We now derive \eqref{uniformbound_u}. From the elliptic estimate \eqref{existence_est2} together with the source estimate \eqref{bound_Fu}, we obtain that
\begin{align} \label{ind_consistency_techeqn2}
\| u_{k+1}& \|_{W^{m+1,p}}  
 \overset{\eqref{existence_est2}}{\leq}   C_e \| F_u(u_k,a_{k+1}) \|_{W^{m-1,p}} \cr
& \overset{\eqref{bound_Fu}}{\leq}  C_e C_0 + C_e \|a_{k+1}\|_{W^{m,p}} \cr 
& \ \ \ \ \ \  + \epsilon\: C_e C_s \big( 1 + \|a_{k+1}\|_{W^{m,p}} + \|u_{k}\|_{W^{m+1,p}} \big) \|u_{k}\|_{W^{m+1,p}} \cr
& \overset{\eqref{hypothesis_induction}}{\leq}   C_e C_0 + 2 C_e^2 C_0 + \epsilon\:4 C_e^2 C_s \big( 1 + 2 C_e C_0 + 4 C_e^2 C_0 \big) C_eC_0,
\end{align}
where we substituted $\|a_{k+1}\|_{W^{m,p}} \leq 2 C_e C_0$ and the induction assumption $\|u_k\|_{W^{m+1,p}} \leq 4\, C_0 C_e^2$ to obtain the last inequality. By \eqref{epsilon_bound}, we have
\beq \nonumber
\epsilon \leq  \frac{1}{4C^2_e C_s(1+ 2C_e C_0 + 4C_e^2 C_0)} ,
\eeq
so that applying the above bound to $\epsilon$ in \eqref{ind_consistency_techeqn2} gives 
\begin{eqnarray} \nonumber
\| u_{k+1}\|_{W^{m+1,p}}  \leq  2 C_0 C_e (1 + C_e) \leq 4 C_0 C_e^2,
\end{eqnarray}
where the last inequality holds since we chose $C_e>1$ initially. This is the sought after bound \eqref{uniformbound_u_again} for $l=1$. 

The bound $\epsilon \leq \epsilon(k+l)$ for $l\in \mathbb{N}$ together with \eqref{uniformbound_a_again} and \eqref{uniformbound_u_again} for $l\in \mathbb{N}$ follow now recursively, which completes the proof of Lemma \ref{Lemma_induction_consistency}.
\QED

\subsection{Decay of the difference of iterates - Proof of Proposition \ref{Lemma_decay}} \label{Sec_decay}

We now prove Proposition \ref{Lemma_decay}, which completes the proof of Theorem \ref{Thm3}. Proposition \ref{Lemma_decay} states that, if $0< \epsilon\leq \epsilon_1$, (i.e. \eqref{epsilon_bound} holds), then there exists a constant $C_d>0$ depending only on $m$, $n$, $p$, $\Omega$ such that \eqref{decay_a} and \eqref{decay_u} hold, i.e.,
\begin{eqnarray} \nonumber
\|\overline{a_{k+1}}\|_{W^{m,p}} &\leq &  \epsilon\: C_d\: \|\overline{u_{k}}\|_{W^{m+1,p}} ,  \cr
\|\overline{u_{k+1}}\|_{W^{m+1,p}}  &\leq & \epsilon\: C_d\:  \|\overline{u_{k}}\|_{W^{m+1,p}} .  
\end{eqnarray} 

\noindent {\it  Proof of Proposition \ref{Lemma_decay}:}
We first establish estimate \eqref{decay_a} on $\|\overline{a_{k+1}}\|_{W^{m,p}}$. By linearity of the Laplacian it is straightforward to extend the  elliptic estimate \eqref{existence_est1} to $\overline{a_{k+1}}$ and obtain
\beq \nonumber
\|\overline{a_{k+1}}\|_{W^{m,p}} \leq C_e \|\overline{F_a(u_{k})}\|_{W^{m-1,p}} ,
\eeq
for the same constant $C_e>0$ as in \eqref{existence_est1}. Applying the non-linear source estimate \eqref{bound_diff_Fa}, we further find that
\beq \label{decay_techeqn0}
\|\overline{a_{k+1}}\|_{W^{m,p}} \leq  \epsilon\: C_e C_a(k) \,  \|\overline{u_{k}}\|_{W^{m+1,p}},
\eeq
where $C_a(k)$ is defined in \eqref{diff_Ca} as 
\begin{eqnarray}
C_a(k)  = C_s \big( 1 +  \| u_k \|_{W^{m+1,p}} + \|  {u_{k-1}}\|_{W^{m+1,p}} \big) .
\end{eqnarray}
Applying the induction hypothesis \eqref{hypothesis_induction} we bound $C_a(k)$ by 
\begin{eqnarray}
C_a(k)  
\leq C_s \big( 1 +  8 C_0 C_e^2 \big),
\end{eqnarray}
which in combination with \eqref{decay_techeqn0} implies the sought after estimate \eqref{decay_a}. 

We now prove estimate \eqref{decay_u} on $\|\overline{u_{k+1}}\|_{W^{m+1,p}}$. By linearity of the Laplacian, the elliptic estimate \eqref{existence_est2} extends to $\overline{u_{k+1}}$, 
\begin{align} \nonumber 
\|\overline{u_{k+1}}\|_{W^{m+1,p}} 
 \leq  C_e \Big(\|\overline{F_u(u_k,a_{k+1})}\|_{W^{m-1,p}}  + \|\overline{dy_{k+1}} \|_{W^{m+1,p}} \Big) ,
\end{align}
where we substituted the boundary conditions \eqref{bdd1} - \eqref{bdd2} for the second term which we bound further by $\|\overline{y_{k+1}} \|_{W^{m+2,p}}$; (all norms are taken over $\Omega$). The source estimate \eqref{bound_diff_Fu} now implies 
\begin{align} \label{decay_techeqn1}
\|\overline{u_{k+1}}\|_{W^{m+1,p}} 
\leq  C_e \Big( C_u(k) \big( \epsilon\: \|\overline{u_{k}}\|_{W^{m+1,p}} + \|\overline{a_{k+1}}\|_{W^{m,p}}  \big)   + \|\overline{y_{k+1}} \|_{W^{m+2,p}}  \Big),
\end{align}
where  $C_u(k)$ is defined in \eqref{diff_Cu} as
\begin{eqnarray} \nonumber
C_u(k) = C_s \big( 1 +  \| u_k \|_{W^{m+1,p}} + \| {u_{k-1}}\|_{W^{m+1,p}} + \| a_{k+1} \|_{W^{m,p}}  \big).
\end{eqnarray} 
Using now induction assumption \eqref{hypothesis_induction} together with $\|a_{k+1}\|_{W^{m,p}} \leq 2 C_0 C_s$,  the bound from Lemma \ref{Lemma_induction_consistency}, we obtain the uniform bound
\begin{eqnarray} \label{decay_techeqn2}
C_u(k) \leq  C_s \big( 1 +  8 C_0 C_e^2 + 2 C_0 C_e \big) .
\end{eqnarray} 
Substituting \eqref{decay_techeqn2} together with estimate \eqref{decay_a} on $\| \overline{a_{k+1}} \|_{W^{m,p}}$ in \eqref{decay_techeqn1} gives us 
\begin{eqnarray} \label{decay_techeqn3}
\|\overline{u_{k+1}}\|_{W^{m+1,p}}  
& \leq & \epsilon  C \|\overline{u_{k}}\|_{W^{m+1,p}} +  C_e \: \|\overline{y_{k+1}} \|_{W^{m+2,p}(\Omega)}.
\end{eqnarray}

It remains to estimate the second term in \eqref{decay_techeqn3}. For this, observe that by linearity of the Laplacian, \eqref{iterate_y} implies that $\overline{y_{k+1}}$ solves
\beq \label{decay_techeqn4b}
\begin{cases} 
\Delta \overline{y_{k+1}} = \overline{\psi_{k+1}}, \cr 
\overline{y_{k+1}}\big|_{\partial\Omega} = 0,
\end{cases}
\eeq
so that the elliptic estimate \eqref{Poissonelliptic_estimate_Lp} yields 
\begin{eqnarray} \label{decay_techeqn4c}
\|\overline{y_{k+1}} \|_{W^{m+2,p}(\Omega)} 
 \leq  C_e \|\overline{\psi_{k+1}} \|_{W^{m,p}(\Omega)} .
\end{eqnarray}
Since the elliptic estimate \eqref{existence_est3} extends to $\overline{\psi_{k+1}}$, we can bound the right hand side in \eqref{decay_techeqn4c} and find
\begin{eqnarray} \label{decay_techeqn4d}
\|\overline{y_{k+1}} \|_{W^{m+2,p}(\Omega)} 
& \overset{\eqref{existence_est3}}{\leq} & C_e^2 \|\overline{F_u(u_k,a_{k+1})} \|_{W^{m-1,p}(\Omega)} \cr
&\overset{\eqref{bound_diff_Fu}}{\leq} & C_e^2 C_u(k) \Big( \epsilon\: \|\overline{u_{k}}\|_{W^{m+1,p}} + \|\overline{a_{k+1}}\|_{W^{m,p}}  \Big)  .
\end{eqnarray}
Substituting \eqref{decay_techeqn4d} back into \eqref{decay_techeqn4b}, and bounding $C_u(k)$ by \eqref{decay_techeqn2} and $\|\overline{a_{k+1}}\|_{W^{m,p}}$ by \eqref{decay_a}, we obtain 
\beq \label{decay_techeqn5}
\|\overline{y_{k+1}} \|_{W^{m+2,p}(\Omega)}    \leq  \epsilon \: C \: \|\overline{u_{k}}\|_{W^{m+1,p}(\Omega)} ,
\eeq
for some suitable constant $C>0$. Finally, using \eqref{decay_techeqn5} to estimate the second term in \eqref{decay_techeqn3}, we obtain for some suitable constant $C_d>0$ that
\begin{eqnarray} \nonumber
\|\overline{u_{k+1}}\|_{W^{m+1,p}(\Omega)}  
& \leq & \epsilon \: C_d \: \|\overline{u_{k}}\|_{W^{m+1,p}(\Omega)} ,
\end{eqnarray}
which is the sought after estimate \eqref{decay_u}. This completes the proof.
\hfill $\Box$

\section{Proof of Theorem \ref{ThmMain}}    \label{Sec_proof_ThmMain}

The proofs in Sections \ref{Sec_source-estimates} - \ref{Sec_decay} complete the proof of Theorem \ref{Thm3}. We now prove our main theorem regarding existence of solutions of the RT-equations \eqref{PDE1} - \eqref{BDD1}, Theorem \ref{ThmMain}, which follows from Theorem \ref{Thm3} together with a rescaling argument to arrange for the smallness assumption \eqref{small_Gamma}, that is, $\Gamma=\epsilon\: \Gamma^*$, and the uniform bound \eqref{Gamma-bound}, i.e.
\begin{align} \nonumber
\|\Gamma^*\|_{W^{m,p}(\Omega)} \: + \: \|d\Gamma^*\|_{W^{m,p}(\Omega)}<C_0,
\end{align} 
which are the incoming assumptions of Theorem \ref{Thm3}. In more detail, given any connection $\Gamma' \in W^{m,p}(\Omega)$ with $d\Gamma'$ bounded in $W^{m,p}(\Omega)$, we define $\Gamma^*$ as the restriction of $\Gamma'$ to the ball of radius $\epsilon$, but with its components transformed as scalars to the ball or radius $1$ (which we take to be $\Omega$), while $\Gamma$ is taken to be the connection resulting from transforming $\Gamma'$ as a connection. The proof below shows that this construction suffices to arrange for assumptions \eqref{Gamma-bound} and \eqref{small_Gamma}.\\

\noindent \emph{Proof of Theorem \ref{ThmMain}.}            
By Theorem \ref{Thm3}, for any connection $\Gamma$ satisfying \eqref{small_Gamma} for $\epsilon < \min(\epsilon_1,\epsilon_2)$ together with the $W^{m,p}$-bound \eqref{Gamma-bound}, there exists $(\Gammati^*,J^*,A^*)$ which solve the rescaled RT-equations \eqref{pde_u} - \eqref{pde_a} with boundary data \eqref{BDD1}. Defining $(\Gammati,J,A)$ by \eqref{ansatz_scaling} as
\beq \nonumber
J=I+\epsilon \, J^* , \hspace{1cm} 
\tilde{\Gamma}=\epsilon\: \tilde{\Gamma}^*, \hspace{1cm} 
A = \epsilon A^*,
\eeq 
Lemma \ref{Lemma_rescaled_RT-eqn} implies that $(\Gammati,J,A)$ solves the RT-equations \eqref{PDE1} - \eqref{PDE4} with boundary data \eqref{BDD1}. It remains to show that, for any connection $\Gamma \in W^{m,p}(\Omega)$ with $d\Gamma \in W^{m,p}(\Omega)$, one can arrange for the hypotheses of Theorem \ref{Thm2}, that is, the scaling $\Gamma=\epsilon\: \Gamma^*$ together with the uniform bound \eqref{Gamma-bound} on $\Gamma^*$ as well as the $\epsilon$-bounds \eqref{epsilon_bound} and \eqref{epsilon_bound_2}, i.e., $\epsilon < \min(\epsilon_1,\epsilon_2)$. 

We now show that, given a connection $\Gamma \in W^{m,p}(\Omega)$ with $d\Gamma \in W^{m,p}(\Omega)$, one can first restrict $\Gamma$ to a small region and then scale the restriction of $\Gamma$ to a large region (which we take to be $\Omega$) such that the resulting $\Gamma$ satisfies the hypotheses of Theorem \ref{Thm2}.  For this, we assume without loss of generality that $\Omega \equiv B_1(0)$ is the ball of radius $1$ and we denote with $B_\epsilon(0)$ the ball of radius $\epsilon$, for $0< \epsilon \leq 1$. Under a coordinate transformation $x \to y\equiv \epsilon x$, (which maps $B_1(0)$ in $x$-coordinates to $B_\epsilon(0)$ in $y$-coordinates), a connection $\Gamma(y)$ given in $y$-coordinates transforms as \cite{HawkingEllis,Weinberg}
\begin{eqnarray} \nonumber
\Gamma(x) ^{\sigma}_{\mu\nu}
&=& \frac{\partial x^\sigma}{\partial y^\gamma} \Big( \frac{\partial y^\alpha }{\partial x^\mu } \frac{\partial y^\beta }{\partial x^\nu } \Gamma(y)^\gamma_{\alpha\beta} + \frac{\partial^2 y^\gamma }{\partial x^\mu \partial x^\nu }   \Big) 
\end{eqnarray}
which for the transformation $x \to y\equiv \epsilon x$ reduces to the scaling
\begin{eqnarray} \label{rescaling_Gamma_1}
\Gamma(x) ^{\sigma}_{\mu\nu}
= \epsilon\: \Gamma(y)^{\sigma}_{\mu\nu}.
\end{eqnarray}

We now arrange for conditions \eqref{Gamma-bound} and \eqref{small_Gamma} on $\Omega = B_1(0)$ in $x$-coordinates and we assume that the connection $\Gamma$ we start with is given in coordinates $y$ on $\Omega$. For this, take $\Gamma(y)$ to be the restriction of $\Gamma$ to $B_\epsilon(0)$ in $y$-coordinate, and define $\Gamma^*(x) \equiv \Gamma(y(x))$. That is, $\Gamma^*$ is the connection $\Gamma(y)$ in $x$-coordinates, defined on $\Omega \equiv B_1(0)$, but with the components of $\Gamma(y)$ transformed as scalar functions---not as connection components. Moreover, the connection $\Gamma(x)$ that results from transforming the restriction of $\Gamma$ to $B_\epsilon(0)$ in $y$-coordinate to $x$-coordinates according to the connection transformation law \eqref{rescaling_Gamma_1} satisfies the sought after scaling \eqref{small_Gamma}. Thus, taking $\Gamma(x)$ as the (initial) connection assumed in Theorem \ref{Thm3}, we only need to show that $\Gamma^*$ satisfies the uniform $W^{m,p}$-bound \eqref{Gamma-bound} in order to verify the hypotheses of Theorem \ref{Thm3}.

To show that $\Gamma^*$ satisfies \eqref{Gamma-bound} for the case $m=1$, we now study the $\epsilon$-scaling of the $W^{1,p}$-norm when the ball of radius $\epsilon$ is scaled up to the unit ball. For this, let $u \in W^{1,p}(B_\epsilon(0))$ be a scalar function, $p>n$. By Morrey's inequality \eqref{Morrey_textbook}, $u$ is H\"older continuous, so the $L^p$-norm of $u$ scales as 
\begin{eqnarray} \label{rescaling_techeqn1}
\|u\|_{L^p(B_\epsilon(0))} &\leq & \|u\|_{L^\infty} {\rm vol}(B_\epsilon(0)) \cr
&\leq & \|u\|_{L^\infty} {\rm vol}(B_1(0)) \: \epsilon^\frac{n}{p}  \ = \ o\big(\epsilon^\frac{n}{p}\big),
\end{eqnarray}
while we have by assumption 
\beq \label{rescaling_techeqn2}
\|Du\|_{L^p(B_\epsilon(0))} = o(1),
\eeq
i.e., bounded by a constant and tending to zero as $\epsilon \to 0$. Now under the transformation $y \to x= \frac{y}{\epsilon}$, (which maps the ball of radius $\epsilon>0$ in $y$-coordinates to the unit ball in $x$-coordinate), we have
\begin{eqnarray} \label{rescaling_techeqn3}
\|u\|_{L^p(B_1(0))} = \epsilon^{-\frac{n}{p}} \|u\|_{L^p(B_\epsilon(0))} 
\overset{\eqref{rescaling_techeqn1}}{\leq} {\rm vol}(B_1(0))  \|u\|_{L^\infty},
\end{eqnarray}
and, since the scaling of first order derivatives cancels the scaling of the measure on $\R^n$ within an error of order $\epsilon^\alpha$, for $\alpha \equiv 1 -\frac{n}{p} >0$, we further have
\begin{eqnarray} \label{rescaling_techeqn4}
\|Du\|_{L^p(B_1(0))} &=&  \Big( \int_{B_1(0)} |D_x u|^p dx  \Big)^\frac{1}{p}   \cr
&=& \Big( \int_{B_1(0)} \epsilon^p |D_{\epsilon x} u(x)|^p \epsilon^{-n} d(\epsilon x) \Big)^\frac{1}{p}   \cr
&=&  \epsilon^{\frac{p-n}{p}} \|D_yu\|_{L^p(B_\epsilon(0))}  \cr
&\leq & \|D_yu\|_{L^p(B_\epsilon(0))} ,
\end{eqnarray}
for all $0< \epsilon \leq 1$, where $D_x$ denotes differentiation with respect to $x$. Combining \eqref{rescaling_techeqn3} - \eqref{rescaling_techeqn4} we obtain
\begin{eqnarray} \label{rescaling_techeqn4b}
\|u\|_{W^{1,p}(B_1(0))} & \leq &  C \big( \|u\|_{L^\infty} + \|D_y u\|_{L^p(B_\epsilon(0))} \big) ,
\end{eqnarray}
where $C>0$ is a constant independent of $\epsilon$. 

Applying now \eqref{rescaling_techeqn4b} to $\Gamma^*$ component-wise, we find for $\Gamma^*(x)=\Gamma(y(x))$, where $x\in \Omega = B_1(0)$, that
\begin{eqnarray} \nonumber
\|\Gamma^*\|_{W^{1,p}(\Omega)} 
&=& \|\Gamma(y(\cdot))\|_{W^{1,p}(B_1(0))}   \cr
&\overset{\eqref{rescaling_techeqn4b}}{\leq} & C\big( \|\Gamma(y)\|_{L^\infty} + \|D_y \Gamma(y)\|_{L^p(B_\epsilon(0))} \big) 
\end{eqnarray}
where $\|\Gamma(y)\|_{L^\infty}$ is the supremum of $\Gamma$ in $y$-coordinates over $B_\epsilon(0)$, so that we can bound the right hand side further by taking the supremum and $L^p$-norm over $\Omega$, namely, 
\begin{eqnarray} \label{rescaling_techeqn5}
\|\Gamma^*\|_{W^{1,p}(\Omega)} 
&\leq & C\big( \|\Gamma(y)\|_{L^\infty(\Omega)} + \|D_y \Gamma(y)\|_{L^p(\Omega)} \big) \cr 
& \overset{\eqref{Morrey}}{\leq} &  2C C_M \|\Gamma(y)\|_{W^{1,p}(\Omega)},
\end{eqnarray}
by Morrey's inequality. Defining now $C_0$ in terms of the initial connection in $y$-coordinates as
\beq \label{def_C0}
C_0 \equiv 2C C_M \big( \|\Gamma(y)\|_{W^{1,p}(\Omega)} + \|d\Gamma(y)\|_{W^{1,p}(\Omega)} \big),
\eeq
which is independent of $\epsilon$, \eqref{rescaling_techeqn5} implies that
\beq \nonumber
\|\Gamma^*\|_{W^{1,p}(\Omega)} \leq C_0. 
\eeq
Likewise, applying \eqref{rescaling_techeqn4b} component-wise to $d\Gamma^*$, we obtain
\begin{eqnarray} \label{rescaling_techeqn6}
\|d\Gamma^*\|_{W^{1,p}(\Omega)} \leq \|d\Gamma\|_{W^{1,p}(\Omega)}  
\leq  C_0.
\end{eqnarray}
Combining \eqref{rescaling_techeqn5} with \eqref{rescaling_techeqn6} gives the sought after bound \eqref{Gamma-bound} for $m=1$. 

The general case $m\geq 1$, follows similarly by applying \eqref{rescaling_techeqn4b} component-wise to higher derivatives, $\partial^l\Gamma^*$ and $\partial^l(d\Gamma^*)$ for $l=0,...,m-1$, (keeping in mind that these terms are H\"older continuous, where $\partial^l$ shall be understood as standard multi-index notation), and defining $C_0$ in \eqref{def_C0} in terms of the $W^{m,p}$-norm of $\Gamma$ and $d\Gamma$. To summarize, we proved that one can always arrange for the smallness assumption \eqref{small_Gamma} - \eqref{Gamma-bound} required in Theorem \ref{Thm3}, by first restricting a given connection to a ball of radius $\epsilon$, and taking the transformation of this connection to the ball of radius $1$ as the starting connection in Theorem \ref{Thm3}, while taking for $\Gamma^*$ the scalar transformed components of the restricted connection.

Finally, observe that the $\epsilon$-bounds \eqref{epsilon_bound} and \eqref{epsilon_bound_2} depend only on the constants $C_M$, $C_0$, $C_s$ and $C_e$, which in turn depend only on $m,n,p$ and $\Omega$. Since $\Omega=B_1(0)$ is kept fixed throughout the argument, we can first choose some $\epsilon$ small enough to satisfy the bounds \eqref{epsilon_bound} and \eqref{epsilon_bound_2}, and then arrange for the scaling \eqref{small_Gamma} for $\Gamma$ by applying the argument \eqref{rescaling_techeqn1} - \eqref{rescaling_techeqn6}. In summary, we proved that the hypotheses of Theorem \ref{Thm3} are satisfied, which completes the proof of Theorem \ref{ThmMain}.
\hfill $\Box$

\section{Applications to the Initial Value Problem in General Relativity}   \label{Sec_appl}

\subsection{Optimal Regularity and the Initial Value Problem}

The Einstein equations $G=\kappa T$ of General Relativity are covariant tensorial equations defined independent of coordinates.  The unknowns in the equations are the metric tensor $g$, and these are coupled to the variables which determine the sources in $T$.   For example, in the case of a perfect fluid, the unknowns are $g_{ij},\rho,p,u_i$, where \cite{HawkingEllis,Choquet}
$$
T=(\rho+p)u^iu^j+pg^{ij}.
$$
The existence of solutions of the Einstein equations are established by PDE methods in coordinate systems in which the Einstein equations take on a solvable form.   The coordinate systems are typically specified by an ansatz for the metric, for example, SSC coordinates for spherically symmetric spacetimes, or harmonic coordinates, wave-gauge coordinates, etc., for the general initial value problem in four dimensions, \cite{HawkingEllis,Choquet}.    Since solutions typically only exist locally in GR, it is important to know whether the breakdown is simply a breakdown of the coordinate system.  This is important both to the theory of the initial value problem in GR, and to numerical relativity.    The question we ask here is:  how do we know the gravitational metric, which is the solution of the equations in a given coordinate system, exhibits its optimal smoothness in the coordinate system in which it is constructed?

For example, assume that one were to construct a solution to the Einstein equations $G=\kappa T$ in a given coordinate system $x$ in which the equations produce unique solutions (locally) within a given smoothness class, starting from initial data.  To make the point, assume the equations produce solutions of optimal smoothness with metric $g\in W^{m+2,p}$, connection $\Gamma\in W^{m+1,p}$,  and $Riem(\Gamma)\in W^{m,p}$. Then application of a transformation $x\to y$ with Jacobian $J\in W^{m+1,p}$ will in general lower the regularity of the whole solution space, lowering the regularity of the metric and its connection $\Gamma$ by one order, but the transformation will preserve the regularity of the curvature tensor $Riem(\Gamma)\in W^{m,p},$  because the connection involves {\it derivatives} of the Jacobian of the coordinate transformation, but the metric and Riemann curvature tensor, being tensors, involve only the undifferentiated Jacobian.\footnote{Alternatively, the anti-symmetric operator $d$ applied to the symmetric leading order term in the formula for the transformed connection, kills the highest order derivatives in the formula for the transformed curvature tensor.}   Therefore, if one were to then express the Einstein equations in the transformed coordinates $y$ in which the metric is one order less smooth than optimal, the resulting existence theory posed in $y$-coordinates, {\it by construction}, would produce the unique transformed solution $g\in W^{m+1,p}$, $\Gamma\in W^{m,p}$, and $Riem(\Gamma)\in W^{m,p}$.   Therefore, and this is the main point, {\it if} we were to construct our solutions in the $y$-coordinates {\it in the first place}, then we would not know that our unique solution was one order below optimal smoothness without knowing about the existence of the inverse transformation $y\to x$.   It is precisely the existence of this transformation from $y$ back to $x$ that is guaranteed by Theorem \ref{Cor1}, because its existence follows from existence for the RT-equations for $\Gamma\in W^{m,p}$, $d\Gamma\in W^{m,p}$,  $m\geq1$, $p>n$.   Theorem \ref{Cor1} tells us that it is sufficient to solve the Einstein equations in a {\it weaker} sense than optimal, by stating that it is sufficient to solve a version of the Einstein equations which only produce metrics and connections one order less smooth than optimal.   If $\Gamma$ and $d\Gamma$ are in $L^{\infty}$, then this is the difference between weak and strong solutions in the true sense of the theory of distributions, \cite{ReintjesTemple1}.

With this in mind, consider as an alternative to solving the RT-equations,  the problems one would encounter in trying to prove that a non-optimal metric is smoothed by one order via application of hyperbolic PDE methods to the standard $3+1$ framework for the initial value problem in GR; an approach we believe to be too complicated in comparison to establishing optimal regularity via the (elliptic) RT-equations.\footnote{See \cite{Nardmann} for another example where the hyperbolic approach appears not feasible, and construction of an auxiliary Riemannian metric is central for extending results on the prescribed scalar curvature problem to Semi-Riemannian smooth manifolds; the results and methods in \cite{Nardmann} are not further related to ours.} The $3+1$ framework is based on foliating spacetime into spatial slices parameterized by a time variable.  We now argue that the $3+1$ framework replaces the problem of smoothing non-optimal metrics in spacetime, to the problem of smoothing the restrictions of the metric and second fundamental form by one order on space-like hypersurfaces as a necessary condition.   To make the point, recall that in wave coordinates the spacetime metric evolves from initial data surfaces by semi-linear wave equations and inherits its spacetime regularity from the regularity of the initial data, c.f. \cite{Choquet,Taylor3}. By this, the current $3+1$ hyperbolic PDE methods for the Einstein equations require the assumption that the induced metric be one derivative more regular than the second fundamental form, and deduce from this that the spacetime metric has the regularity of the induced metric on the Cauchy surface. For this to yield optimal regularity, the second fundamental form must be in addition one order more regular than the curvature as a necessary condition. Now the formula for the second fundamental form accounts for the embedding of the induced metric, and correspondingly its formula involves the connection coefficients from the ambient spacetime, so the second fundamental form, in general, inherits the regularity of the spacetime connection.  Thus without a procedure for finding a gauge condition and a Cauchy surface such that the induced metric and induced second fundamental form both have one more order of regularity than they exhibited in the original non-optimal spacetime coordinate system, the $3+1$ framework will estimate a non-optimal solution as being one order less regular than it really is.  Fixing this within the $3+1$ framework appears to us to be a formidable problem.   The difficulty is that although the induced metric on a $3$-surface is positive definite, and might be regularized using harmonic coordinates for that metric as in \cite{DeTurckKazdan},\footnote{Note that (positive definite) Riemannian metrics always exhibit optimal regularity in harmonic coordinates \cite{DeTurckKazdan}, because the regularity for the Laplacian can be deduced from the source terms, without requiring boundary data.}   the fact that the formula for the second fundamental form involves the spacetime connection, means the problem of regularizing the second fundamental form on a Cauchy hypersurface in a coordinate gauge that also regularizes the metric on that surface is as formidable as the problem of regularizing non-optimal spacetime metrics in the first place.   The main point is this:   Without a procedure for simultaneously regularizing the metric and second fundamental form on Cauchy surfaces, or a proof demonstrating that this holds on a Cauchy surface under the gauge condition assumed for a $3+1$ analysis, the Cauchy problem estimates non-optimal solutions as one order less regular than they really are, and in this sense, the Cauchy problem is incomplete in each Sobolev space.

\subsection{Application of the RT-equations in Spherically Symmetric Spacetimes}

We apply Theorem \ref{Cor1} to give a new theorem establishing the optimal smoothness of spherically symmetric solutions generated by the Einstein equations $G=\kappa T$ in Standard Schwarzschild Coordinates (SSC) with arbitrary source terms $T$.  The issues around optimal regularity addressed by the RT-equations are represented nicely in SSC coordinates because three of the four Einstein equations $G=\kappa T$ are first order in the metric, and thus metric solutions are only one order smoother than the curvature tensor.   We  begin with a discussion of the central issue involved.

The fact that the Einstein equations admit coordinate systems in which the metric is one degree less smooth than optimal, leads one to anticipate that the Einstein equations might be easier to solve at this lower level of smoothness.\footnote{Indeed, for elliptic equations, the Lax-Milgram Theorem is an example in which it is easier to establish the gain of one derivative in $u$ over $f$ in $\Delta u=f$, but the second derivative gain requires the development of elliptic regularity theory, \cite{Evans}.}    In certain cases, the Einstein equations might actually take their simplest form in  coordinate systems which produce only one metric derivative above the curvature tensor--because in coordinates where the metric is one order less smooth,  the equations need impose {\it fewer constraints}.   We now show that this is {\it precisely} what happens in spherically symmetric spacetimes in SSC, the example we now discuss in detail.

Consider then the case of time dependent spherically symmetric spacetimes in which the gravitational metric takes the general form 
\begin{eqnarray}\label{spheregeneral}
ds^2=-B(t,r)dt^2+\frac{dr^2}{A(t,r)}+E(t,r)dtdt+C(t,r)d\Omega^2,
\end{eqnarray}
where
$$
d\Omega^2=d\theta^2+\sin^2{\theta}d\theta^2
$$
is the standard line element on the unit sphere.    Then generically, when $C_r\neq0$,  (the most general case is not of interest here),  there exists a coordinate transformation to coordinates in which the metric takes the Standard Schwarzschild Coordinate form \cite{Weinberg}
\begin{eqnarray}
ds^2=-B(t,r)dt^2+\frac{dr^2}{A(t,r)}+r^2d\Omega^2,
\end{eqnarray}
and this represents the coordinates in which the Einstein equations (arguably) take their simplest form.\footnote{The authors invite the reader to put the metric ansatz into MAPLE to compute the Einstein equations in general case (\ref{spheregeneral}), to see that the equations are {\it significantly} more complicated in general coordinate systems than in SSC.}   In SSC, the Einstein equations reduce to a ``locally inertial'' formulation derived by Groah and Temple in \cite{GroahTemple} as follows.  

According to \cite{GroahTemple},  three of the four Einstein equations determined by $G=\kappa T$ are first order in $A$ and $B$, and one is second order.    The first order equations are equivalent to,\footnote{In \cite{GroahTemple}, the SSC metric ansatz is taken to be $ds^2=A(t,r)dt^2+B(t,r)dr^2+r^2d\Omega^2$, so to recover the formulas from \cite{GroahTemple}, make the substitutions $A\rightarrow\frac{1}{B}$, $B\rightarrow A$}  
 
\begin{eqnarray}\label{firstorder1}
\left\{-r\frac{A_r}{A}+\frac{1-A}{A}\right\}&=&\frac{\kappa B}{A}T^{00}r^2=\frac{\kappa}{A}T^{00}_Mr^2\\\label{firstorder2}
\frac{A_t}{A}&=&\frac{\kappa B}{A}T^{01}r=\kappa\sqrt{\frac{B}{A}}T^{01}_Mr\\\label{firstorder3}
\left\{r\frac{B_r}{B}-\frac{1-A}{A}\right\}&=&\frac{\kappa }{A^2}T^{11}r^2=\frac{\kappa }{A}T^{11}_Mr^2,
\end{eqnarray}
and the the two conservation laws $Div\, T=0$ are equivalent to
\begin{eqnarray}\label{conservation_law00}
\{T^{00}_M\}_{,0}+\left\{\sqrt{AB}T^{01}_M\right\}_{,1}=-\frac{2}{r}\sqrt{AB}T^{01}_M,\ \ \ \ \ \ \ \ \ \ \ \ \ \ \ \ \ \ \ \ \ \ \ \ \ \ \ \ \ \ \ \ \ \ 
\end{eqnarray}
\begin{eqnarray}\label{conservation_law01}
\{T^{01}_M\}_{,0}+\left\{\sqrt{AB}T^{11}_M\right\}_{,1}=
-\frac{1}{2}\sqrt{AB}\left\{ \frac{4}{x}T^{11}_M+\frac{(\frac{1}{A}-1)}{r}(T^{00}_M-T^{11}_M)\right.
\\\nonumber\left.+\frac{2\kappa r}{A}(T^{00}_MT^{11}_M-(T^{01}_M)^2)-4rT^{22}\right\},
\end{eqnarray}
where $T^{\alpha\beta}_M$ is the Minkowski stress tensor defined by, (c.f. \cite{GroahTemple}), 
$$
T^{00}_M=BT^{00},\ \ 
T^{01}_M=\sqrt{\frac{B}{A}}\,T^{01},\ \ 
T^{11}_M=\frac{1}{A}T^{11},\ \ T^{22}_M=T^{22},
$$
and we employ the standard notation 
$$\frac{\partial}{\partial t}\left\{\cdot\right\}=\left\{\cdot\right\}_{,0}=\left\{\cdot\right\}_{,t},\ \ \ \frac{\partial}{\partial r}\left\{\cdot\right\}=\left\{\cdot\right\}_{,1}=\left\{\cdot\right\}_{,r}.$$
By the Bianchi identities,  equations (\ref{firstorder1}) - (\ref{conservation_law01}) follow from $Div\, T=0$ which follows as an identity from $G=\kappa T$.   In \cite{GroahTemple} it was shown that the Einstein equations $G=\kappa\, T$ for metrics in SSC are  equivalent to the system (\ref{firstorder1}), (\ref{firstorder3}), (\ref{conservation_law00}), (\ref{conservation_law01}), in the weak sense when $T\in L^{\infty}$.   In addition, the system closes when an equation of state $p=p(\rho)$ is imposed, and the first order equation (\ref{firstorder2}) follows as an identity, (c.f. \cite{GroahTemple}).

The SSC equations (\ref{firstorder1}), (\ref{firstorder3}), (\ref{conservation_law00}), (\ref{conservation_law01}) were introduced in \cite{GroahTemple} to prove the first existence theorem for shock wave solutions of the Einstein equations using the Glimm scheme,   (c.f.\cite{SmollerTemple,Israel,ReintjesTemple1,VoglerTemple,LeFlochStewart}).  Groah and Temple remarked that the equations could only be solved in coordinates in which the metric appeared to be singular at shock waves, (in the sense that, although no delta function sources appear in the $L^{\infty}$ curvature tensor, the metric is only Lipschitz continuous, and this is only one derivative smoother than the curvature).   It is still an open question whether these $C^{0,1}$ metric solutions of $G=\kappa T$ can always be smoothed one order to $C^{1,1}$ by coordinate transformation, and based on this, authors in \cite{ReintjesTemple1,ReintjesTemple_wave},  posed the problem of {\it Regularity Singularities}.   

As an application of Theorem \ref{Cor1}, note that if $T\in W^{m,p}$, $m\geq1$, $p>n=4$, then solutions of (\ref{firstorder1}), (\ref{firstorder3}), (\ref{conservation_law00}), (\ref{conservation_law01}), would in general have $(A,B)\in W^{m+1,p}$,  $\Gamma\in W^{m,p}$, and since $G=\kappa T$, also $G\in W^{m,p}.$    Putting the full Riemann curvature tensor into a computer algebra system (MAPLE or Mathematica) one sees by inspection that the terms of lowest regularity in $G$ match the terms of lowest regularity in $Riem(\Gamma)$, so in general, $Riem(\Gamma)\in W^{m,p}$.  For such solutions of the SSC equations, we have that $\Gamma$ and $d\Gamma$ have the same regularity $W^{m,p}$, and the metric $g\in W^{m+1,p}$ is only one derivative more regular.   Thus solutions of the SSC equations with $T\in W^{m,p}$, $m\geq1$, $p>4$, is an example that fits the assumptions of Theorem \ref{Cor1}.   The result is a new regularity result for solutions of the SSC equations which we record in the following theorem:

\begin{Thm}
Assume $T\in W^{m,p}$, $m\geq1$, $p>4$, and let $g\equiv(A,B)$ be a solution of the SSC equations (\ref{firstorder1}), (\ref{firstorder3}), (\ref{conservation_law00}), (\ref{conservation_law01}) satisfying
$$
g\in W^{m+1,p},\ \ \Gamma\in W^{m,p},\ \ d\Gamma\in W^{m,p},
$$
in an open set $\Omega$.   Then for each $q\in\Omega$ there exists a coordinate transformation $x\to y$ defined in a neighborhood of $q$, such that, in $y$-coordinates, $g\in W^{m+2,p},$ $\Gamma\in W^{m+1,p}$, $Riem(\Gamma)\in W^{m,p}$.
\end{Thm}

\appendix

\section{Proof of elliptic estimate \eqref{Poissonelliptic_estimate_Lp}}   \label{Sec_appendix_ell}

For completeness, we now give a proof of estimate \eqref{Poissonelliptic_estimate_Lp} of Theorem \ref{Thm_elliptic_reg} stated in Section \ref{Sec_Prelim}.\\

\noindent {\bf Theorem 2.1. (Elliptic Regularity):} 
{\it  For $m\geq 1$, $1<p<\infty$, let $f\in W^{m-1,p}(\Omega)$ and $u_0 \in W^{m+1,p}(\Omega)$, which we assume to both be scalar functions. Assume $u \in W^{m+1,p}(\Omega)$ solves the Poisson equation $\Delta u = f$ with Dirichlet data $u_0$ in the sense that $u - u_0 \in W^{1,p}_0(\Omega)$.\footnote{The space $W^{1,p}_0(\Omega)$ denotes the closure of $C_0^\infty(\Omega)$, the space of smooth functions with compact support, with respect to the $W^{1,p}$-norm.}  Then there exists a constant $C>0$ depending only on $\Omega$, $m,n,p$ such that 
\beq \label{Poissonelliptic_estimate_Lp_app}
\| u \|_{W^{m+1,p}(\Omega)} \leq C \Big( \| f \|_{W^{m-1,p}(\Omega)} +  \| u_0 \|_{W^{m+1,p}(\Omega)} \Big).
\eeq }

\Proof
Assume that $w \in W^{m+1,p}(\Omega)\cap W^{1,p}_0(\Omega)$ solves $\Delta w =f$, Lemma 9.17 in \cite{GilbargTrudinger} then gives us the estimate
\beq \label{GilTru_1}
\|w\|_{W^{2,p}(\Omega)} \leq C \|f\|_{L^p(\Omega)}
\eeq
for some constant $C>0$ depending only on $\Omega$, $m,n,p$. Estimate \eqref{GilTru_1} directly extends to higher derivatives by differentiation. Namely, using that $D^\alpha w$ solves $\Delta D^\alpha w = D^\alpha f$, where $\alpha$ is a muti-index and $D^\alpha$ the corresponding derivative, we obtain
\begin{eqnarray} \label{GilTru_2}
\|w\|_{W^{m+1,p}(\Omega)} 
&\leq & \sum\limits_{0\leq |\alpha|\leq m-1}\|D^\alpha w\|_{W^{2,p}(\Omega)}  \cr
&\leq & C \sum\limits_{0\leq |\alpha|\leq m-1} \|D^\alpha f\|_{L^p(\Omega)}  \cr
& = &  C \|f\|_{W^{m-1,p}(\Omega)},
\end{eqnarray}
which proves \eqref{Poissonelliptic_estimate_Lp_app} in the special case $u_0=0$.  

To extend this to general Dirichlet data, $u_0 \neq 0$, let $v$ be the solution of the Laplace equation with Dirichlet data $u_0$, that is, $v$ solves $\Delta v=0$ and $v-u_0 \in W^{1,p}_0(\Omega)$ and can be constructed via Green's representation formula. By uniqueness of solutions to the Poisson equation it follows that $u=w+v$, since $\Delta u = \Delta w=f$ together with the correct Dirichlet data $u-u_0 \in W^{1,p}_0(\Omega)$. Now, using the triangle inequality twice gives us
\begin{eqnarray} \label{GilTru_3}
\|u\|_{W^{m+1,p}}  
& \leq &  \|w\|_{W^{m+1,p}}  + \|v-u_0\|_{W^{m+1,p}}  + \|u_0\|_{W^{m+1,p}}. 
\end{eqnarray}
We can apply estimate \eqref{GilTru_2} to the first two terms, since $w$ and $v-u_0$ both vanish on the boundary in the sense that $w,\: v-u_0 \in W^{1,p}(\Omega)$, which gives us 
\begin{eqnarray} \label{GilTru_4}
\|w\|_{W^{m+1,p}}    & \leq &  C \|f\|_{W^{m-1,p}(\Omega)},  
\end{eqnarray}
and
\begin{eqnarray} \label{GilTru_5}
\|v-u_0\|_{W^{m+1,p}(\Omega)} & \leq & C \|\Delta(v-u_0)\|_{W^{m-1,p}(\Omega)} \cr
& \leq & C \|u_0\|_{W^{m+1,p}(\Omega)},
\end{eqnarray}
where $\Delta(v-u_0) = \Delta u_0$ and $\|\Delta u_0\|_{W^{m-1,p}(\Omega)} \leq \|u_0\|_{W^{m+1,p}(\Omega)}$ imply the last inequality. Substitution of estimates \eqref{GilTru_4} and \eqref{GilTru_5} into \eqref{GilTru_3} yields now the sought after estimate \eqref{Poissonelliptic_estimate_Lp_app} and completes the proof.
\QED

\section*{Conclusions}  
Authors began the study of Regularity Singularities by asking whether Lipschitz shock waves proven to exist in Standard Schwarzschild coordinates, might actually be one order smoother in other coordinate systems in which the Einstein equations are too complicated to solve.   This has led us to a much more general theory of non-optimal solutions of the Einstein equations, and the authors now conjecture that without resolving the problem of optimal regularity, the existence theory for the initial value problem in GR is incomplete in each Sobolev Space.   Although fundamental to the theory of the Einstein equations, this appears to be a new point of view on the initial value problem for the Einstein equations.   Extending this theory to the case of shock waves in which the gravitational metric is only Lipschitz continuous, is the topic of authors current research.

\section*{Acknowledgements}

The authors thank Jos\'e Natario and the Instituto Superior T\'ecnico in Lisbon for supporting this research and funding two visits of the second author. The authors thank Craig Evans for suggesting reference \cite{Dac}, and for pointing out the Calderon-Zygmund singularities. We thank John Hunter, Pedro Gir\~ao, Steve Shkoller and Kevin Luli for helpful discussions. M.R. is grateful to IMPA (Rio de Janeiro, Brazil) for their hospitality in February 2018, where parts of this paper were written.  Finally, we thank Joel Hass for writing a nice public science article about our research program in the 2018 UC-Davis yearly Mathematics Newsletter.

\providecommand{\bysame}{\leavevmode\hbox to3em{\hrulefill}\thinspace}
\providecommand{\MR}{\relax\ifhmode\unskip\space\fi MR }
\providecommand{\MRhref}[2]{%
  \href{http://www.ams.org/mathscinet-getitem?mr=#1}{#2}
}
\providecommand{\href}[2]{#2}

\end{document}